\documentclass[aps,pra,10pt,twocolumn,groupedaddress,notitlepage,showpacs,floatfix]{revtex4-1}
\pdfoutput=1
\usepackage{graphicx,graphics,epsfig,subfigure,times,bm,bbm,amssymb,amsmath,amsfonts,amsthm,mathrsfs,MnSymbol}
\usepackage{gensymb}
\usepackage[matrix,frame,arrow]{xypic}
\usepackage[pdfstartview=FitH]{hyperref}
\usepackage{subfigure}
\hypersetup{
    colorlinks=true,       
    linkcolor=red,          
   citecolor=magenta,        
    filecolor=magenta,      
    urlcolor=cyan,           
    runcolor=cyan
}
\usepackage[pdftex]{color}
\usepackage{braket}
\usepackage{enumerate}
\usepackage[normalem]{ulem}
\usepackage[usenames,dvipsnames]{xcolor}
\usepackage{multirow}
\usepackage{mathtools}
\DeclarePairedDelimiter{\ceil}{\lceil}{\rceil}

\definecolor{orange}{rgb}{1,0.5,0}

\newcommand{\bes} {\begin{subequations}}
\newcommand{\ees} {\end{subequations}}
\newcommand{\bea} {\begin{eqnarray}}
\newcommand{\eea} {\end{eqnarray}}

\definecolor{gold}{rgb}{0.85,.66,0}

\newcommand{\beq}{\begin{equation}}

\newcommand{\eeq}{\end{equation}}

\newcommand{\ignore}[1]{}

\def\a{\alpha}

\def\g{\gamma}
\def\d{\delta}

\def\s{\sigma}

\def\>{\rangle}
\def\<{\langle}

\newcommand{\ident}{\mathbb{I}}

\def\s0{I}
\def\sx{\sigma^x}

\def\sz{\sigma^z}

\newcommand{\ig}[1]{}
\newcommand{\ak}{\texttt{akmaxsat}\,}

\begin{document}

\title{MAX 2-SAT with up to 108 qubits}
\author{Siddhartha Santra}
\author{Gregory Quiroz}
\affiliation{Department of Physics and Center for Quantum Information Science \&
Technology, University of Southern California, Los Angeles, California
90089, USA}
\author{Greg Ver Steeg}
\affiliation{Information Sciences Institute, University of Southern California, Marina del Rey, California, USA}
\affiliation{Department of Computer Science and Center for Quantum Information Science \&
Technology, University of Southern California, Los Angeles, California
90089, USA}

\author{Daniel A. Lidar}
\affiliation{Departments of Electrical Engineering, Chemistry, and Physics, and Center for Quantum Information Science \& Technology, University of Southern California, Los Angeles, California 90089, USA}
\begin{abstract}
We experimentally study the performance of a programmable quantum annealing processor, the D-Wave One (DW1) with up to $108$ qubits, on maximum satisfiability problem with $2$ variables per clause (MAX 2-SAT) problems. We consider ensembles of random problems characterized by a fixed clause density, an order parameter which we tune through its critical value in our experiments.  We demonstrate that the DW1 is sensitive to the critical value of the clause density. The DW1 results are verified and compared with \texttt{akmaxsat}, an exact, state-of-the-art algorithm. We study the relative performance of the two solvers and how they correlate in terms of problem hardness. We find that the DW1 performance scales more favorably with problem size and that problem hardness correlation is essentially non-existent. We discuss the relevance and limitations of such a comparison.
\end{abstract}
\maketitle

\section{Introduction}
Adiabatic quantum computation (AQC) is a model of solving computational problems, in particular hard optimization problems, by evolving a closed system in the ground state manifold of an adiabatic Hamiltonian $H(t)$ with $t\in[0,t_f]$ \cite{FarhiAQC:00,Farhi:01}. The ground state of the beginning Hamiltonian $H_B=H_\textrm{ad}(0)$ is assumed to be easily prepared, while the ground state of the problem Hamiltonian $H_P=H_\textrm{ad}(t_f)$, represents the solution to the computational problem. AQC has been proven to be polynomially equivalent to standard, closed-system, circuit model QC \cite{Aharonov:04,kempe:1070,Siu:2005:062314,Oliveira:05,PhysRevLett.99.070502}, but so far it is unclear whether this equivalence extends to the open system, non-zero temperature setting. There is some theoretical evidence of inherent robustness of open system AQC \cite{PhysRevA.65.012322,SarandyLidar:05,PhysRevA.71.032330,PhysRevA.75.062313,PhysRevA.79.022107,TAQC} and scalability using currently available technology. A case in point are the D-Wave processors, comprised of superconducting rf SQUID flux qubits \cite{Harris2010}. Recent experimental evidence \cite{Johnson:2011ys,DWave-16q,q-sig,q100,q100-comment} suggests that the first commercial generation D-Wave One (DW1) ``Rainier" processor (with up to $128$ qubits) implements physical quantum annealing (QA) \cite{Brooke30041999}, a non-zero temperature, non-universal form of AQC, whose algorithmic performance has been extensively discussed in the literature \cite{finnila_quantum_1994,Kadowaki1998,Santoro,RevModPhys.80.1061,PhysRevLett.101.170503,sqa1,Battaglia:2005fk,morita:125210}. Quantum annealing can also be understood as the quantum-mechanical version of the simulated annealing
(SA) \cite{Kirkpatrick1983} algorithm for optimization problems. While SA employs the slow annealing
of (classical) thermal fluctuations to converge on the ground state manifold, QA additionally uses quantum
fluctuations. There is extensive numerical \cite{Santoro,RevModPhys.80.1061,PhysRevLett.101.170503} as well as analytical \cite{morita:125210} evidence which shows that
QA can be more efficient than SA for the problem of finding ground states of classical Ising-type
Hamiltonians. 

In this work we experimentally study the performance of physical QA, using the DW1 processor, on MAX 2-SAT optimization problems (maximum satisfiability problem with two variables per clause) \cite{mezmont}. We examine both the scaling with problem size and the classical phase transition in problem hardness as a function of the clause density, i.e., the ratio of the number of clauses to variables. 
The clause density is related to computational complexity, is associated with rigorous bounds, and is a natural order parameter for random MAX 2-SAT, as the problem exhibits a ``hardness" phase transition at a critical value $\a_c=1$ \cite{Coppersmith,FernandezdelaVega2001131}. 
We present evidence for this transition in our DW1 experiments. Thus the clause density serves as a tunable hardness parameter for analyzing performance that is specific to the MAX 2-SAT problem.

\begin{figure}[t]
 \centering
\vspace{-1.0cm}
\hspace*{0.3cm}\includegraphics[width=\linewidth]{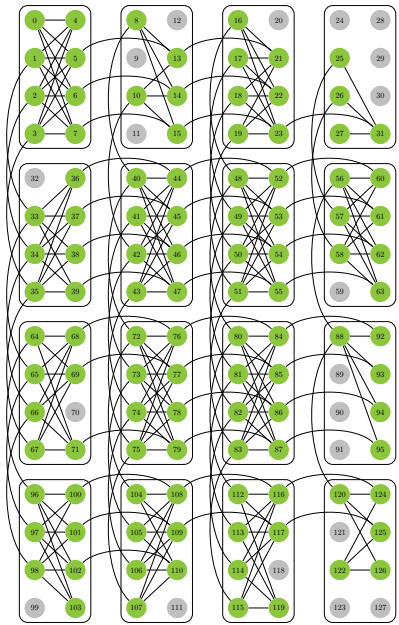}
\vspace{-2.0cm}
 \caption{(color online) The D-Wave
One Rainer DW1 processor consists of $4\times4$ unit cells of eight qubits (circles), connected by programmable inductive couplers (lines). The $108$ green (grey) circles denote functional (inactive) qubits. Most qubits connect to six other qubits. In the ideal case, where all qubits are functional and all couplers are present (as in the central four unit cells), one obtains the non-planar ``Chimera" connectivity graph of the DW1 processor \cite{Choi1,Choi2}.}
  \label{conn}
\end{figure}

One might hope to be able to detect 
a quantum speedup by comparing physical QA to highly optimized classical solvers. 
While recent work \cite{McGeoch} attempted to show that the latest generation of D-Wave processors (the D-Wave Two ``Vesuvius" processor, with up to $512$ qubits) could outperform the best classical solvers on random instances  of MAX 2-SAT
\footnote{In more detail, the McGeoch and Wang (MW) study \cite{McGeoch}, working with the DW2, used a $439$ qubit subgraph of Chimera and considered three problems: (1) Chimera-structured QUBO instances (this is actually an ensemble of uniform samples from the Ising model on Chimera with $J_{ij},h_j \in \{-1,1\}$ \cite{Selby}), (2) Weighted Max 2-SAT, (3) the Quadratic Assignment Problem. Their main conclusion is that in their experiments the DW2 (together with a software layer called Blackbox) outperformed the software against which it was tested. In the case of problem (1), the DW2 is reported as outperforming its nearest rival (CPLEX), amongst those tried, by a factor of $3600$. The times recorded by MW were for the first point that CPLEX found the optimal solution, and not the time at which it proved it optimality. However, several researchers have reported classical implementations for all three problems which outperform the DW2 and in particular the MW benchmarks \cite{Selby,Puget}. Our ensemble of MAX 2-SAT problems differs from the weighted MAX 2-SAT problems considered by MW, since we used uniform weights with the additional constraint of a fixed clause density. In addition, unlike MW's case (2), our MAX 2-SAT problem ensemble inherits the native Chimera graph structure along with the connectivity contraints of the processor by design, which eliminates the need for using the Blackbox layer (that uses conventional heuristics along with hardware queries), resulting in a more transparent comparison. While we do not tune the implementation of \texttt{akmaxsat} to the structure of processor-compatible problems, this lets us compare its performance on truly random and processor-compatible problems without any bias. Comparison of our results to the native (Chimera-structured) case (1) studied by MW is not straightforward because the resulting Ising ensemble does not correspond to MAX 2-SAT with a fixed clause density. Moreover, our focus is not on the absolute time to solution (which is somewhat arbitrary anyway since it depends on variables such as the type of processor, its clock speed, and the number of cores), but rather on the scaling with problem size.}, 
concurrent results already demonstrated a classical stochastic solver outperforming the DW1 processor for an ensemble of random Ising spin glass problems with a native embedding on the Chimera graph~\cite{q100} (see Fig.~\ref{conn}). 
Nevertheless, the competitive nature of the results along with the mounting evidence of quantum phenomena~\cite{Johnson:2011ys,DWave-16q,q-sig,q100,q100-comment} suggests the intriguing but currently unproven possibility that the D-Wave quantum annealing architecture may some day be capable of outperforming any classical solver for some ensembles of problems, though it seems inevitable that some form of error correction will eventually be required \cite{jordan2006error,PhysRevLett.100.160506,PhysRevA.86.042333,PhysRevLett.108.080501,2012arXiv1208.6371Y}. Our study attempts to shed light on this possibility by studying a random ensemble of MAX 2-SAT problems characterized by a given clause density $\a$. We verify the empirical solutions of the MAX 2-SAT problem obtained by the DW1 using \texttt{akmaxsat} \cite{akmaxsat}, a state-of-the-art, exact branch and bound algorithm, which we also use as a performance benchmark. We find that the DW1 and \ak exhibit not only distinct scaling, but also very different sensitivities to problem hardness. In fact, we  show that over the random ensemble that is characterized by a fixed clause density and is compatible with the Chimera graph, the DW1 has a scaling with problem size that is better than \texttt{akmaxsat}'s, and there is no correlation between the two solvers in terms of problem hardness. 

However, we hasten to point out that our comparison is not unproblematic. The same work that most definitively established that the DW1 performs quantum annealing \cite{q100} also found that simulated annealing \cite{Kirkpatrick1983} was significantly faster than the DW1 on its ensemble of random Ising spin glass instances. While our ensemble is different, this does suggest that stochastic classical algorithms such as simulated annealing, rather than exact and deterministic ones such as \texttt{akmaxsat}, are the better benchmark. 
Moreover, Ref.~\cite{q100} also found that other exact, deterministic classical solvers scale better than \texttt{akmaxsat} for its ensemble of random spin glass problems \footnote{Specifically, the biqmac algorithm \cite{biqmac} used in the spin glass server \cite{sgserver}, exact belief propagation using bucket sort \cite{dechter1999bucket} and a related divide-and-conquer algorithm.}. Our study does therefore certainly not settle the question of a quantum speedup for physical QA, but should rather be seen as a first indication that MAX 2-SAT is an interesting candidate for such a speedup, when perceived through the lens of fixed clause density ensembles. Follow-up studies using the $512$-qubit D-Wave Two will shed more light on the scaling question.

The structure of the paper is as follows: in Section~\ref{sec:2SAT-background} we provide pertinent background on the MAX 2-SAT optimization problem and its phase transition as a function of the clause density. In Section~\ref{sec:2SAT-Ising} we briefly review the DW1, describe the procedure for mapping MAX 2-SAT instances into Ising problems, and define the restricted ensemble of DW1-compatible problems. Section~\ref{sec:results} presents our results: we compare the MAX 2-SAT success probabilities and time to solution using the DW1 processor and \texttt{akmaxsat}, discuss the evidence for a transition at $\alpha_c$, and study the problem hardness correlation between the DW1 and \texttt{akmaxsat}. We present our conclusions and discuss scope for future work in Section~\ref{sec:concl}. Various supplementary details are presented in the Appendices.

\section{MAX 2-SAT Background}
\label{sec:2SAT-background}

In this section we briefly review pertinent theoretical background concerning the random MAX 2-SAT problem and its solution methods. We focus on random ensembles characterized by a fixed clause density.

\subsection{2-SAT and MAX 2-SAT}
\label{subsec:2sat_probs}
Many multivariate problems of practical interest involve determining values of variables $\{x_i\}_{i=1}^N$, collectively called a configuration $x$, that extremise the value of some objective. Satisfiability (SAT) problems are one such class of problems defined in terms of $N$ Boolean variables and a set of $M$ constraints between them where each constraint takes the special form of a clause. An example of a  problem instance of a 2-SAT problem, written in Conjunctive Normal Form (CNF), and also called a formula, is:
\begin{align}
F({x})&:=
{(x_1\lor x_2)\land(x_1\lor \neg x_4)
\land\cdots \land(x_N\lor x_4)}
\ ,
\label{satinst}
\end{align}
which is the logical AND of $M$ clauses, where each clause itself is a logical OR of exactly two literals, defined as a variable or its negation.  A clause that evaluates to {\sc TRUE} ({\sc FALSE}) is called SAT (UNSAT). The 2-SAT problem is a \emph{decision problem} of determining whether there exists a collection of Boolean values for $x$ such that Eq.~\eqref{satinst} evaluates to {\sc TRUE}, in which case the formula is said to be satisfiable. A related \emph{optimization problem} known as MAX 2-SAT is to find the variable assignment which maximizes the number of satisfied clauses. While 2-SAT is in the complexity class P, i.e., it admits a polynomial (in fact linear \cite{lin-2SAT}) time solution, MAX 2-SAT, like 3-SAT, is NP-complete \cite{Papadimitriou:book}.

\subsection{Clause density and phase transition}
\label{sec:PT}

In this work we are interested in random MAX 2-SAT, where 2-SAT problem instances are generated by choosing uniformly at random, with fixed clause density, from among all possible clauses, without clause repetition (the same clause may not appear twice). The clause density is 
\beq
\a=M/N .
\eeq
As the number of clauses grows at a fixed number of variables, it becomes harder to satisfy all clauses. Indeed, the probability of satisfiability versus the clause density $\a=M/N$ exhibits a phase transition at a critical value $\a_c=1$ in the thermodynamic limit ($N\rightarrow \infty$), whose finite-size scaling has also been studied \cite{Bollobas:2001,Goerdt1996469}. Thus, the clause density is an order parameter.
An intuitive explanation for the specific value of $\a_c$ is that for each clause only one of the two variables needs to be {\sc TRUE}. Therefore when there are $N$ variables and $N$ clauses, one essentially uses up one variable to satisfy each clause.
 
The phase structure of MAX 2-SAT has also been extensively studied. Coppersmith \textit{et  al.} proved that MAX 2-SAT exhibits a phase transition at the same critical clause density as 2-SAT \cite{Coppersmith}. Namely, in the large $N$ limit, to the left of the critical clause density the maximum fraction of clauses that are satisfiable is almost always $1$, while to the right  a fraction of all clauses are unsatisfiable. In the large clause density limit, the fraction of clauses that are satisfiable is almost surely the same as that expected from a random assignment, $3/4$. 
Near the phase transition, we have the most uncertainty about the correct fraction of satisfied clauses.
For k-SAT with $k>2$, this type of uncertainty near the phase transition has been linked to the appearance of an easy-hard-easy pattern of computational complexity for backtracking solvers~\cite{monasson}. 
The relevance of the phase structure for the difficulty of the decision version of MAX 2-SAT has been empirically confirmed by Shen and Zhag~\cite{shen2003empirical}. 
The finite-size scaling window of the MAX 2-SAT phase transition has a clause density width of $\Theta(N^{-1/3})$ \cite{Coppersmith}.

These and other results~\cite{mezmont} suggest that it is natural to explore random MAX 2-SAT ensembles characterized by a fixed clause density, and this is our focus here.

\subsection{Polynomial time approximation scheme}
\label{sec:PTAS}
Many NP-complete problems can be \emph{approximately} solved to arbitrary precision in polynomial time. 
A $\rho$-polynomial time approximation scheme ($\rho$-PTAS) for MAX 2-SAT is an algorithm that provides an assignment of variables that provably satisfies a number of clauses within at least a fraction $\rho$ of the maximum number of clauses that can be satisfied for any formula. Goemans and Williamson~\cite{maxcut} demonstrated a $\rho$-PTAS for MAX 2-SAT with $\rho \approx 0.87$, and an improved version of their result achieves $\rho \approx 0.94$~\cite{lewin}.  On the other hand, it has also been shown that no $\rho$-PTAS exists for $\rho > 21/22 \approx 0.95$ unless P$=$NP~\cite{inapprox}. Thus, in the worst case, it is not only difficult to find an assignment that satisfies the maximum number of clauses, it is also difficult to find an assignment that comes close. We shall return to the $\rho$-PTAS issue from an experimental perspective in Section~\ref{sec:concl}.

\subsection{Optimized classical numerical solvers}
Optimized exact classical MAX 2-SAT solvers have been extensively studied and regularly compete in an annual competition~\cite{maxsatcomp}. Here by ``exact" we mean that the solver is guaranteed to eventually return a correct answer. The basic idea behind the most successful exact solvers is to combine a branch and bound algorithm that searches the (exponentially large) tree of possible assignments~\cite{dpl} with heuristics to improve performance. Improvements come about in two ways. First, branches of the search space are avoided by intelligently upper bounding the maximum number of clauses that can be satisfied in that branch. Second, heuristics are used to simplify a formula, reducing the number of clauses or variables. In this work we benchmark the DW1 processor against a recent MAX SAT competition winner, \texttt{akmaxsat} \cite{akmaxsat}, that incorporates all of these techniques. We motivate this choice in Section~\ref{sec:results}.

\section{Ensembles of 2-SAT problems and their restriction to the DW1 processor}
\label{sec:2SAT-Ising}

We focus here on the average complexity for an ensemble of MAX 2-SAT problems characterized by a fixed clause density $\a$. The behavior of algorithms with respect to an ensemble may be taken to signify the \emph{typical} behavior when given a specific problem instance. We note that it is not known whether this typical behavior implies anything about the worst case complexity, i.e., MAX 2-SAT is not known to be random self-reducible, unlike  certain NP problems \cite{Feigenbaum:1993,Ajtai:1999}. 

\subsection{Quantum annealing using the DW1}

The DW1 implements the quantum annealing Hamiltonian:
\begin{equation}
H(t)=A(t) H_B + B(t) H_P\ ,\quad t\in[0,t_f]\ ,
\label{adbH}
\end{equation}
where the ``annealing schedules" $A(t)$ and $B(t)$ (shown in Appendix~\ref{app:DW}) are, respectively, monotonically decreasing and increasing functions of time, satisfying $A(0)\gg\textrm{max}(k_BT,B(0))$ and $B(t_f)\gg A(t_f)$.
 The beginning and problem Hamiltonians implemented on the DW1 correspond to a transverse-field, non-planar Ising model, i.e.,
\bes
\begin{align}
H_B &=\sum_{j\in V}\sx_j \\
H_P & =\sum_{j\in V}h_j \sigma^{z}_j+\sum_{(i,j)\in E}J_{ij}\sz_i\sz_j
\label{eq:Hp}
\end{align}
\ees
where $\sigma^{x(z)}_j$ represent the spin-1/2 Pauli matrices for the $j$th qubit. Thus to solve MAX 2-SAT problems using the DW1 we map these problems to the Ising model. In the DW1 the $N$ rf SQUID flux qubits occupy the vertices $V$ of the so-called ``Chimera" graph (see Fig.~\ref{conn}), with maximum degree $6$, and are coupled inductively along the edges $E$ of this graph. The local fields $h_j$ and the couplers $J_{ij}$ are programmable and once chosen they define a ``problem instance". Each ``annealing run" corresponds to evolving $H(t)$, with a preprogrammed and fixed set of local fields and couplers, from $t=0$ to a predetermined annealing time $t_f$, followed by a projective measurement of all qubits in the computational basis, i.e., the eigenbasis of the Ising Hamiltonian $H_P$. Each such measurement results in a spin configuration $\{s_1,\dots,s_N\}$, where $s_j = \pm 1$ is the eigenvalues of $\sz_j$. By repeating these annealing runs many times one builds up statistics of spin configurations for a given problem instance. The processor can then be reprogrammed to generate statistics for a new problem instance.

\subsection{Mapping MAX 2-SAT to equivalent Ising problems}
\label{sec:mapping}
In order to solve instances of MAX 2-SAT on the DW1, we must construct the problem Hamiltonian $H_P$ of Eq.~\eqref{eq:Hp}, such that the ground state configuration encodes the satisfying assignment for the problem instance. Following the prescription of  \cite{FarhiAQC:00} for the conversion of SAT problems to finding the ground state(s) of quantum Hamiltonians, as a first step we transform from Boolean to binary variables, letting {\sc TRUE}=0 and {\sc FALSE}=1, so that the truth table of the OR function becomes the multiplication of the binary variables. We next identify the binary variables $\{x_j\}_{j=1}^{N}$ of a 2-SAT formula with the $\pm1$ eigenstates of the Pauli spin operator $\sigma_j^z$ acting on qubit $j$, i.e.,
\beq
\sigma_j^z\ket{x_j} = (-1)^{x_j}\ket{x_j}\ , \quad x_j\in\{0,1\}\ .
\eeq
We also define variables $v_j^k \in\{-1,0,1\}$, where the indices $j=\{1,2,...,N\}$ and $k=\{1,2,...,M\}$ label the variables and clauses respectively, with $v_j^k=-1$ $(+1)$ if $x_j$ appears negated (unnegated) in the $k$th clause and $v_j^k=0$ for all clauses that $x_j$ does not appear in.  Each two-variable clause, $\Omega_k,~k\in\{1,2,\dots,M\}$, in an arbitrary 2-SAT formula $F=\Omega_1\wedge\Omega_2\wedge...\wedge\Omega_M$ is then translated into a corresponding 2-local term in the problem Hamiltonian of the form
\begin{equation}
H_{\Omega_k}=\frac{\ident-v^k_{j_1}\sz_{j_1}}{2}\frac{\ident-v^k_{j_2}\sz_{j_2}}{2}\ .
\label{eq:Homega}
\end{equation}
It is easy to check that in this manner if $\{x_{j_1},x_{j_2}\}\in\{0,1\}^2$ violate the clause then $H_{\Omega_k}$ is associated with an energy penalty of $1$, and zero otherwise. Rather than taking the logical AND of all clauses as in the original 2-SAT problem, the problem Hamiltonian is now constructed as  
\begin{equation}
H_P=\sum_{\Omega_k\in F} H_{\Omega_k}\ ,
\label{totalh}
\end{equation}
i.e., the sum of the energies of all $M$ clauses contained in the 2-SAT instance $F$. This means that the ground state of $H_P$ corresponds to the bit assignment that violates the minimal number of clauses, i.e., the ground state is the solution to the MAX 2-SAT problem for the problem instance $F$. 
A generic computational basis state of the system can be written as $\ket{\psi}=\ket{x_1x_2\dots x_N}$.
In the case that $\{x^*_j\}$ is a satisfying assignment for the formula $F$ we have the correspondence
\begin{align}
&F|_{\{x^*_j\}}=1\implies H_P\ket{\psi^*}=0 \ ,
\end{align}
where $\ket{\psi^*}=\ket{x^*_1x^*_2\dots x^*_N}$,  while all non-satisfying assignments correspond to computational states with positive energy. In the case of MAX 2-SAT the ground state might have a positive energy $E_{\textrm{min}}>0$ and the question becomes to determine the assignment $\ket{\psi}$ such that $H_P\ket{\psi}=E_{\textrm{min}}\ket{\psi}$.

Written out in detail, Eq.~\eqref{totalh} becomes
\begin{align}
H_P 
& = \frac{1}{4}\sum_{\Omega_k\in F} 2\ident - v^k_{j_1}\sz_{j_1} - v^k_{j_2}\sz_{j_2}+ v^k_{j_1}v^k_{j_2}\sz_{j_1}\sz_{j_2}\ ,
\label{eq:H_P-map}
 \end{align}
and upon equating with the problem Hamiltonian in Eq.~\eqref{eq:Hp}, after rescaling by a factor of $4$ and dropping the constant term, we obtain the local fields $h_j$ and the couplings $J_{ij}$ in terms of the parameters of the given MAX 2-SAT instance:
\begin{align}
h_{j_i} = -\sum_k v^k_{j_i}  \ , \quad J_{j_1 j_2} &= \sum_k v^k_{j_1}v^k_{j_2} \ ,
\label{eq:handJ}
 \end{align}
where $i\in\{1,2\}$ and the indices $j_1, j_2$ are the qubit indices on the Chimera graph.

\subsection{Restricted ensemble of DW1 processor-compatible 2-SAT problems}
\label{subsec:restrict_probs}

The DW1 process-compatible problem instances must satisfy a number of constraints, namely $N\leq 108$ and the DW1 Chimera graph connectivity. To account for these constraints we generated restricted ensembles $\mathcal{E}_{\textrm{DW}}(N,\a)$ with $13$ different numbers of variables and $20$ different clause density values, as follows:
\begin{itemize}
\item $N\in \{16, 24, 32, 39, 46, 53, 60, 67, 75, 80, 87, 98, 108\}$, and $\a=[0.1,2.0]$ in increments of $0.1$.
\item We define DW1 processor-compatible problem instances as those instances whose clauses are formed by two literals $x_{j_1},x_{j_2}$ which correspond to qubits $j_1,j_2$ on the Chimera graph $G=(E,V)$ that are active as well as coupled, i.e., $\{j_1,j_2\}\in V$ and $(j_1,j_2)\in E$. Recall that not all qubits are active (see Fig.~\ref{conn}).
\item When $\a <1/2$ there are more variables than can fit into the clauses. For $\a <1/2$, any variable that did not appear in a clause was not used in $H_P$. 
\item At each value of $N$ and $\a$ we generated an ensemble, $\mathcal{E}_{\textrm{DW}}(N,\a)$, of $500$ DW1 processor-compatible random 2-SAT problem instances. We excluded all instances involving identical clauses. In our ensembles we applied negation to each of the two variables representing the qubits uniformly at random.
\end{itemize}
We thus have a total of $13\times 20=260$ such ensembles $\mathcal{E}_{\textrm{DW}}(N,\a)$, with a total of $260\times 500=130,000$ DW1 processor-compatible problems.

To cover a range of interesting clause density values we used a maximum value of $2\a_c=2$, thus ensuring that our instances were all well within the range of the the finite-size scaling window of the phase transition, whose width is $\Theta(N^{-1/3})$ \cite{Coppersmith}: four our range $16 \leq N\leq 108$ we have $0.40 \geq N^{-1/3} \geq 0.21$. 
The maximum value of $\a$ supported by the DW1 processor is discussed in Appendix~\ref{app:max-alpha}.
We note that having enforced an equal probability for negated and unnegated variables somewhat restricts the hardness as it is known that an unbalanced probability of negation can lead to harder instances \cite{Austrin:2007:BMM:1250790.1250818}.

To test how well our DW1-restricted instances approximate an unrestricted random ensemble, we also generated ensembles, $\mathcal{E}(N,\a)$, of $1000$ random 2-SAT problems with a given number of variables and clause density and no constraint on the literals that comprise a single clause, except that no clauses are repeated. A comparison between $\mathcal{E}_{\textrm{DW}}(N,\a)$ and $\mathcal{E}(N,\a)$ is presented in the next section. 

\section{Experimental and Numerical Results}
\label{sec:results}

In this section we report on our experimental and numerical results for the ensembles of MAX 2-SAT problems described above. A complete description of the settings under which we ran the  \texttt{akmaxsat} algorithm is given in Appendix~\ref{app:ak}.

We compare the scaling of the solution time required by \texttt{akmaxsat} to the empirical probability of the correct ground state found by the DW1 processor, from which we compute an extrapolated time required to achieve a certain solution threshold accuracy. This is, of course, not entirely an ``apples-to-apples" comparison, since it compares an exact algorithm with a probabilistic machine. Moreover, there exist faster stochastic classical algorithms \cite{q100}. However, to check the proposed DW1 solutions for correctness an exact algorithm is required, and we decided to use \texttt{akmaxsat} due to its excellent performance on the benchmark problem sets of MAX 2-SAT used at the MAXSAT-2009 and MAXSAT-2010 evaluations for state-of-the-art MAX SAT solvers \cite{maxsatcomp}. We make no claims here as to the significance of our results in the larger context of whether experimental quantum annealing can outperform \emph{all} classical algorithms. Rather, we focus on the scaling comparison with a state-of-the-art exact classical algorithm, and on whether there is any correlation in problem hardness between this classical solver and the DW1. We find that the DW1 has a scaling that is better than that of \texttt{akmaxsat}, a performance gap which increases with the clause density, and that there is rapidly decreasing correlation between the DW1 and \texttt{akmaxsat} for the hardness of problem instances, as $\a$ increases. In addition we closely examine the behavior in the vicinity of the critical clause density.

\begin{figure}[tr]
 \centering
 \includegraphics[width=\columnwidth]{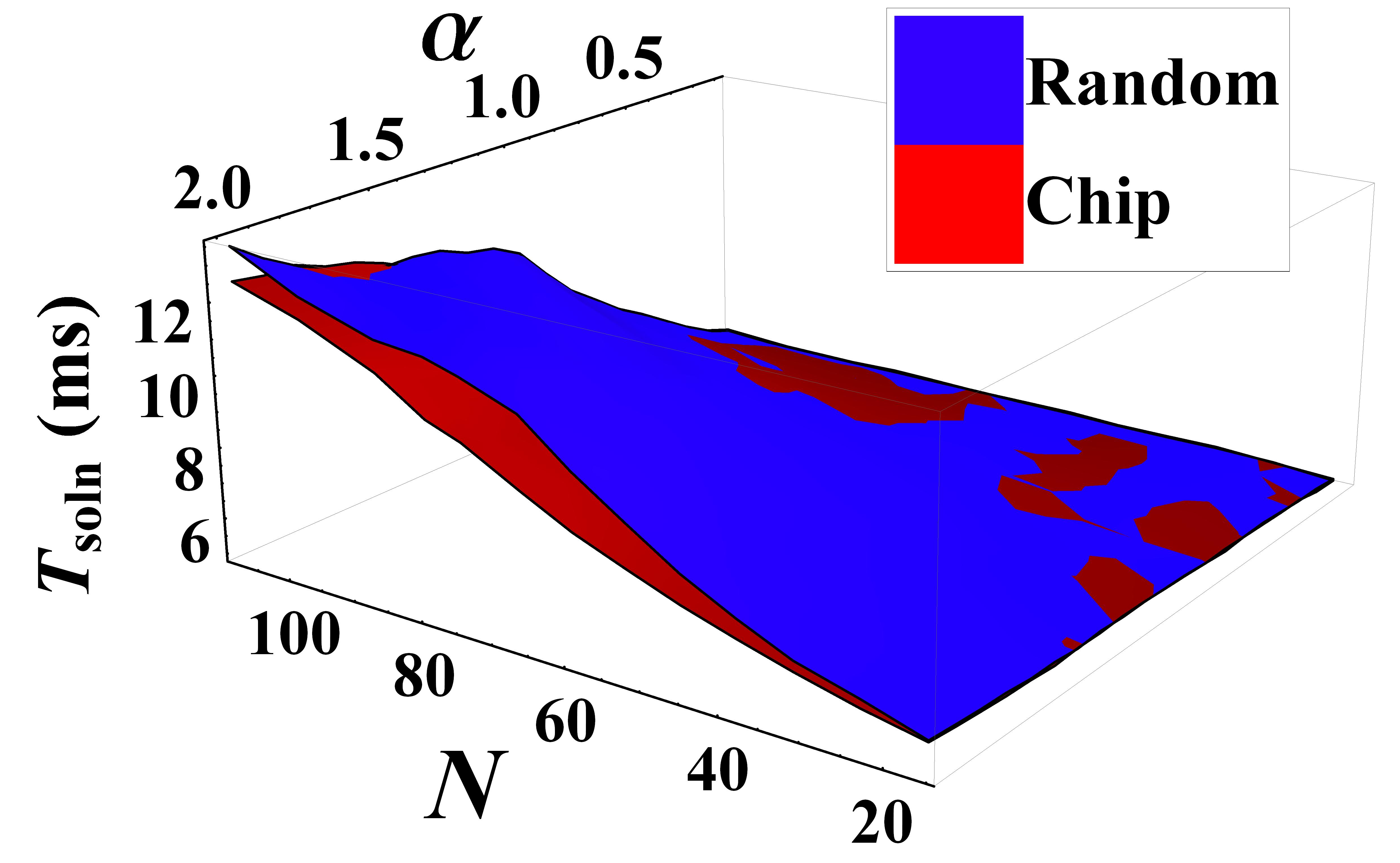}
 \caption{(color online) Computational effort comparison between an ensemble of random (blue surface) and DW1 processor-compatible (red surface) MAX 2-SAT instances for \texttt{akmaxsat}.  The time to solution is generally greater for the random ensemble; however, substantial differences between $T_\textrm{soln}$ for DW1 processor-compatible and random instances are not observed for a majority of the $\a$ and $N$ values considered. All results are averaged over $500$ instances of each ensemble and $10$ \texttt{akmaxsat} runs per data point.}
  \label{fig:ak_time}
\end{figure}

\subsection{Comparison of \texttt{akmaxsat} for the random and DW1-compatible ensembles}
\label{subsec:classical}

As noted above, the DW1 processor-compatible problem instances are restricted by the Chimera graph connectivity (see Fig.~\ref{conn}). To test how the resulting ensemble $\mathcal{E}_{\textrm{DW}}(N,\a)$ compares with an unrestricted random ensemble $\mathcal{E}(N,\a)$, we analyzed the performance of \texttt{akmaxsat} for instances drawn from each ensemble. 

In Fig.~\ref{fig:ak_time}, we present results for such an analysis for the time to solution as a function of $\a$ and $N$ for each ensemble. The clause density $\a$ and problem size $N$ are varied as described in Sec.~\ref{subsec:restrict_probs} with 1000 instances generated for each ensemble for a given $\a$ and $N$. The \texttt{akmaxsat} solver is deterministic, yet the solution time for a given instance may deviate due to variations in the initial starting point of the algorithm; hence, all instances are averaged over $10$ algorithm implementations. (Another view of this data is offered in Appendix~\ref{app:t-vs-a}; see Fig.~\ref{fig:t-vs-a}.)

We find that the solution time $T_\textrm{soln}$ is generally somewhat larger for the unrestricted random ensemble $\mathcal{E}(N,\a)$, implying (unsurprisingly) that on average the unrestricted instances are somewhat harder for \texttt{akmaxsat} than the DW1 processor-compatible case.  The differences between the solution times of the two ensembles are relatively small, differing by at most $2$ms, around $N=67$ with $\a=2.0$. Solution times are essentially identical for $\a\lesssim1$ and $N\lesssim40$, with small regions of the $(\a,N)$-space where the $\mathcal{E}_{\textrm{DW}}(N,\a)$ ensemble requires somewhat larger average solution times. The solution times differ somewhat more for $N>40$ and $\a>1$. It is important to remember in any case that hardness is also dependent upon the solution method. Indeed, we show below that instances which are hard for \texttt{akmaxsat} need not be hard for the DW1. Moreover, we show later (see Fig.~\ref{fig:psat}) that according to another hardness measure, processor-compatible instances are harder than random ones.

\subsection{Experimental Results}
\label{sec:exp-res}

In this subsection we describe our DW1 results. Details about the experimental settings are given in Appendix~\ref{app:DW}.

\begin{figure}[tl]
 \centering
 \includegraphics[width=\columnwidth]{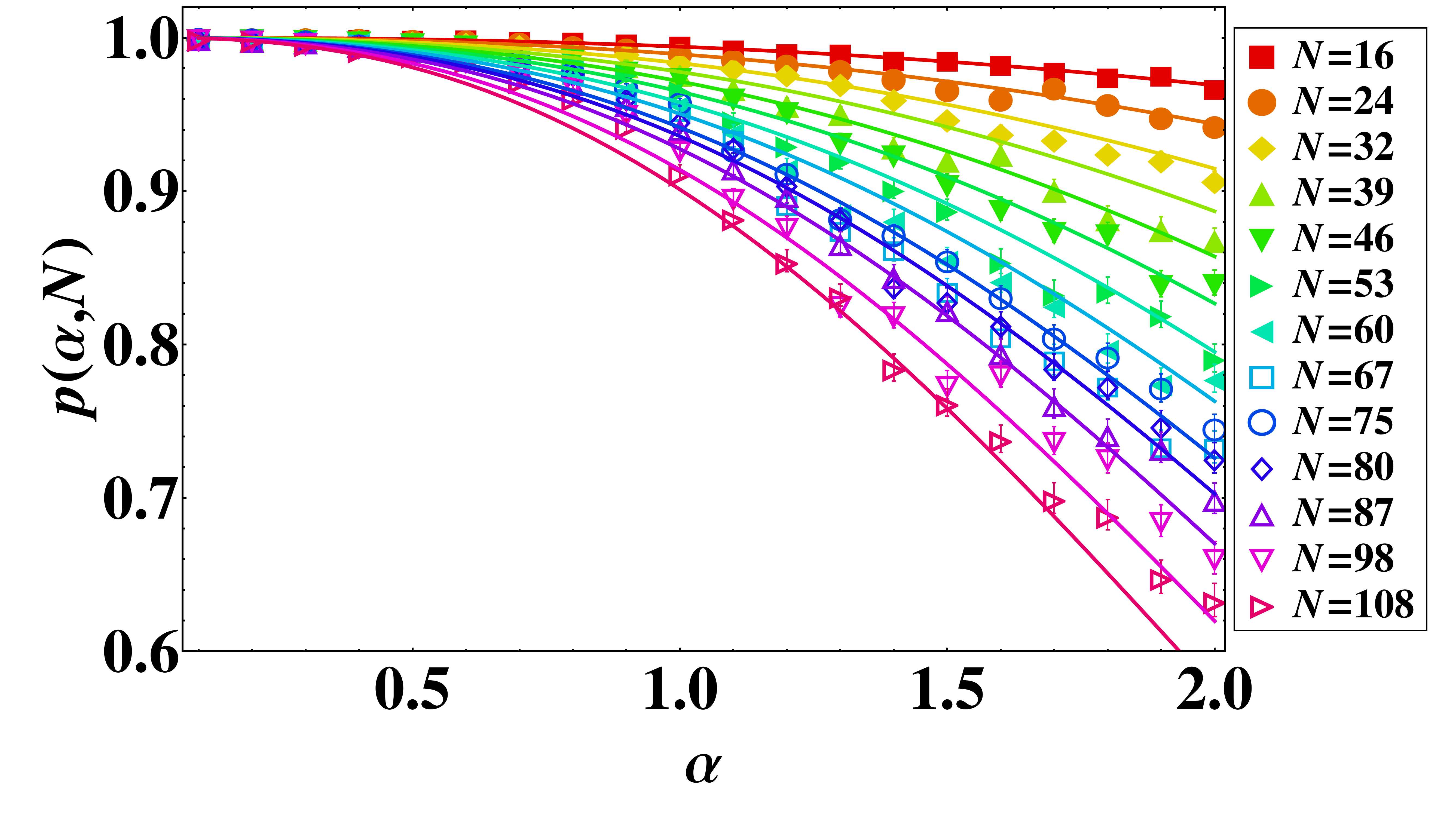}
 \caption{(color online) Data points are the mean experimental success probability of obtaining a correct MAX 2-SAT solution for the DW1, as a function of the clause density $\a$ for various values of $N$. Solid lines are a fit using Eq.~\eqref{eq:pfit} (see text). The probability decreases as the number of clauses increases for a given $N$.  The $\a$-dependence becomes more pronounced as the problem size $N$ increases, indicating that problems become harder. All data points are averages over $500$ random DW1 processor-compatible instances with each instance averaged over $100$ annealing runs. Error bars denote standard error. Note the absence of a specific feature at the critical clause density $\a_c=1$. However, see Fig.~\ref{fig:sorted_inst}.}
\label{fig:DW1 processor_prob}
\end{figure}

\begin{figure}[tr]
 \centering
 \includegraphics[width=0.9\columnwidth]{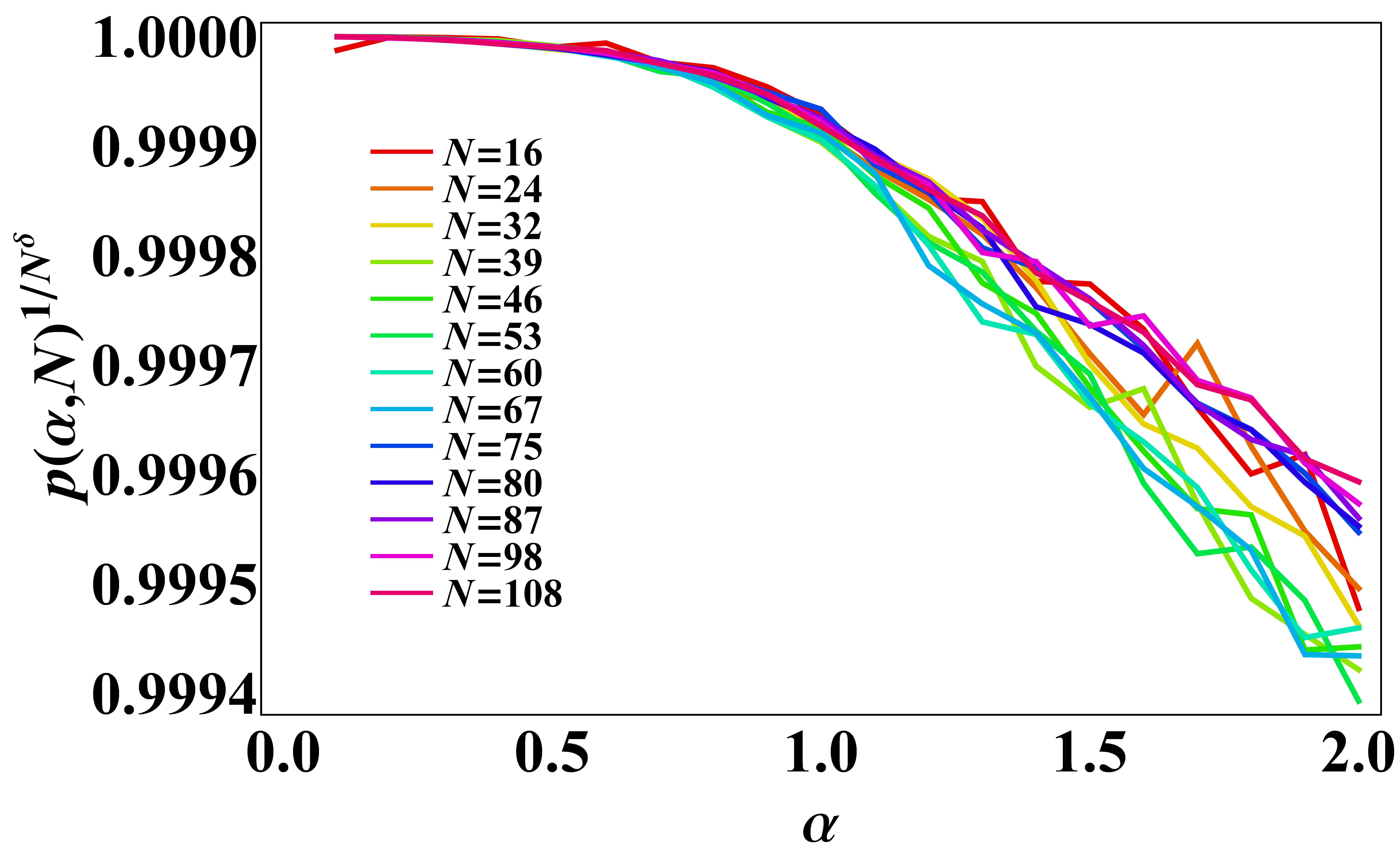}
 \caption{Data collapse analysis for the mean experimental success probability. Shown is the function $p(\a,N)^{N^{-3/2}}$ evaluated at the experimental $p(\a,N)$ values shown in shown in Fig.~\ref{fig:DW1 processor_prob}. While we found that $\d=3/2$ works well, we do not have an explanation for this particular value.}
\label{fig:dc_PvsA}
\end{figure}

\subsubsection{Mean success probability as a function of $N$ and $\a$}
In Fig.~\ref{fig:DW1 processor_prob}, we present experimental DW1 processor results for the mean success probability as a function of $\a$ for various $N$ values. The average for each data point is over the number of annealing runs and instances. Each set of data points includes a least squares fit using the following function as an ansatz:
\begin{equation}
p(\a,N)=\exp\left(-A\a^{\g}N^\d\right),
\label{eq:pfit}
\end{equation}
where we normalized $p(0,0)=1$, and we find $A=9.28[6]\times10^{-5}$, $\g=2.40[9]$ (the number in square brackets is the round-off of the remaining digits), and $\d=3/2$ for all $\a$ and $N$ considered. To test  the universality of this ansatz we show a data collapse in Fig.~\ref{fig:dc_PvsA}, which indicates a reasonable fit to the data, yielding a good collapse for larger values of $N$ ($N\geq75$) and slightly overestimating the success probability at low $N$ values, as can also be seen from Fig.~\ref{fig:DW1 processor_prob}. Note that the success probability decreases more rapidly with increasing $\a$ than with increasing $N$.

Several fundamental factors contribute to the decrease of the success probability: the inherent increase in problem difficulty as $N$ increases results in a smaller ground state gap, which in turn enhances coupling to the finite-temperature environment in the form of thermal excitations, and non-adiabaticity due to the finite annealing time \cite{Albash:12}. Added to this is the qubit approximation to the rf SQUID \cite{Amin:13}. Control errors such as miscalibration and the finite digital-to-analog (DAC) converter resolution contribute as well. A larger number of these occur as both $N$ and the number of clauses increase. A detailed discussion of the effect of control errors and their contribution to the observed success probabilities is presented in Appendix~\ref{app:rescale}.

\begin{figure*}
\centering
\includegraphics[scale=0.07]{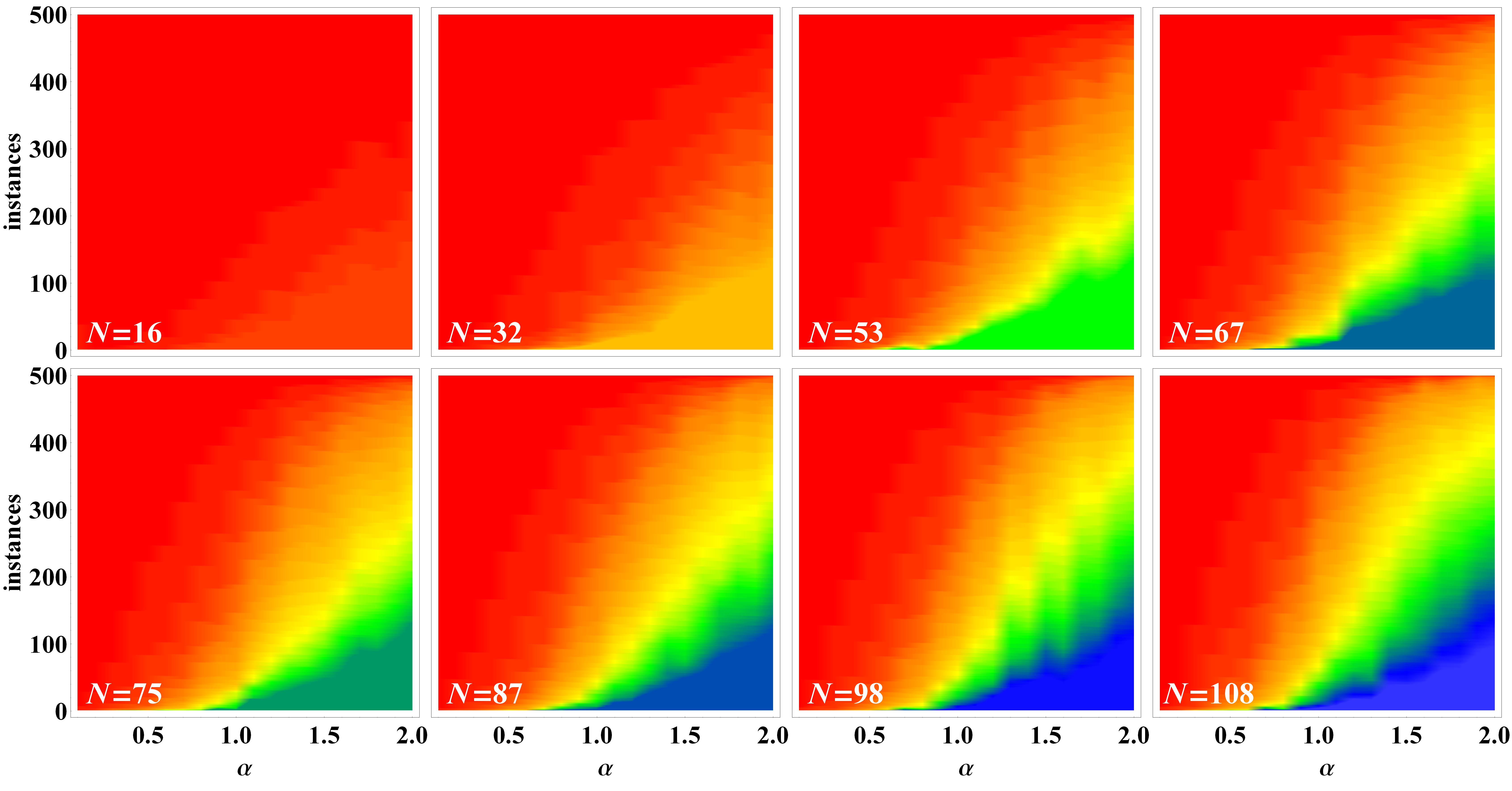}
\raisebox{0.25\height}{\centering\includegraphics[scale=0.09]{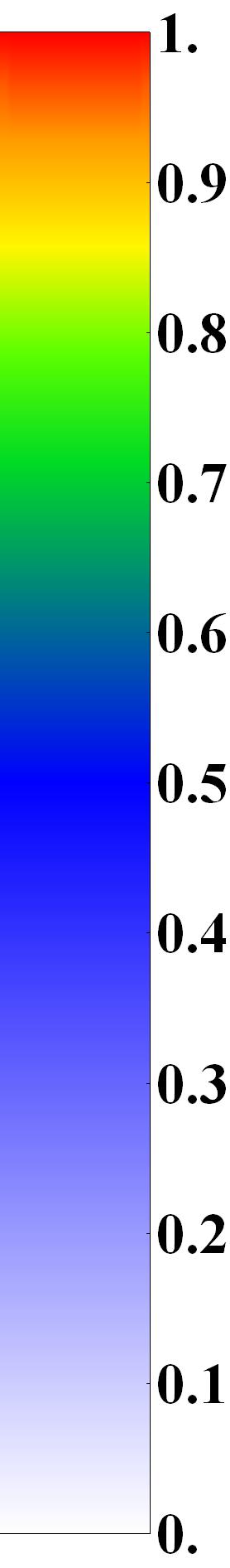}}\caption{(color online) Probability of obtaining a correct MAX 2-SAT solution for each of the $500$ DW1 processor-compatible instances sorted from least to greatest probability for $\a\in[0.1,2.0]$ and various $N$ values. In agreement with Fig.~\ref{fig:DW1 processor_prob}, problem instances become harder (i.e., have a lower success probability) with increasing $\a$ and $N$. Note the  hardness transition around $\a_c = 1$.}
\label{fig:sorted_inst}
\end{figure*}

\subsubsection{Sorted success probabilities as a function of $N$ and $\a$}
A fine-grained characterization of the ensemble of problem instances for fixed $N$ is given in Fig.~\ref{fig:sorted_inst}, where we present contour plots of the success probability for each of the $500$ problem instances as a function of $\a$. That is, for each value of $\a$ we first sort the instances by success probability, and then plot the probabilities of the sorted instance set for that value of $\a$ (thus the instance number varies as one moves horizontally through each panel). The predominance of the color red in the panels indicates that most problem instances had a probability of success close to unity. As suggested by the mean case results shown in Fig.~\ref{fig:DW1 processor_prob}, the success probabilities decrease with increasing $\a$ and $N$. Perhaps most interesting is the appearance of a soft transition around $\a \approx 1$. That is, for all $N$ most of the problems have success probability very close to $1$ for $\a < 1$, while lower success probabilities appear for $\a > 1$. This can be interpreted as an experimental indication of the critical clause density.

\begin{figure*}[t]
\includegraphics[width=2\columnwidth]{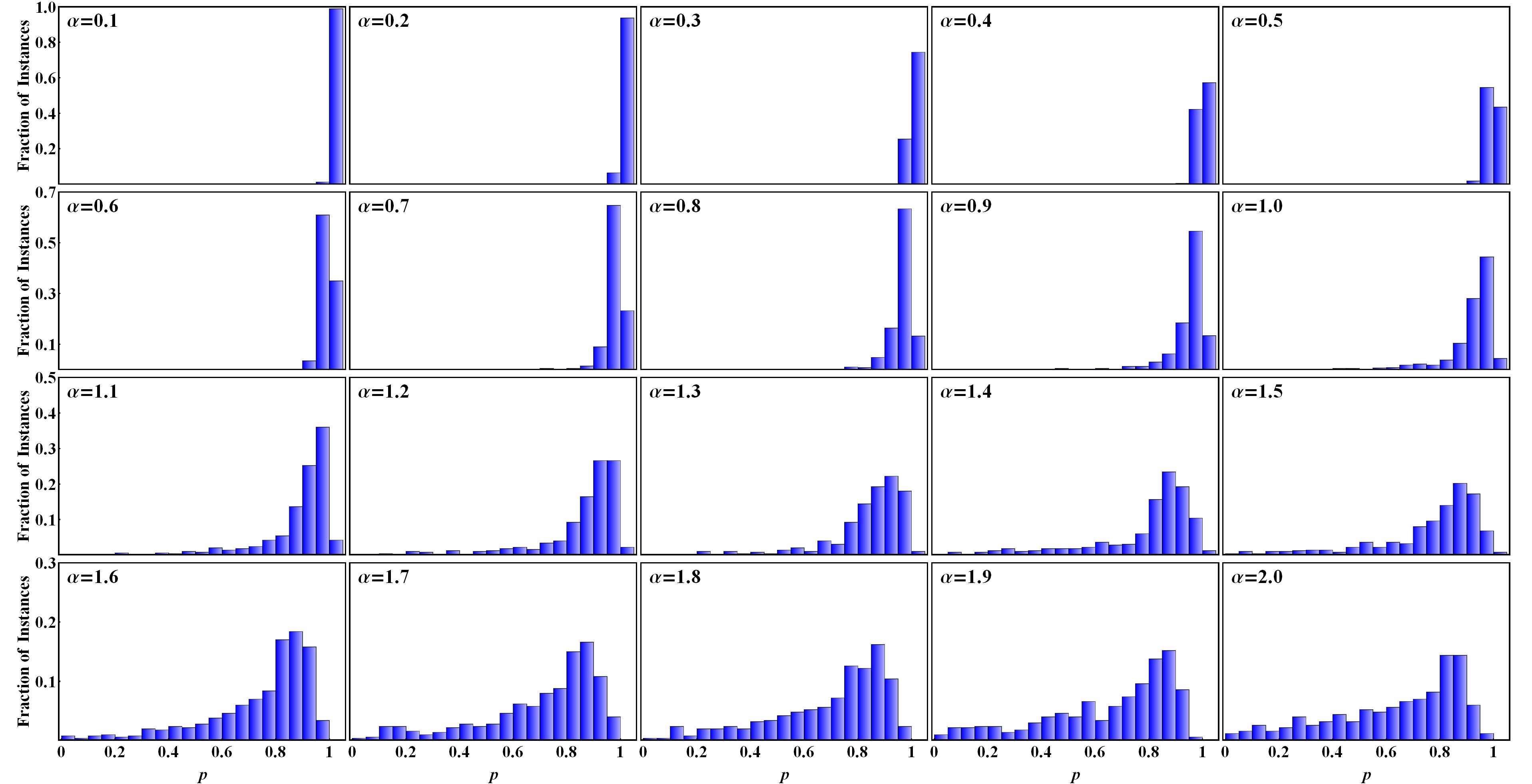}
\caption{(color online) Histogram of success probabilities $p$ for each of the 500 problem instances with $N=108$ and $\a \in \{0.1,\dots,2.0\}$. Each bar is the fraction of instances (relative to $500$) with a given empirical success probability.}
\label{fig:unimodhist}
\end{figure*}

We take a closer look at the ensembles of problem instances with $N=108$ in Fig.~\ref{fig:unimodhist}, where we show the success probability distribution for all values of $\a$ we studied. The distribution has a peak that moves from $p=1$ to $p=0.9$ as $\a$ increases, and develops a broad wing at lower success probabilities. Its unimodality is a feature that persists for all values of $\a,N$ we tried \footnote{We note that Ref.~\cite{q100} found that for an ensemble of Ising spin-glass problems the histogram of success probabilities solved using the DW1 was bimodal. This result agreed with predictions of a simulated quantum annealer, but differed from that of classical simulated annealing. 
The difference is most likely primarily due to the different annealing time, $5\, \mu$s in Ref.~\cite{q100} compared to our $1$ms. While we did not vary the annealing time in the present study, Ref.~\cite{q100} observed that a longer annealing time resulted in increasingly large fraction of problems being solved with high probability, thus shifting some of the weight of the distribution from the `hard' (low success probability) problems to the `easy' ones. Moreover, the ensemble of Ising spin glass problems considered in Ref.~\cite{q100} differed from ours in several important ways.  Their coupling strengths were randomly allowed to be only $\pm 1$ with no fractional couplings. Another difference was that the couplings between two neighboring spins were completely independent of the local fields, unlike our Eq.~\eqref{eq:handJ}.}.

\subsubsection{Transition at the critical clause density}
\label{subsec:transition}

While the phase transition discussed in Section~\ref{sec:PT} becomes sharp in the limit $N\rightarrow \infty$, analytic bounds provide a more nuanced characterization for finite $N$. The bounds of Coppersmith \textit{et al}. \cite{Coppersmith} provide useful intuition but should be interpreted cautiously as they only apply for finite, but very large $N$. In the so-called ``finite-scaling window,'' $\a \in [1-N^{-1/3},1+N^{-1/3}]$, the probability that a random formula is satisfied is bounded away from $1$ and $0$. Additionally, the average number of clauses that cannot be satisfied is a small constant fraction of $N$. Another, softer transition occurs as we increase the clause density. The average fraction of clauses that cannot be satisfied goes as $1/4 - O(1/\sqrt{\a})$ (\cite{Coppersmith}, Theorem~4). In the limit of large clause density, a random assignment (which fails to satisfy each clause with probability $1/4$) is nearly optimal, however, our experiments are far from this parameter regime.

\begin{figure} 
   \centering
   \includegraphics[width=1\columnwidth]{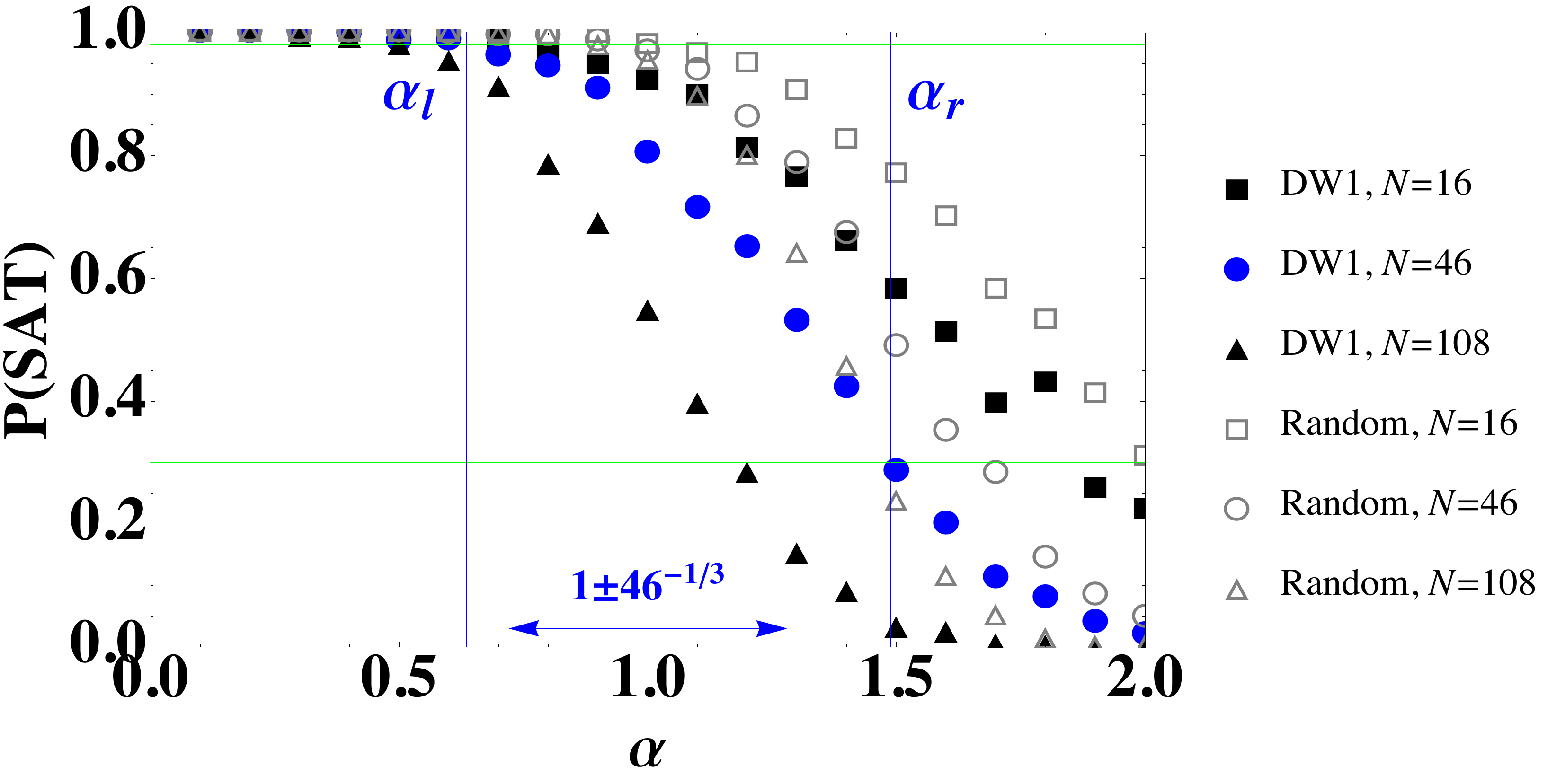} 
   \caption{ Probability of a random formula being satisfiable as a function of $\a$ for DW1-compatible (solid) and random (hollow) instances. An example of how to estimate the width of the scaling window is highlighted in blue for the DW1 processor instances at $N=46$ (solid circles). }
   \label{fig:psat}
\end{figure}

\begin{figure} 
   \centering
   \includegraphics[width=1\columnwidth]{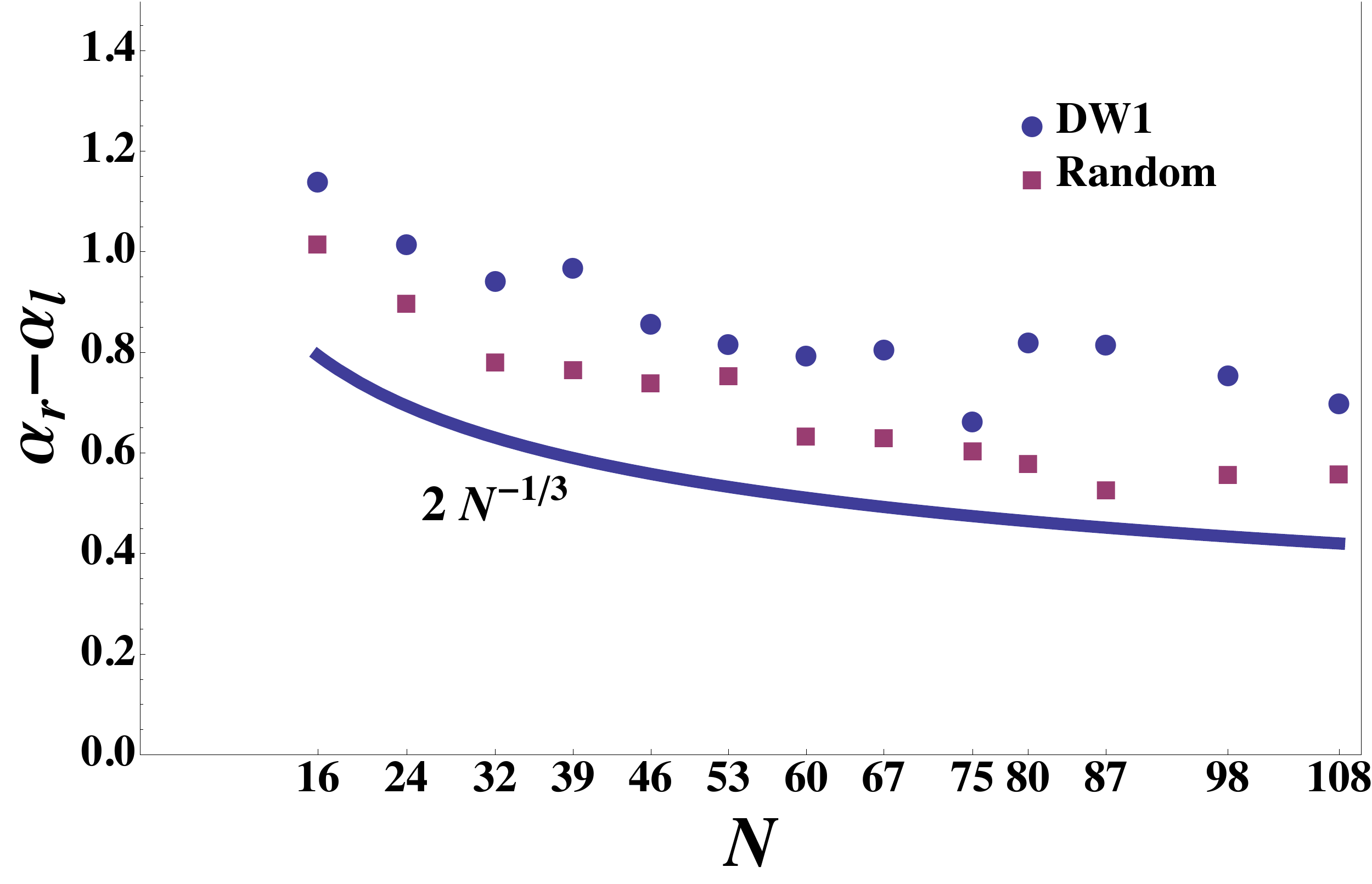} 
   \caption{Comparison of the width of the finite-scaling window estimated empirically for random and DW1-compatible instances.}
   \label{fig:empiricalwindow}
\end{figure}

We can empirically estimate the width of the finite-scaling window by estimating the width of the region in which the probability of a formula being satisfied is far from 0 or 1 \cite{shen2003empirical}. 
Fig.~\ref{fig:psat} compares the probability a formula is satisfiable for DW1-compatible and random instances for various $N$ as a function of $\a$. 
Analytically, the width of the finite-scaling window should be $2 N^{-1/3}$. 
We estimate the value of the clause density $\a_l$ ($\a_r$) at which the probability of a random formula being satisfiable drops below $0.98$ ($0.3$), as demonstrated in Fig.~\ref{fig:psat}. Then we plot our empirical estimate of the width of window, $\a_r-\a_l$ as a function of $N$ in Fig.~\ref{fig:empiricalwindow}. We confirm that the scaling of the size of this window is indeed proportional to $N^{-1/3}$. Note that the observed width of the scaling window is larger than the analytic predictions (valid only for asymptotically large $N$) for both random instances and processor-compatible instances.
Referring again to Figs.~\ref{fig:psat} and \ref{fig:empiricalwindow}, we note that the DW1 processor-compatible instances have a slightly wider window whose center is shifted slightly towards lower values of $\a$. In other words, at a given value of $\a$ a processor-compatible instance will tend to be harder to satisfy. This is in interesting contrast to the situation for the empirical success probabilities we found for \ak in Fig.~\ref{fig:ak_time}.

\begin{figure} 
   \centering
   \includegraphics[width=1\columnwidth]{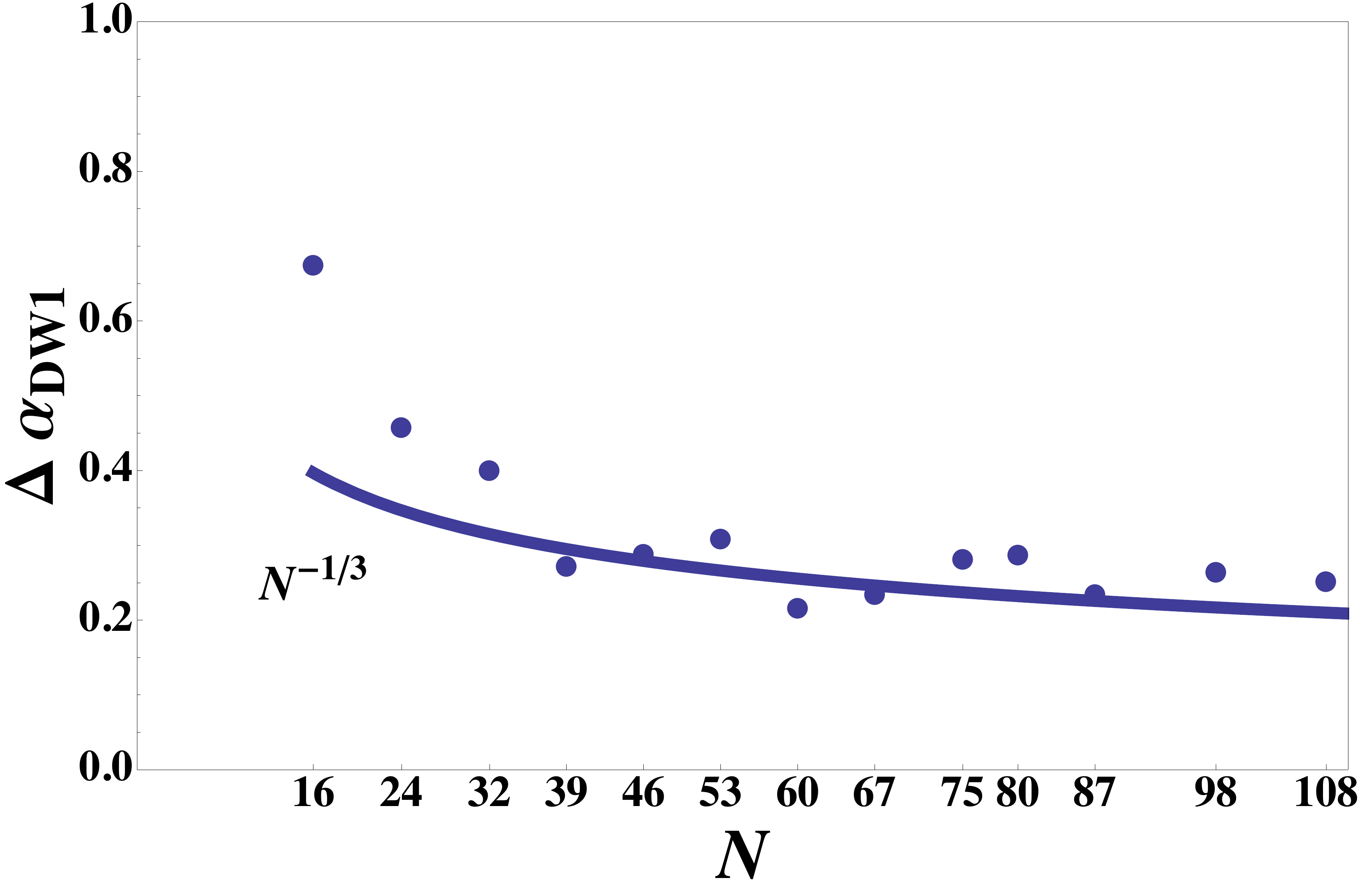} 
   \caption{An estimate of the finite scaling effects for solving instances on DW1.}
   \label{fig:dwempiricalwindow}
\end{figure}

We now show that the phase transition has computational consequences for both \texttt{akmaxsat} and DW1. Again, we use $p(\a,N)$, the probability that DW1 produced the correct solution, as a proxy for problem difficulty.
In Fig.~\ref{fig:dwempiricalwindow}, we consider a plot analogous to Fig.~\ref{fig:empiricalwindow}, except now we estimate for each $N$ the value, $\Delta \a_{\textrm{DW1}} = \a_{\textrm{DW1},r}-\a_{\textrm{\textrm{DW1}},l}$ where $\a_{\textrm{DW1},r}$ ($\a_{\textrm{DW1},l}$) indicates the value of $\a$ at which the probability of successfully solving a problem for DW1 falls below the threshold $0.97$ ($0.99$).  
The threshold $0.97$ was set high enough that the $p(\alpha,N)$-$\alpha$ curve passes through it for every $N$.
The largest minimum success probability was $0.967$ for $N=16$, as seen in Fig.~\ref{fig:DW1 processor_prob}. For very small $N$, many factors including control errors may affect this curve. While it seems plausible that the $N^{-1/3}$ scaling will continue to hold for larger $N$, solving larger problems on next-generation processors (such as the $512$ qubit DW2) will be required to verify this hypothesis.

\begin{figure} 
   \centering
   \includegraphics[width=1\columnwidth]{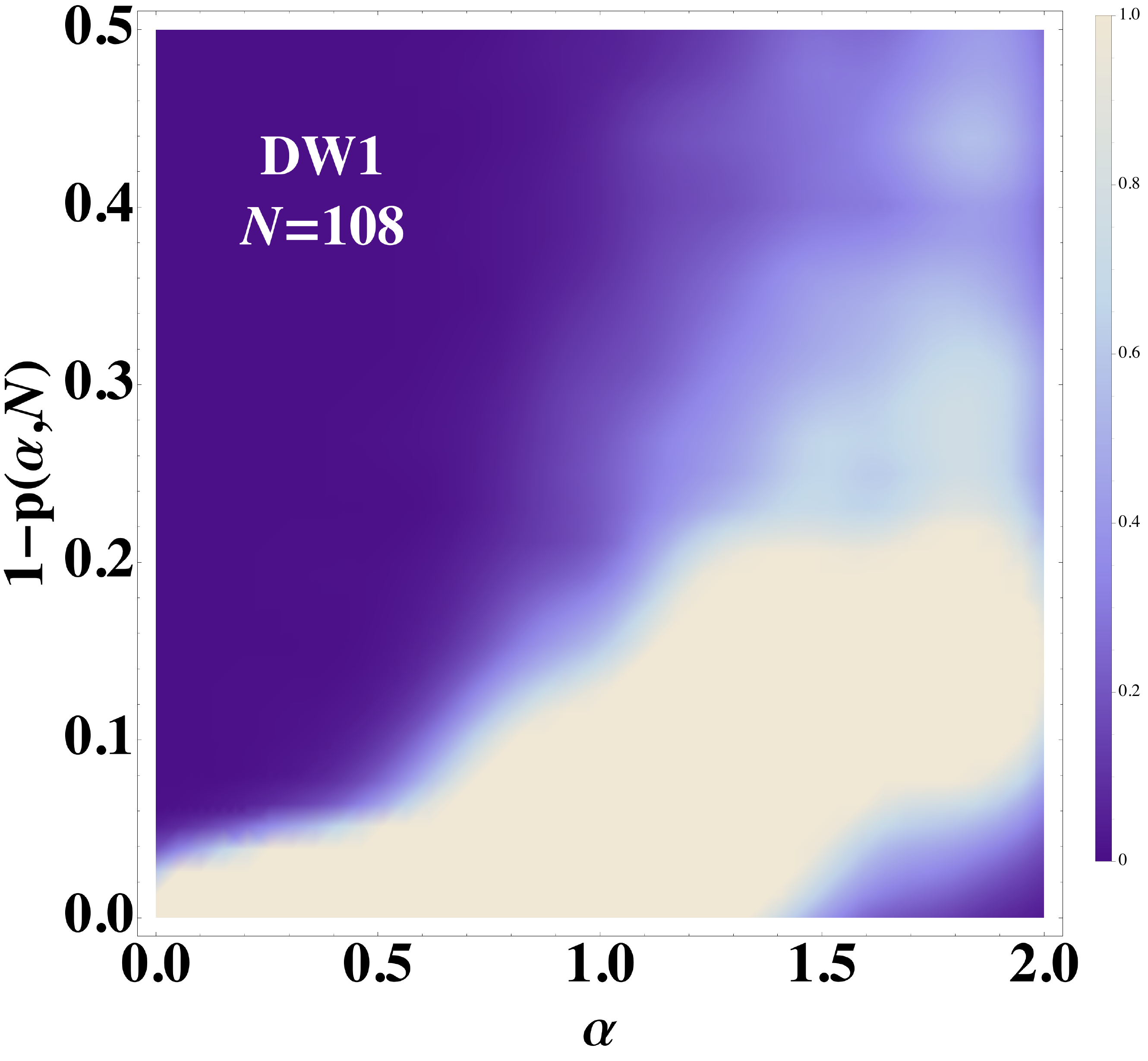} 
   \caption{A density histogram showing the distribution of failure probabilities for random instances as a function of clause density $\a$ for DW1.}
   \label{fig:dw1density}
\end{figure}

\begin{figure} 
   \centering
   \includegraphics[width=1\columnwidth]{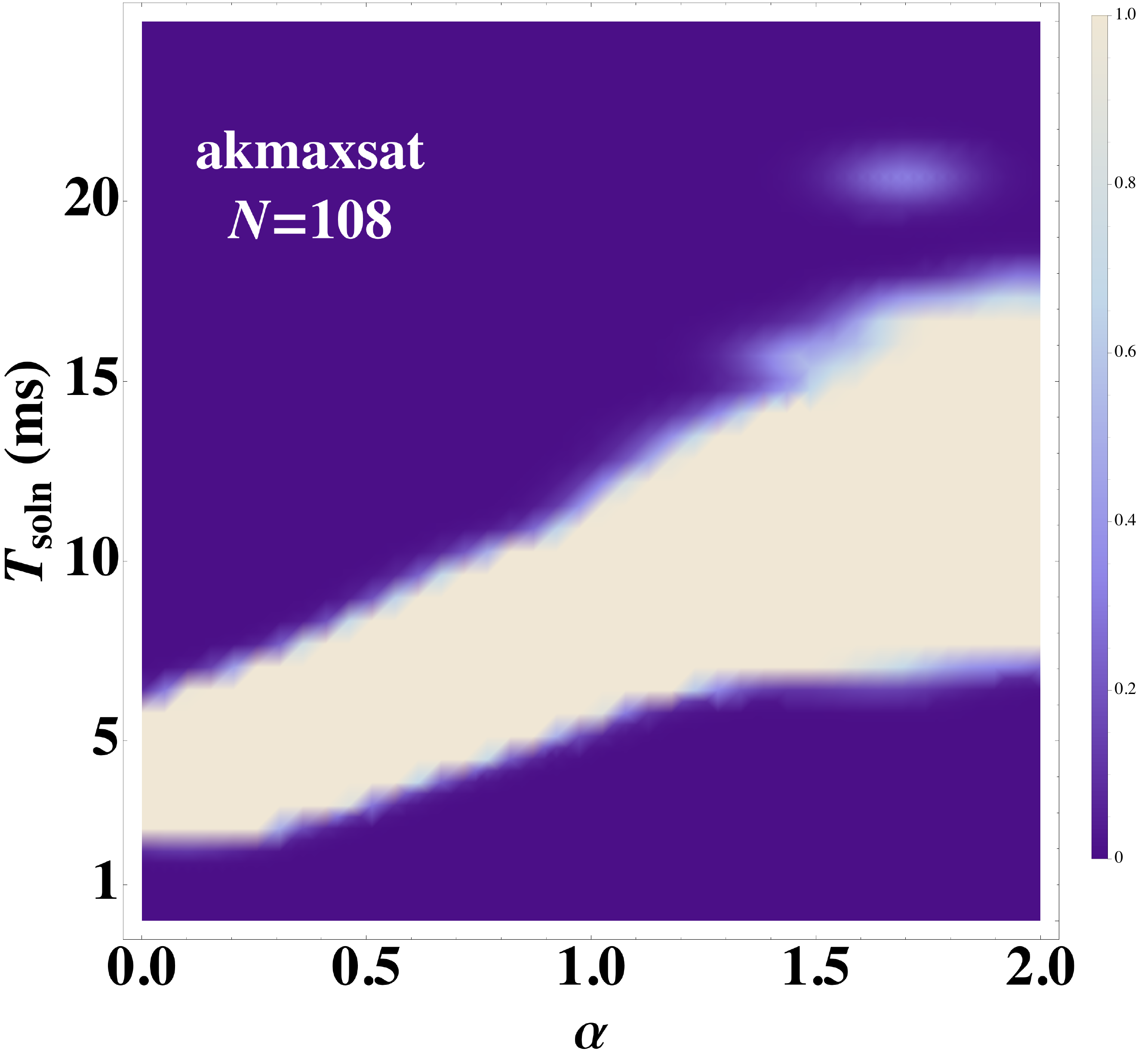} 
   \caption{A density histogram showing the distribution of the time to solution $T_{\textrm{soln}}$ for random processor-compatible instances as a function of clause density $\a$ for \texttt{akmaxsat}.}
   \label{fig:akdensity}
\end{figure}

We also see the effect of the phase transition in Fig.~\ref{fig:dw1density}, in which we compare $p(\a,N=108)$ to $\a$ using a density histogram. A sharp change in the number of difficult instances clearly occurs around $\a = \a_c = 1$. Furthermore, the variance in problem difficulty is much higher to the right of the transition. 
We compare the time to solution for \texttt{akmaxsat} to $\a$ in Fig.~\ref{fig:akdensity}. While the difficulty clearly increases with clause density, we see again that the variance in difficulty increases at the phase transition. While the appearance of difficult instances is linked to the phase transition in both cases, surprisingly the instances that are difficult appear unrelated, as we will see in Sec.~\ref{sec:corr}. 
We discuss further evidence for a transition at $\a_c$ in Section~\ref{sec:corr} (see in particular Fig.~\ref{fig:corr_plots}).


\subsection{Scaling of the time to solution: DW1 vs \texttt{akmaxsat}}
\begin{figure}[t]
 \centering
 \includegraphics[width=\columnwidth]{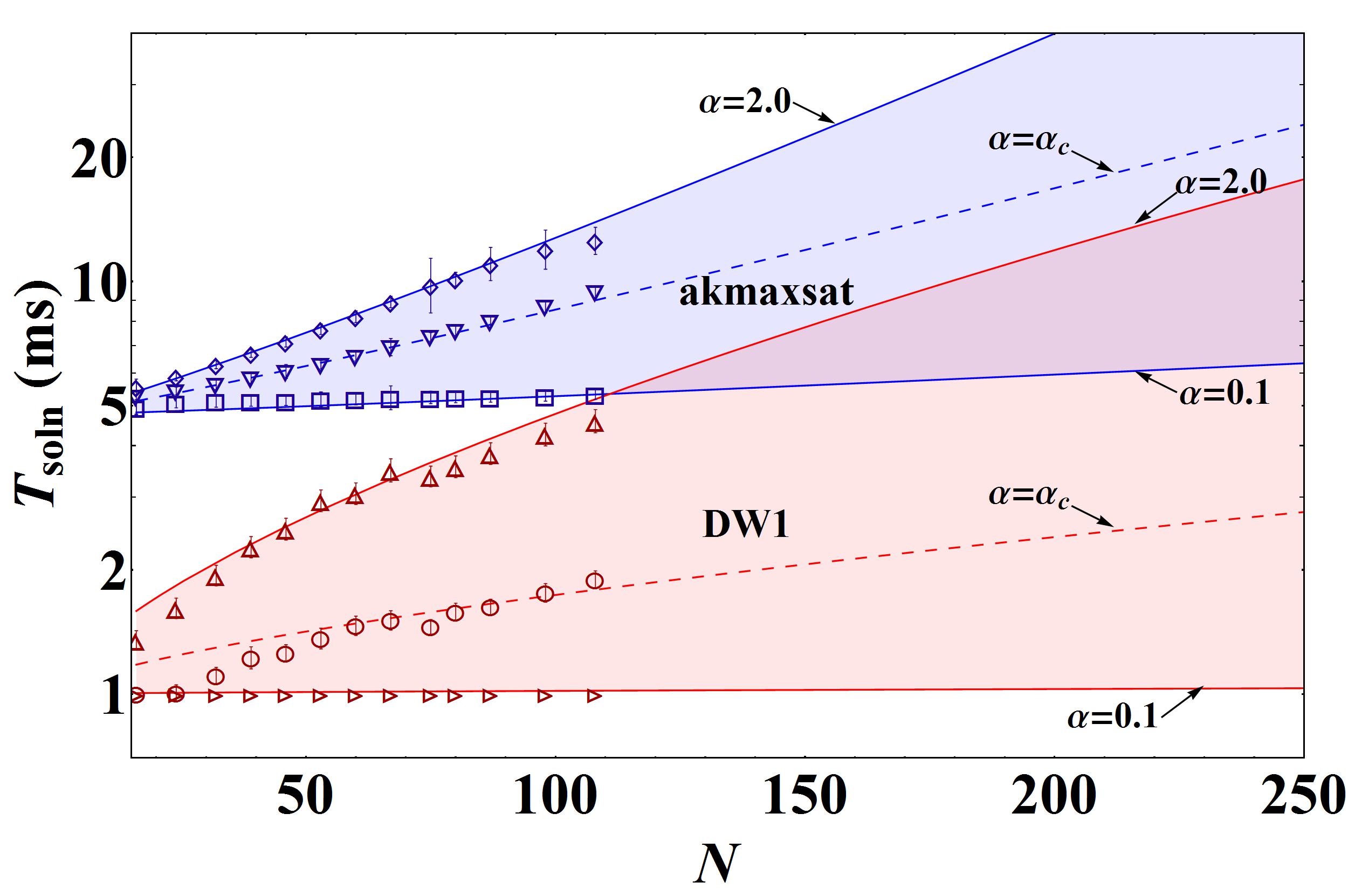}
\caption{(color online) Computational effort comparison between \texttt{akmaxsat} and the DW1 as a function of the problem size $N$ and for different clause densities $\a$. Results are shown along with best fits for the entire data set, to the function in Eq.~\eqref{eq:exp_fit}. Shown are (1) \texttt{akmaxsat} with $\a=0.1$ (squares), $\a=\a_c=1$ (down-triangles), and $\a=2.0$ (diamonds); (2) DW1 with $\a=0.1$ (right-triangles) $\a=\a_c=1$ (circles), and $\a=2.0$ (up-triangles). Remaining values of $\a$ are summarized by the shaded regions. Note that while extrapolations for \texttt{akmaxsat} can be deemed reliable, this is not the case for the DW1 data, as argued in the text. Thus the DW1 extrapolations shown should be considered as merely suggestive for a limited range of $N$ values whose upper bound we do not know at present.}
  \label{fig:DW1 processor_vs_ak}
\end{figure}

Given a success probability $p$, assuming no correlation between successive annealing runs on the DW1 (an assumption that is satisfied to a good approximation \cite{q100}), the probability of not getting a single correct answer in $k$ runs is $(1-p)^k$. We define $p_{\textrm{desired}}$ as the threshold probability of getting one or more correct answers, i.e., $p_{\textrm{desired}}=1-(1-p)^k$. For a run with annealing time $t_f$ the  time required to reach the ground state at least once in $k$ runs with probability $p_{\textrm{desired}}$ is: 

\begin{eqnarray}
T_\textrm{soln}(p_{\textrm{desired}})&=&t_f k(p_\textrm{desired})\nonumber\\
&=&t_f \ceil[\Bigg]{\frac{\log(1-p_{\textrm{desired}})}{\log(1-p)}},
\label{eq:time}
\end{eqnarray}
where $t_f = 1$ms for our experiments.

Taking $p$ as the mean success probability reported in Fig.~\ref{fig:DW1 processor_prob}, the extrapolated time to solution for $p_{\textrm{desired}}=0.99$ for $\a=0.1$ (right-triangles) and $\a=2.0$ (up-triangles) is plotted in Fig.~\ref{fig:DW1 processor_vs_ak} versus problem size $N$ (for complete data see Appendix~\ref{app:collapse}). Note that for $\a=0.1~\forall N$, the DW1 obtained the correct solution with $p\geq99\%$ on average. Clearly, this subensemble of problems was too easy for the chosen value of the annealing time, since this translates into a single repetition ($k=1$), i.e., a constant time to solution equal to the (unoptimized) annealing time of $1$ms. On the other hand, as $N$ grows the time to solution must of course eventually start to grow as well. This illustrates the danger of extrapolating to large $N$ values from experimental data based on a single (unoptimized) annealing time. In fact this conclusion also applies to the other values of $\a$ shown, since for a fixed annealing time the time to solution must eventually blow up as a function of $N$, due to the inevitable reduction in success probability resulting from restricting the time to settle on an optimal solution in an increasingly growing configuration space \cite{q100}. This extrapolation caveat does not apply to our \texttt{akmaxsat} data, since there we simply let the algorithm run until it finds a solution.

\subsubsection{Extrapolation for the fixed-$\a$ ensemble}
With the just-stated caveat in mind, we present a best fit to the entire DW1 data set, and separately for the \texttt{akmaxsat} timing dataset. We find that the data is well fit by the following function:
\begin{equation}
T_\textrm{soln}(\a,N)=A\exp\left(B\a^{\g}N^{\d}\right)
\label{eq:exp_fit}
\end{equation}
where the values of the various parameters are given in Table~\ref{tbl:fitparams}. The numerical values were obtained from a least squares fit followed by a data collapse of the data shown in Fig.~\ref{fig:DW1 processor_vs_ak} for both \texttt{akmaxsat} and the DW1. More details regarding the data collapse are given in Appendix~\ref{app:collapse}, where we also present a second fit with more free parameters; this does not change our conclusions below.
 
Comparing the numerical values of the exponents $\g$ and $\d$ of, respectively, the clause density and the number of variables in Table~\ref{tbl:fitparams} we see that while \texttt{akmaxsat} has the smaller value of the clause density exponent, the DW1 has the smaller exponent for the number of variables. This explains the better scaling for the DW1 as a function of $N$ seen in Fig.~\ref{fig:DW1 processor_vs_ak}, in spite of the fact that the DW1 has a much larger value of the exponent $B$.

\begin{figure}[t]
\centering
\includegraphics[width=1.07\columnwidth]{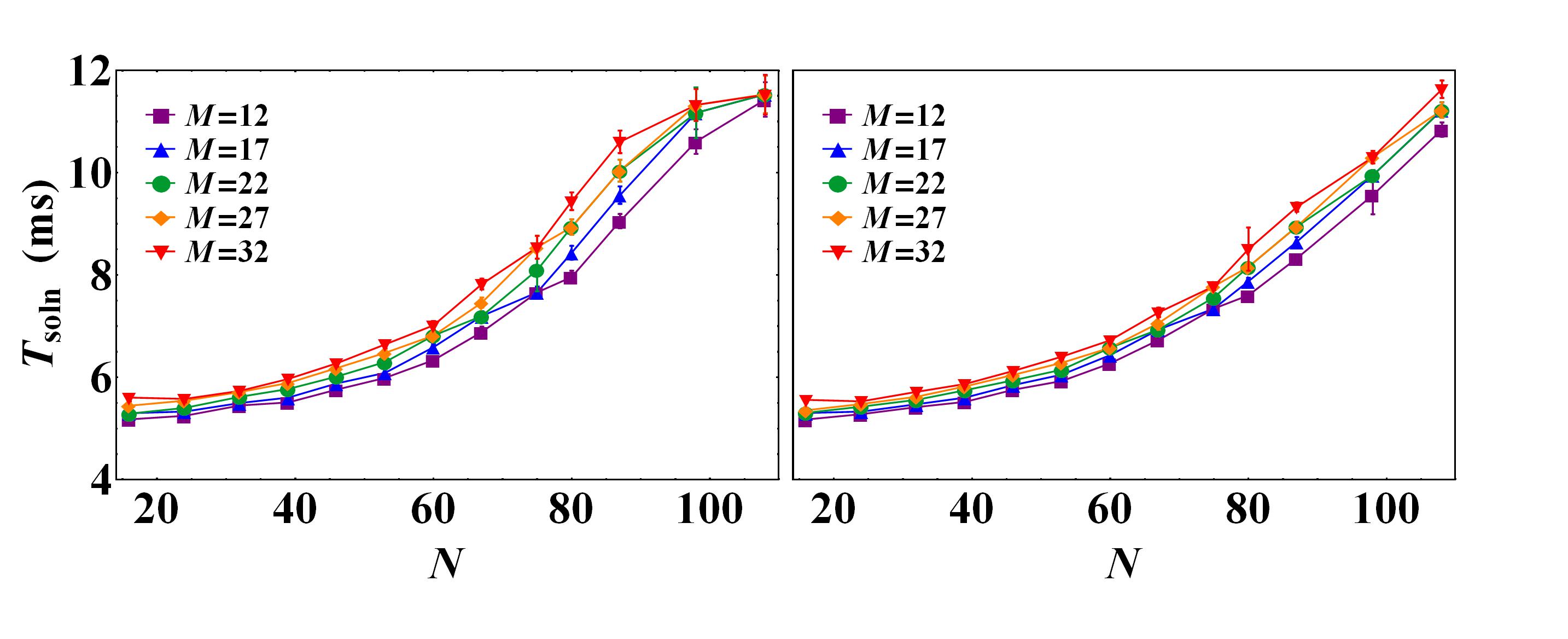}
\caption{(color online) Solution time versus $N$ for fixed $M$ for the unrestricted (left) and DW1 processor-compatible (right) ensembles.}
\label{fig:const-M}
\end{figure}

\begin{figure*}[t]
\centering
{\includegraphics[width=1.05\columnwidth]{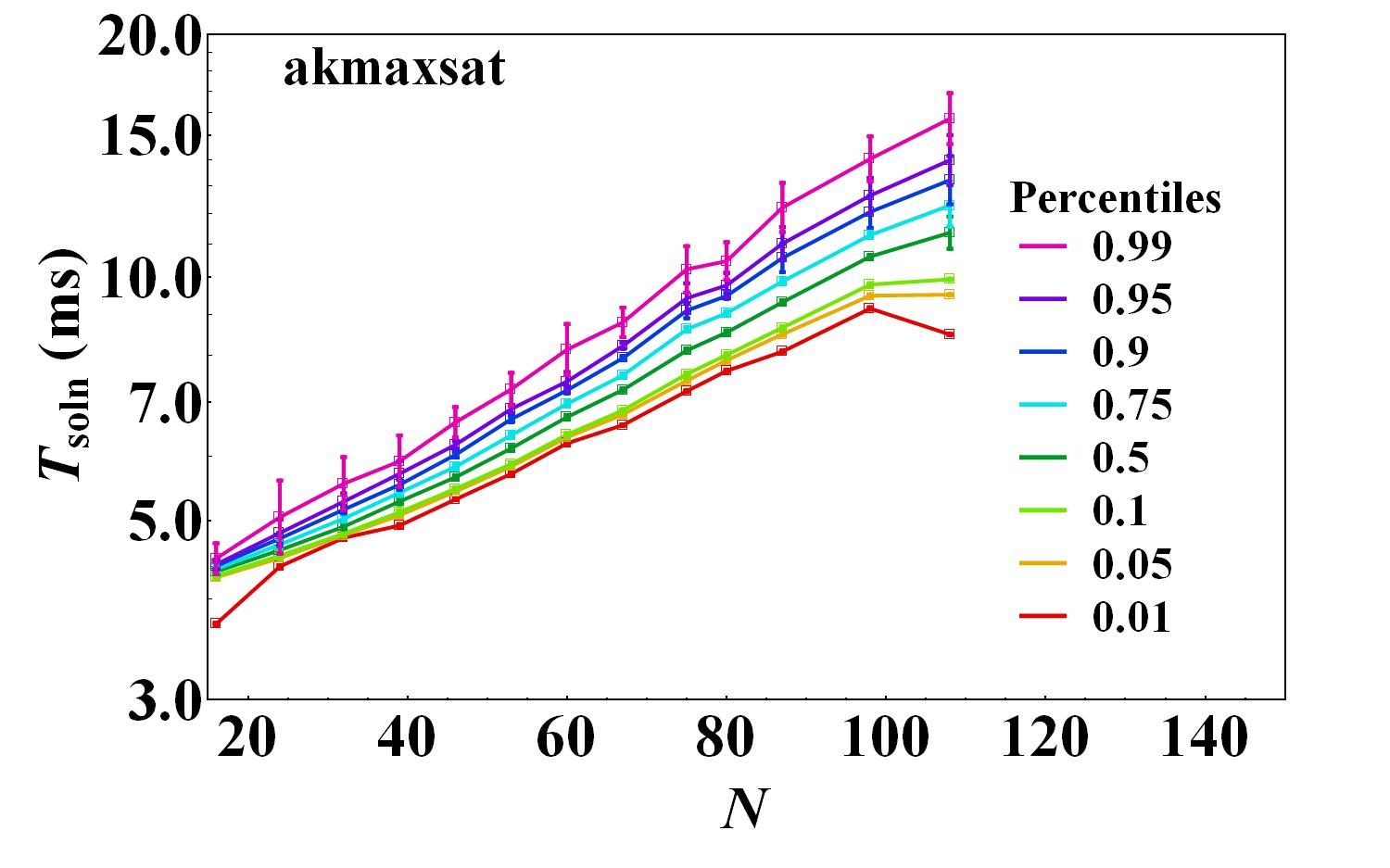}}\hspace*{-0.2cm}
{\includegraphics[width=1.05\columnwidth]{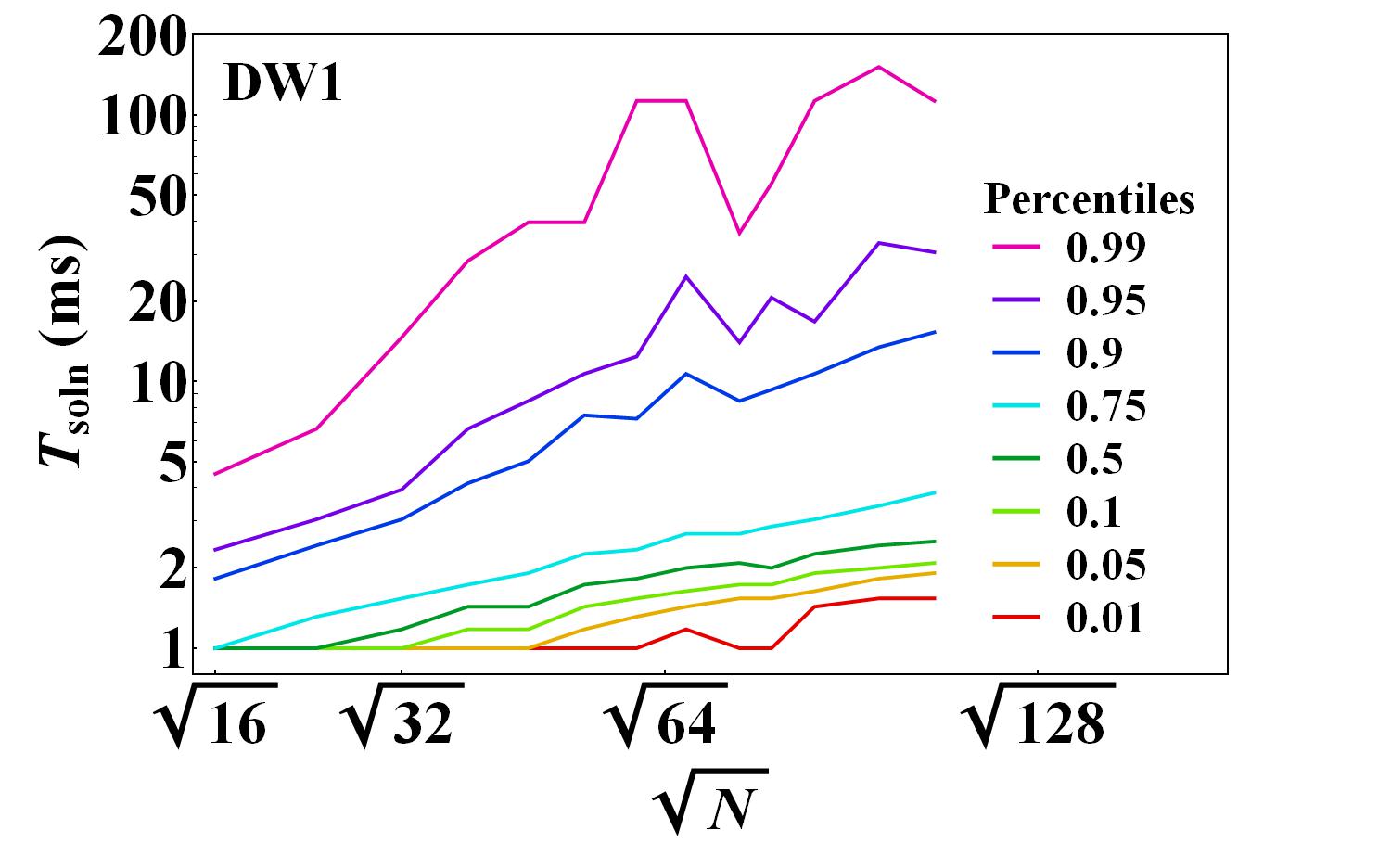}}\hspace*{-0.2cm}
\caption{(color online) Scaling plots of the time to solution at $\a=2.0$ for different percentiles of the distribution of problem hardness, (a) Log-linear for \texttt{akmaxsat}, (b) Log-square-root for the DW1. In the \texttt{akmaxsat} case problem instances were sorted and binned by their required solution time. The percentiles then indicate the cutoff value of the solution time for the corresponding set of instances, with shortest and longest solution times indicated by the $0.01$ and $0.99$ percentiles, respectively.  In the DW1 case we computed the percentiles from the histograms shown in Fig.~\ref{fig:unimodhist}. The easiest and hardest sets of problem instances are indicated by the $0.01$ and $0.99$ percentiles, respectively. While the performance of \ak appears fairly uniform across the different percentiles, the performance of the DW1 is significantly better for the lower (i.e., easier) percentiles, with a noticeable deterioration seen at the $90$th percentile.}
\label{fig:quant_plots}
\end{figure*}

\begin{table}[tr]
\begin{tabular}{ccccccc}
 & $A$ & $B$ & $\g$ & $\d$ &  $\d-\g$ & $R^2$ \\ \hline
AK & $4.750$ & $0.003[6]$ & $3/4$ & $1.106[8]$ & $0.356[8]$ & $0.999[4]$\\
DW1 &  $1.0$ & $0.026[0]$ & $3/2$ & $0.663[7]$ & $-0.836[2]$ & $0.991[4]$
\end{tabular}
\caption{Numerical values for the parameters appearing in Eq.~\eqref{eq:exp_fit}. The last column is the coefficient of determination of the fit, with $R^2=1$ denoting a perfect fit. Note that $A$ has units of msec and all other coefficients are dimensionless.}
\label{tbl:fitparams}
\end{table}

\subsubsection{The limit of large $N$ and constant $M$}
\label{sec:const-M}

Departing momentarily from our emphasis on the fixed-$\a$ ensemble, note that Eq.~\eqref{eq:exp_fit} can also be written as
$A\exp\left(B M^{\g}N^{\d-\g}\right)$. While viewed in this way \texttt{akmaxsat} has a better scaling with the number of clauses $M$ since it has the smaller $\g$ value, our fit yields a negative value for the DW1 exponent $\d-\g$ of $N$, while \texttt{akmaxsat}'s exponent is positive (see Table~\ref{tbl:fitparams}). Now consider the limit of large $N$ and small $\a$, while keeping the number of clauses $M$ constant. In this limit the probability of repeated variables vanishes, so that it becomes possible to satisfy each clause independently. A parallel processor capable of updating all clauses simultaneously would therefore solve the MAX 2-SAT problem in constant time in this limit. This is indeed the prediction of Eq.~\eqref{eq:exp_fit} for the DW1 time to solution, given that $\d-\g<0$:
\begin{equation}
\lim_{N \rightarrow \infty}T_\textrm{soln}^\textrm{DW1}(\textrm{const}/N,N)=A=\textrm{const}.
\label{eq:exp_fit2}
\end{equation}
In contrast, the predicted scaling of the time to solution for \texttt{akmaxsat} diverges with $N$ even in this limit, which clearly shows its suboptimality. An independent check of this is shown in Fig.~\ref{fig:const-M}, where we plot the time to solution for \texttt{akmaxsat}, for the unrestricted ensemble $\mathcal{E}(N,\a)$ and the DW1-compatible ensemble $\mathcal{E}_{\textrm{DW}}(N,\a)$. It can be seen that indeed, \texttt{akmaxsat} solution times do not seem to converge to a constant as $N$ increases, and only a mild improvement is seen as $M$ decreases.

The fact that the DW1 seems capable of ``recognizing" that the fixed-$M$, large $N$ limit is easy, while \texttt{akmaxsat} does not, is interesting. It suggests that the DW1 naturally acts as a parallel processor, making ``cluster moves" that simultaneously find SAT solutions for multiple clauses.

\subsubsection{Time to solution for different levels of problem hardness}

In Fig.~\ref{fig:quant_plots} (a) we plot, for $\a=2$, the time taken by \texttt{akmaxsat} to solve the problems at $8$ different percentiles of the $500$ instances at each $N$, and compare this in Fig.~\ref{fig:quant_plots} (b) to the estimated time to solution for DW1 for the same percentiles requiring $p_{\textrm{desired}}=.99$. Note that  the same percentile value in the two figures generally represent different, possibly overlapping sets of problem instances. In Fig.~\ref{fig:quant_plots}(a) lower and upper percentiles correspond to shorter and longer solution times, respectively, with the easiest (hardest) problems being in the $0.01$ ($0.99$) percentile. The scaling for \texttt{akmaxsat} is approximately exponential (note the logarithmic time axis), with the higher percentiles having larger exponents. The solution time for the hardest ($0.99$) percentile is approximately twice that for the easiest ($0.01$)  for $N=108$. As seen in Fig.~\ref{fig:quant_plots}(b), the range of DW1 solution times varies significantly more between different percentiles than for \texttt{akmaxsat}.  Disregarding fluctuations due to the control errors and the small sample size, the scaling of the DW1 solution times appears to be more favorable than for \texttt{akmaxsat} for all percentiles, and to match an exponential of $\sqrt{N}$ rather than $N$, in agreement with the scaling of the tree-width of the Chimera graph \cite{Choi1,Choi2,q100}. Once again we point out  that extrapolation to larger $N$ values are unreliable due to our suboptimal annealing time. We also note that the procedure whose results are shown in Fig.~\ref{fig:quant_plots} corresponds to first computing the percentiles, then estimating the time to solution; in Appendix~\ref{app:comp} we show that the order of these operations does not change the results.

\begin{figure*}[t]
\includegraphics[width=\linewidth]{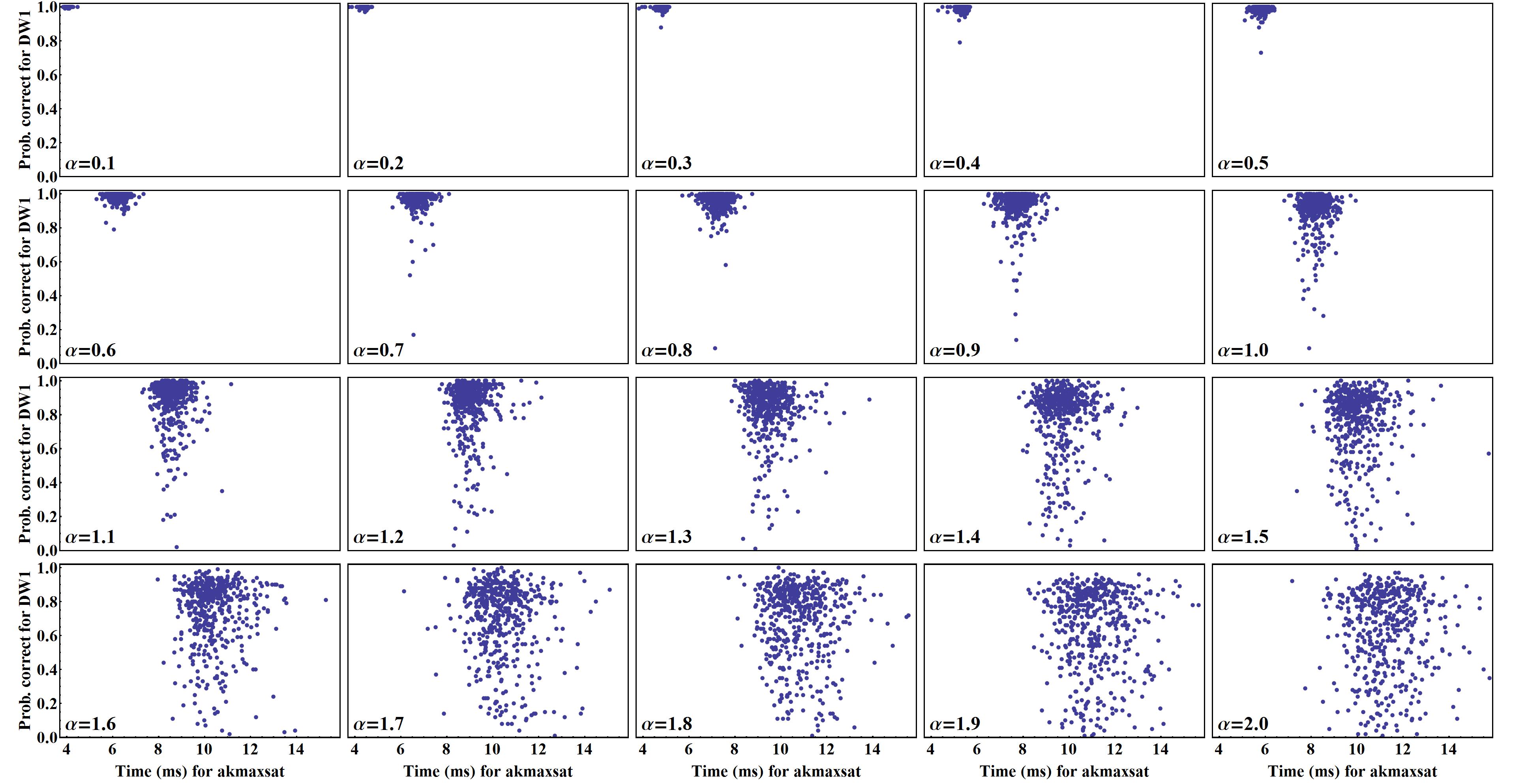}
\caption{Scatter plot of success probability of DW1 vs. time taken by \texttt{akmaxsat} for each problem instance with $N=108$ and $\a\in\{0.1, 0.2, \ldots, 2.0\}$ ($500$ instances per $\a$ value). While for very small clause densities all problems are easy for the DW1 and take roughly the same amount of time for \texttt{akmaxsat}, as $\a$ approaches $\a_c =1$ from below harder problems start to appear for the DW1 and \texttt{akmaxsat} timings start to spread. Above $\a_c$ there is a significant spread in both the DW1 success probabilities and the \texttt{akmaxsat} timings. As a result the correlation between these two variables steadily diminishes 
as $\a$ grows.}
\label{fig:corr_plots}
\end{figure*}

\subsection{Are the DW1 and \texttt{akmaxsat}  correlated?}
\label{sec:corr}

In Fig.~\ref{fig:corr_plots}, we compare the solution time required by \texttt{akmaxsat} to the probability $p$ of finding a correct solution by the DW1 for all values of $\a \in\{0.1,\dots,2.0\}$, for $N=108$. For $\a < \a_c =1$ the \texttt{akmaxsat} results are strongly clustered around their mean and the clear majority of problem instances are easy for the DW1. For $\a \geq 1$ hard problem instances start to appear and the \texttt{akmaxsat} timing results spread around their mean. The DW1  probabilities are much more scattered; while they still cluster somewhat near $p=1$, there is an increasingly large spread across the entire range of possible values of $p$. The correlation between the the DW1 success probability and the \texttt{akmaxsat} time to solution thus steadily diminishes with $\a$; in particular we see that for $\a > \a_c$ there is a large spread of $p$ values for any fixed solution time.

\begin{figure*}[t]
\includegraphics[width=\linewidth]{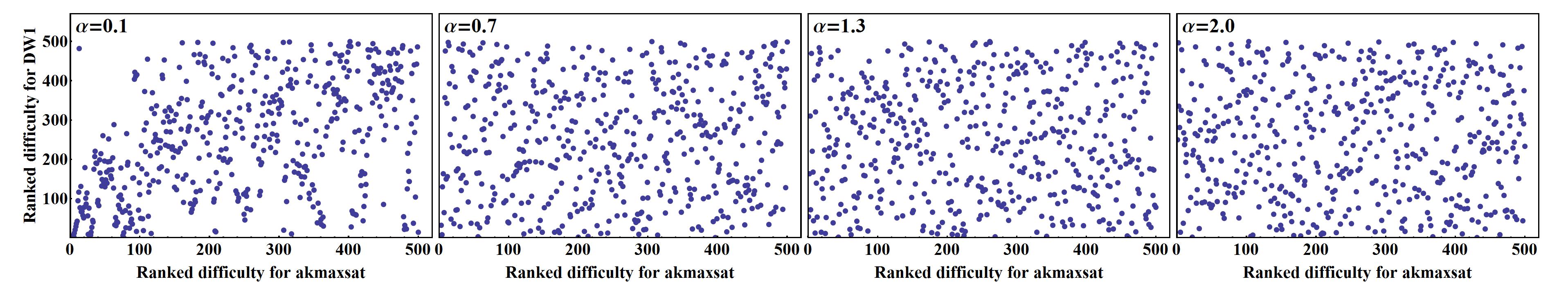}
\caption{Ranked difficulty comparison (copula plot) between DW1 and akmaxsat for $N=108$ and representative values of $\a\in\{0.1, 0.7, 1.3, 2.0\}$ (other values yield very similar looking plots). The instances used here are the same as in Fig.~\ref{fig:corr_plots}. Clearly there is essentially no correlation for problem difficulty, even for small values of the clause density..}
\label{fig:copula_plots}
\end{figure*}

A more direct measure of correlations is given in Fig.~\ref{fig:copula_plots}. Shown in this figure is a ranked difficulty comparison between the DW1 and \texttt{akmaxsat}, where we rank the difficulty of each instance from $1$ (easiest) to $500$ (hardest) for both solvers. Perfect correlation would then be evidenced as all instances falling on the $45{\degree}$ diagonal. While for $\alpha=0.1$ there is a slight tendency for clustering of the instances along this diagonal, such evidence of correlation is entirely absent for all higher values of the clause density.

If it is true that QA offers an advantage for certain ensembles of problems, then we also expect that for the random ensemble (which is, in a sense, a uniform average over possible ensembles), we should see specific problem instances for which QA offers an advantage over classical solvers. That we see no correlation in problem difficulty between QA and one particular classical solver is an interesting, though not conclusive, piece of evidence in this direction.

\section{Conclusion and Future Work}
\label{sec:concl}

In this work we undertook a study of the performance of the DW1 quantum annealing processor on MAX 2-SAT problems, and compared it to \texttt{akmaxsat}, a competitive exact classical solver. We focused on the random problem ensemble characterized by a fixed clause density, as the latter is a well defined order parameter which allows ``tuning" the problem hardness. After showing how MAX 2-SAT problems can be mapped to the DW1, we studied three main experimental questions: 
\begin{enumerate}
\item Is there experimental evidence for a hardness transition at or near the critical clause density? 
\item Is there a correlation between the DW1 and \texttt{akmaxsat} in terms of problem hardness?
\item How does the time to solution scale for the two solvers, and is there evidence of a significant difference? 
\end{enumerate}

Our answer to the first question is a qualified ``yes". Within the limitations of our relatively small number of variables we did indeed find evidence for a significant decrease in DW1 success probability starting around the critical clause density. Moreover, we found that the width of the finite-size scaling window follows the theoretical expectation. This suggests that QA is sensitive to the most essential feature of problem hardness. Our answer to the second question is a resounding `no'. This means that within the ensemble of hard random MAX 2-SAT problems there are likely to be found problems for which QA has an advantage over exact classical solvers, and vice versa. However, one cannot exclude on the basis of our data that there might exist strong correlations between QA as encapsulated by the DW1 and stochastic classical solvers. Indeed, Ref.~\cite{q100} showed that in terms of random spin glass problem instance hardness, \emph{simulated} quantum annealing correlates very well with the DW1 (essentially as well as the DW1 correlates with itself). Since simulated quantum annealing has an efficient classical implementation using quantum Monte Carlo algorithms (by mapping to a classical spin problem in one extra dimension), it is undoubtedly desirable to follow up our work with a simulated quantum annealing study of the same set of MAX 2-SAT problem instances. This remark and more applies also to the third question: we found that the DW1 scaling with problem size is clearly better than \texttt{akmaxsat}'s over the range of problem sizes and clause densities we studied, and this is encouraging for QA, but at the same time additional study with classical stochastic solvers such as simulated annealing are needed in order to establish whether the DW1 advantage persists.

There are several other interesting directions for future research (see also Ref.~\cite{q100}). We focused exclusively on the probability of finding the actual ground state, meaning that even a single-qubit error disqualifies a state as a correct solution; this criterion could be relaxed and one could instead focus on the distribution of excited states  or Hamming distances from the ground state. The connection between problem hardness and the minimum energy gap between the ground and lowest excited state encountered during the annealing evolution is another question of great interest. 
Finally, it is obviously interesting to extend the results presented here to larger problem sizes and clause densities using the DW2 processor and its successors.

We conclude with a suggestion for future research that is related to the $\rho$-PTAS discussed in Section~\ref{sec:PTAS}, which highlights a particularly interesting aspect of MAX 2-SAT. Recall that a $\rho$-PTAS is an algorithm that provides an assignment of variables that provably satisfies a number of clauses within at least a fraction $\rho$ of the maximum number of clauses that can be satisfied for any formula, and that no $\rho$-PTAS exists for $\rho > 21/22$ unless P$=$NP~\cite{inapprox}. This is a rather tight bound and it is tempting to try to probe it empirically. Of course an experiment cannot satisfy the conditions of rigorous proof required by the definition of a $\rho$-PTAS, but suppose the data is interpreted as a means to estimate an empirical $\rho$, as follows: when the processor finds an incorrect solution one counts the number of satisfied clauses $n_e$ for the excited state it found and compares to the correct solution for that instance, i.e., the true maximal number of clauses $n_t$ that can be satisfied. The ratio $\rho'=n_e/n_t$ is the empirical $\rho$ for that instance. One can then analyze the distribution of empirical $\rho$ values over all instances, and compare it to existing classical bounds. 
Note that $\rho'$ cannot be used in a straightforward manner to infer anything about the P versus NP question, since even if $\rho'>21/22$ the inapproximability result states that this violates P$\neq$NP only if the inequality can be achieved in poly$(N)$ time. Even if we find that  $\rho'$ appears to be constant as $N$ increases, it could be that we had not picked the ``worst case" distribution, and if we had, it would reduce $\rho'$  below $21/22$ (at least asymptotically). 
Still, we believe this is an interesting question. We suspect that instances near $\a_c$ are ``hard to approximate", and if one considers the output of the best $\rho$-PTAS available~\cite{lewin}, $\rho'$ for random instances will be distributed in the range $[0.94,1]$, while we expect that QA will yield a distribution of $\rho'$ values peaked closer to $1$. The question for the future is then whether this may be used to infer anything about the asymptotic computational efficiency of QA.

\acknowledgments
We thank Sergio Boixo, Amit Choubey, Matthias Troyer, and Zhihui Wang for useful discussions and valuable input. The authors gratefully acknowledge funding by the Lockheed Martin Corporation, by ARO-MURI grant W911NF-11-1-0268, and by ARO-QA grant number W911NF-12-1-0523.

\appendix

\section{Experimental settings}
\label{app:DW}

\begin{figure}[t] 
\centering
\includegraphics[width=3in]{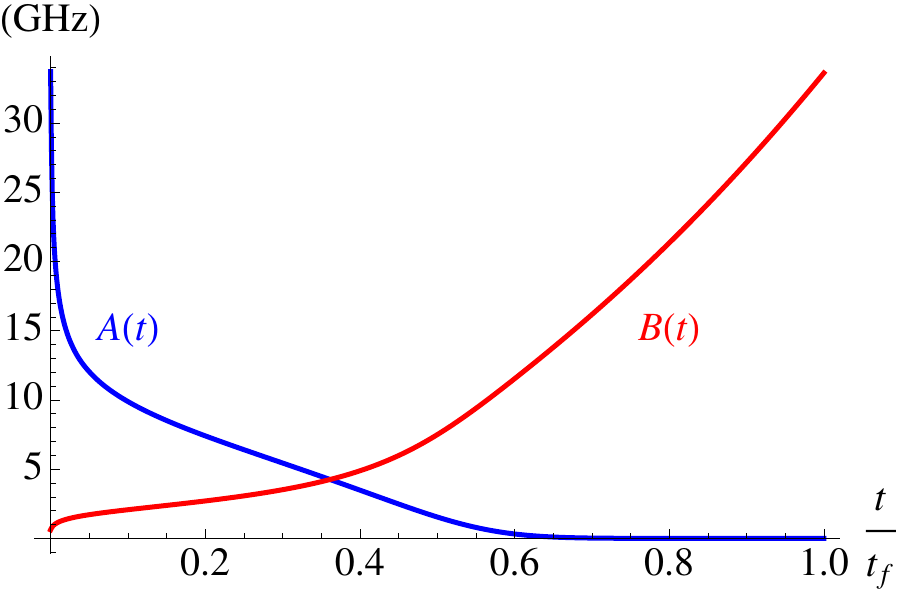} 
\caption{The experimental annealing schedules  $A(t)$ and $B(t)$ appearing in the system Hamiltonian $H(t)$ [Eq.~\eqref{adbH}].}
   \label{fig:A_B}
\end{figure}

Our experiments were performed using the D-Wave One Rainier processor at the USC Information Sciences Institute, comprising $16$ unit cells of $8$ superconducting flux qubits each, with a total of $108$ functional qubits.  The couplings are programmable superconducting inductances.  Fig.~\ref{conn} is a schematic of the device, showing the allowed couplings between the qubits which form a ``Chimera'' graph \cite{Choi1,Choi2}.  The qubits and unit cell, readout, and control have been described in detail elsewhere \cite{Harris2010,Berkley:2010zr,Johnson:2010ys}. The processor performs a quantum annealing protocol to find the ground state of a classical Ising Hamiltonian, as described by the transverse Ising Hamiltonian in Eq.~\eqref{adbH}.  The initial energy scale for the transverse field is 33.7GHz (the $A$ function in Fig.~\ref{fig:A_B}), ensuring that the initial state is to an excellent approximation a uniform superposition in the computational basis, with any deviations mainly due to control errors resulting in non-uniformity in the values of the local transverse fields. The final energy scale for the Ising Hamiltonian (the $B$ function) is $33.6$GHz, about $15$ times the experimental temperature of $17$mK $\approx 2.3$GHz.  

We performed $100$ runs 
for each problem instance on the DW1. Each run returns a state measured in the computational basis (eigenvectors of $\sigma^z$), i.e., a proposed solution. Applying $H_P$ as given in Eq.~\eqref{eq:H_P-map} to this state yields the corresponding energy. The success probability, $p(\a,N)$, is defined as fraction of times (out of $100$) the measured state is the ground state, i.e., is the correct solution as verified against the guaranteed-correct solution returned by \texttt{akmaxsat}. 

We used default settings for the DW1, including programming and thermalization times of 1ms, and an annealing time of 1ms, which we did not attempt to optimize. Moreover, we did not average over different choices of subsets of active qubits, nor did we consider different ``gauges'', i.e., reassignments of qubit up/down values which leave the spectrum invariant \cite{q-sig}. Thus our results may have been affected by systematic flux and coupler biases. However, removing such biases via averaging and gauges would have only improved the DW1 performance, so that our results can be viewed as lower performance bounds.

\section{Maximum possible value of the clause density}
\label{app:max-alpha}
What is the maximum $\a$ allowing for a meaningful ensemble of problems to be generated?

Consider a single edge between two nodes $x_1,x_2$. Then the largest formula we can generate has $4$ clauses:
\begin{align}
 F=(x_1\lor x_2)\land(\neg x_1\lor x_2)\land(x_1\lor \neg x_2)\land(\neg x_1 \lor \neg x_2)
\label{F} 
\end{align}
Of course this problem instance is UNSAT, but is in principle allowed. If we only allow $M=3$ clauses then in the example above there are $\binom{N}{M}=4$ possible formulae.

More generally, given a total of $C$ edges between a total of $N$ variables, we can use each edge to generate $4$ clauses as in Eq.~\eqref{F}, and thus have one unique formula with $M_{\max} = 4C$ clauses. Again, this particular formula will be UNSAT, but is allowed. 

Now, for any $n\leq N$ we can always choose a subgraph (of the Chimera graph) that has $c$ edges between the $n$ nodes and contributes up to $4$ unique clauses. Thus $\a_{\max}(n)=4c/n$, but for any $n \leq N$ there are $\binom{N}{n}$ choices of the subgraph, each of which might have a different number of edges. We claim that the ratio $c/n$ is maximized for $n=N$ which lets us use all $c=C$ edges.
To see this suppose we start with all $N=108$ qubits and $C=255$ edges. Removing any qubit would result in the loss of at most $6$ edges, and $5$ edges on average given the actual connectivity of the Chimera graph. Thus we should check whether 
$(C-5)/(N-1) < C/N$, which holds at $N=108$ and $C=255$.
Also, for any number of qubits $n<N$ and edges $c<255$ between them, we can check that the same inequality holds: $(c-5)/(n-1) < c/n$.
This means that the ratio $c/n$ increases with $n$ and thus will be maximal at the end point $n=N=108$.

This establishes that the maximum possible clause density is $\a^{*}_{\max} = 4C/N$ corresponding to one unique UNSAT problem.

However, a unique formula is unsuitable when an ensemble is desired, as in our case. Thus consider clause densities $3C/N <\a < 4C/N$. For formulae with $\a$ in this range we note that there will necessarily be a combination of the type of Eq.~\eqref{F}). This can be proven by the pigeon hole principle: Let $\a=\frac{3C+c}{N}$. If we choose $3$ possible clauses from each of the $C$ edges the formula may still be SAT; however we may choose the remaining $c$ clauses only from the unchosen remaining clauses from $c$ edges. This means that the problem is guaranteed to be UNSAT. 

This brings us to the range $\a<3C/N$. In this range we can have ensembles with $\binom{4C}{3C-a}$ unique formulae corresponding to problems with $\a=\frac{3C-a}{N}$. In practice we chose $\a_{\max} = 2$, which guaranteed that some of our problem instances were SAT and that we had large enough ensembles (at least $500$ instances for each value of $\a$ and $N$).

\begin{figure}[t]
\includegraphics[width=1\columnwidth]{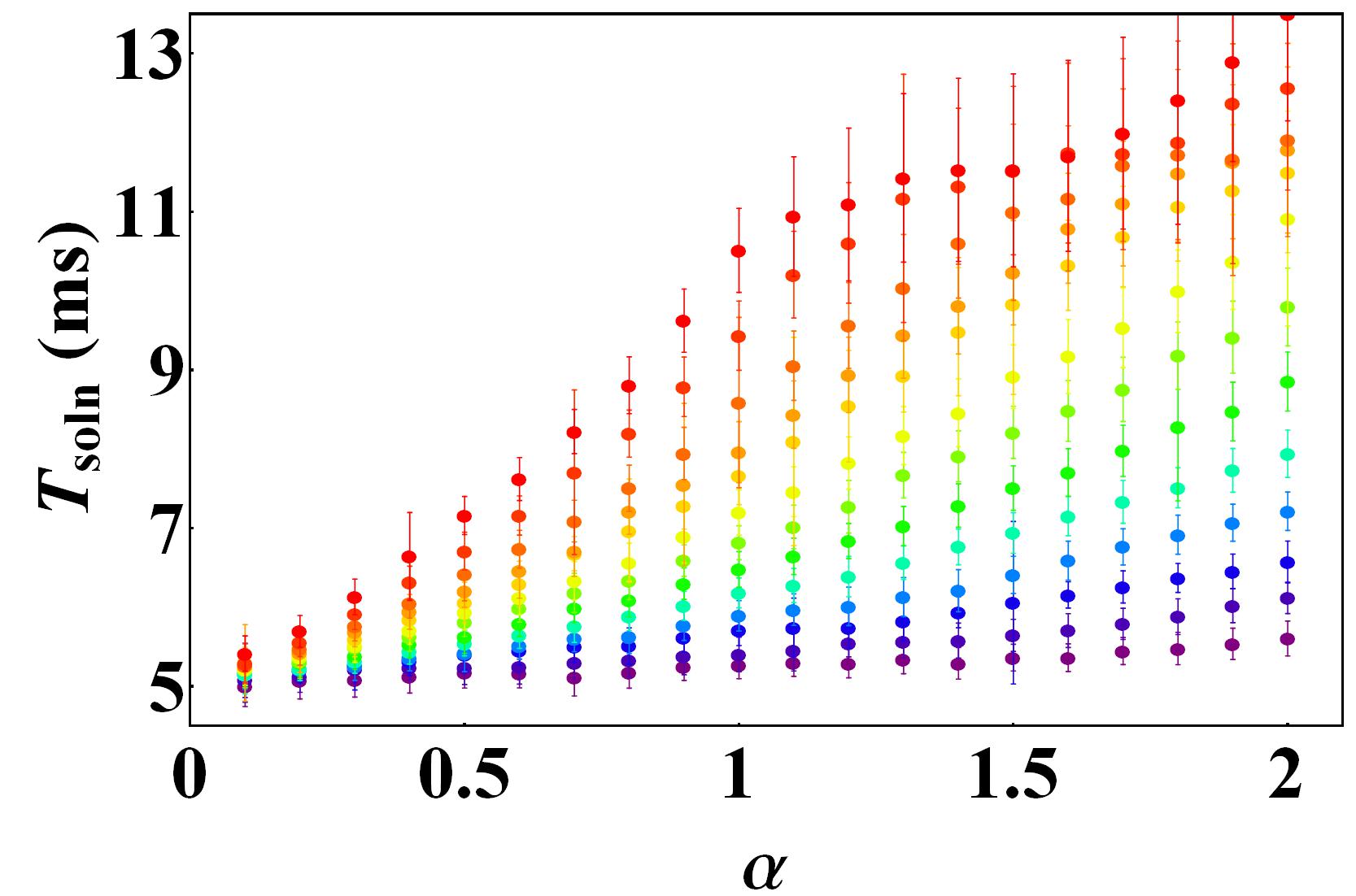}
\includegraphics[width=1\columnwidth]{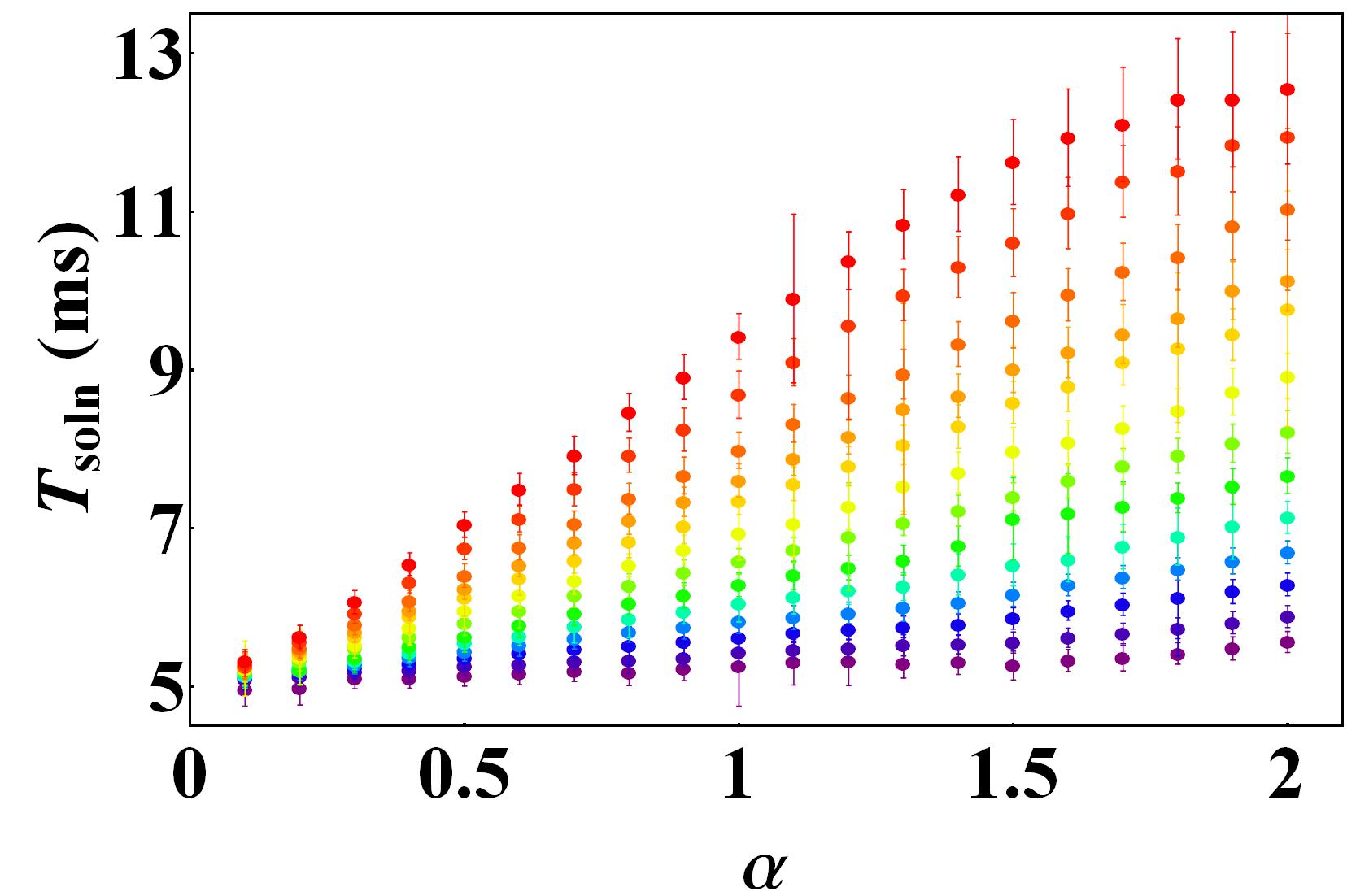}
\caption{Solution time vs $\a$ for akmaxsat using random (top) and processor-compatible (bottom) instances for all $N$ values, increasing from bottom (purple, $N=16$) to top (red, $N=108$). }
\label{fig:t-vs-a}
\end{figure}

\begin{figure*}[t]
\centering
\includegraphics[width=2\columnwidth]{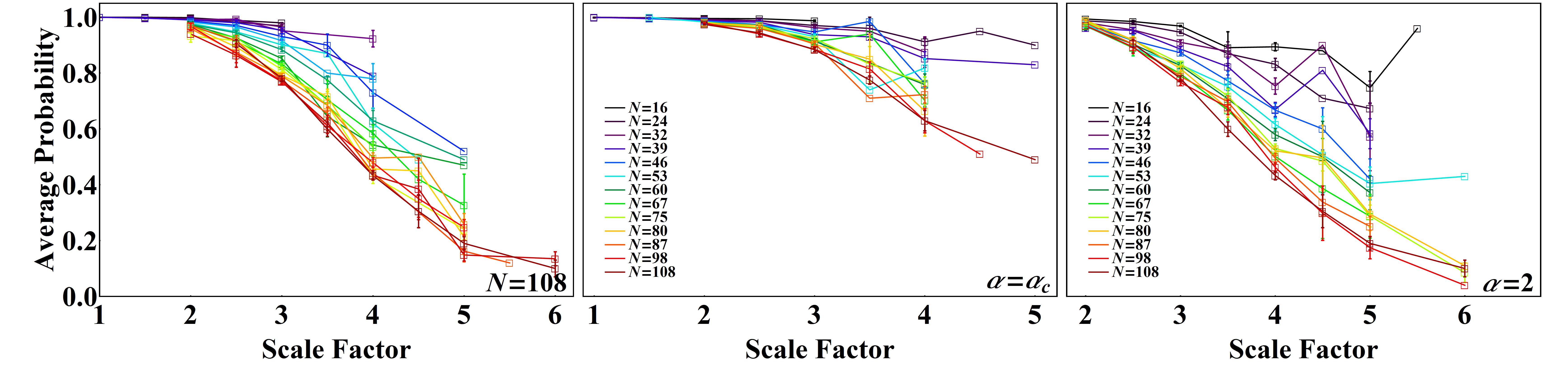}
\caption{Left: Probability of obtaining the correct solution for DW1 versus the scale factor determined by $\max(|h_i|/2,|J_{ij}|)$ for $N=108$ and clause densities ranging from $\alpha=0.1$ (black, top) to $\alpha=2.0$ (red, bottom) in increments of 0.1. Middle: Probability of obtaining the correct solution for DW1 versus the scale factor for $\alpha=1.0$. Right: the same for $\alpha=2.0$.}
\label{fig:scale-vs-alpha-N}
\end{figure*}

\begin{figure*}[t]
\includegraphics[width=1\columnwidth]{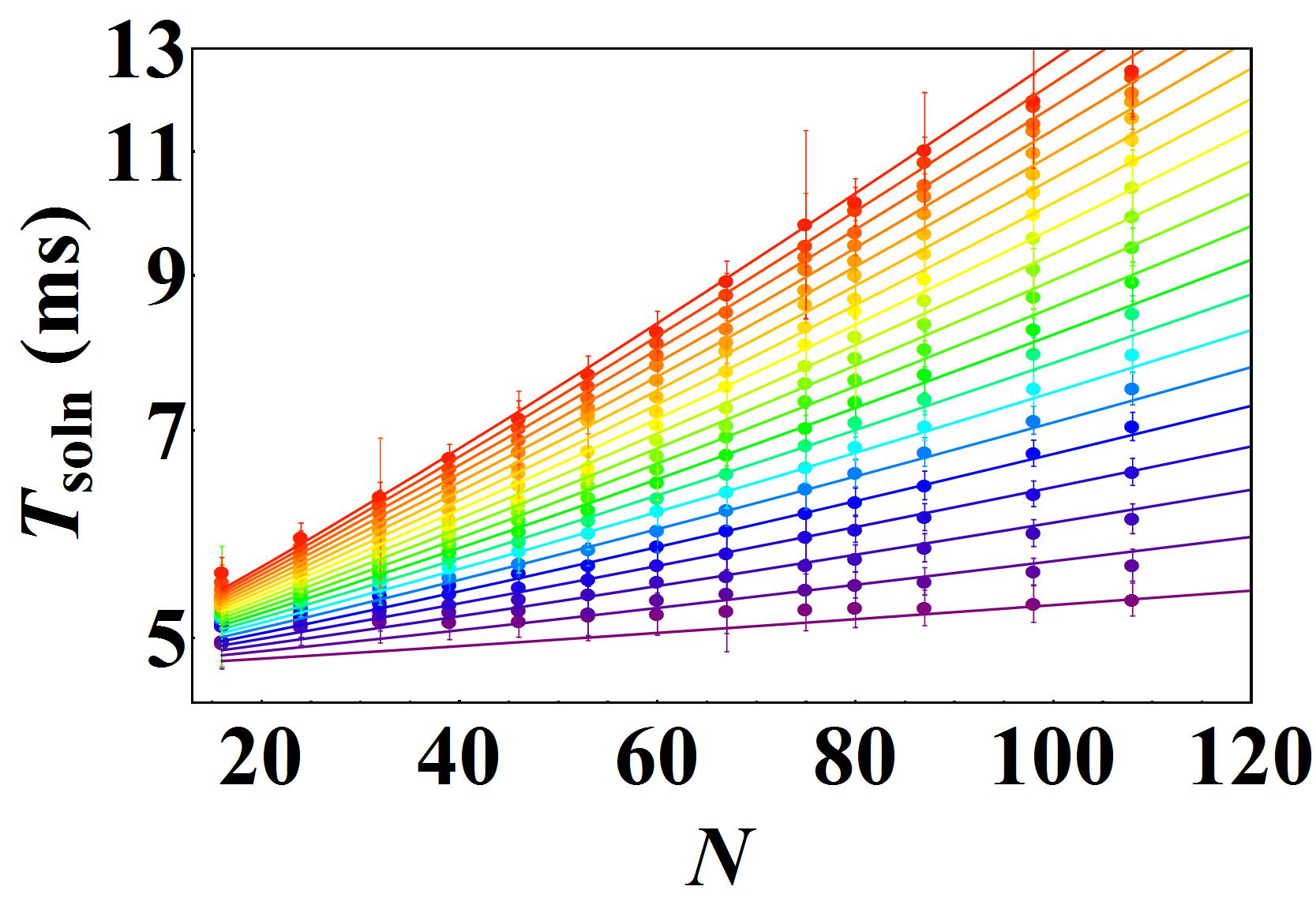}
\includegraphics[width=1\columnwidth]{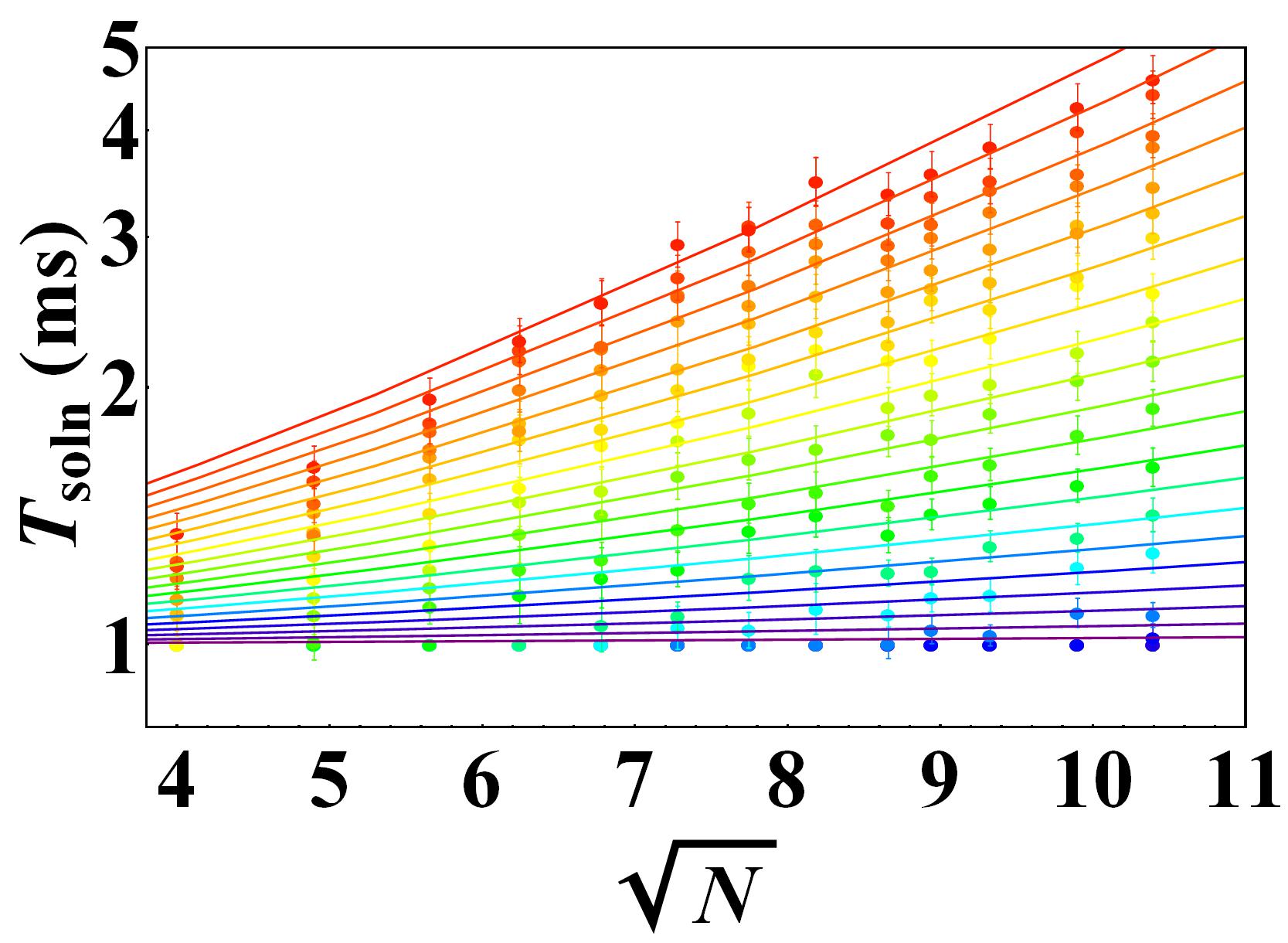}
\caption{Solution time vs $N$ for akmaxsat (left) and DW1 (right). Color variations denote $\a$ from $\a=0.1$ (purple, bottom) to $\a=2.0$ (red, top) with corresponding best fits. The fit shown is the one in Eq.~\eqref{eq:exp_fit} with the fit parameters resulting from the data collapse.}
\label{fig:t-vs-N_dc}
\end{figure*}

\begin{figure*}[t]
\includegraphics[width=0.9\columnwidth]{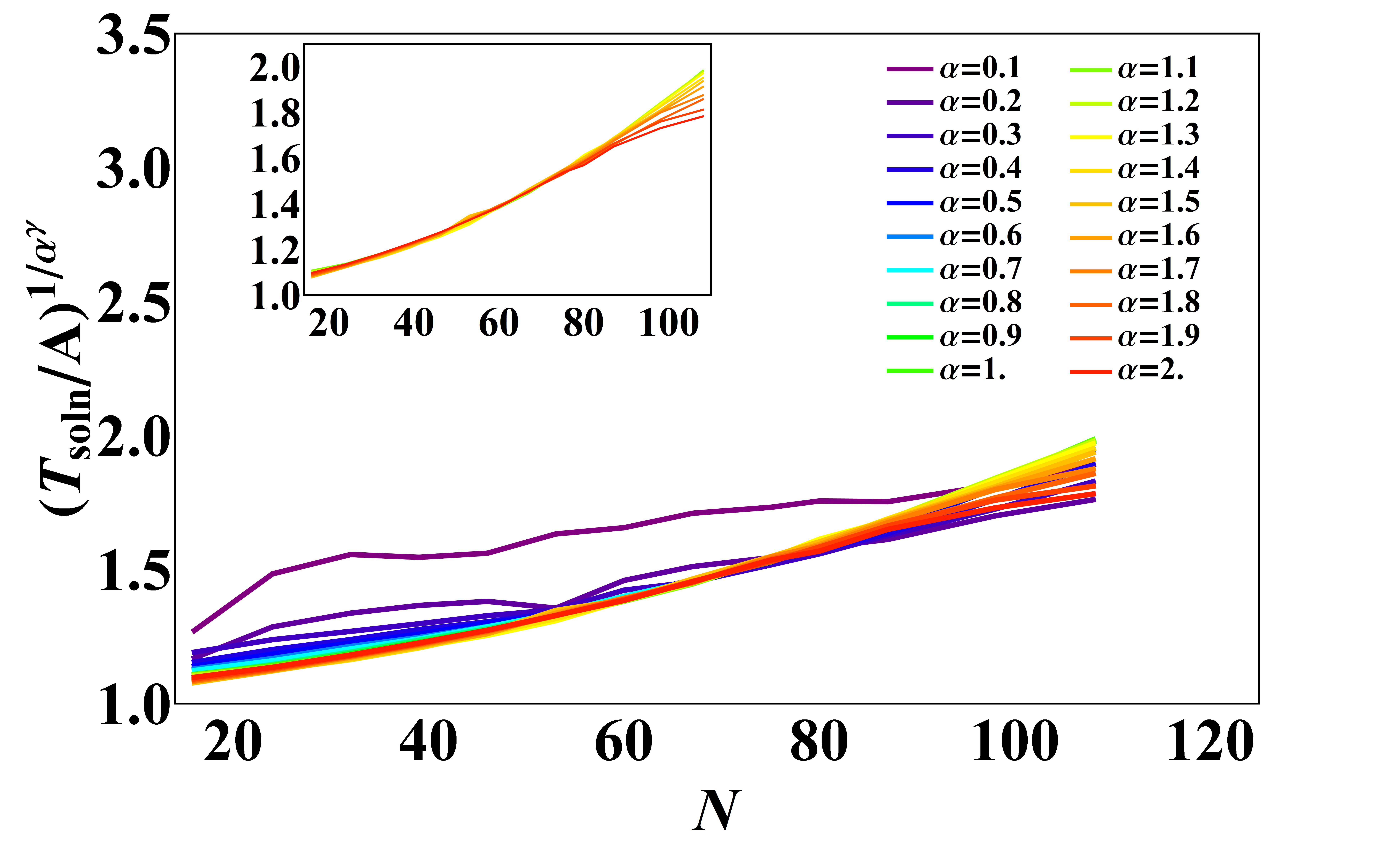}
\includegraphics[width=0.9\columnwidth]{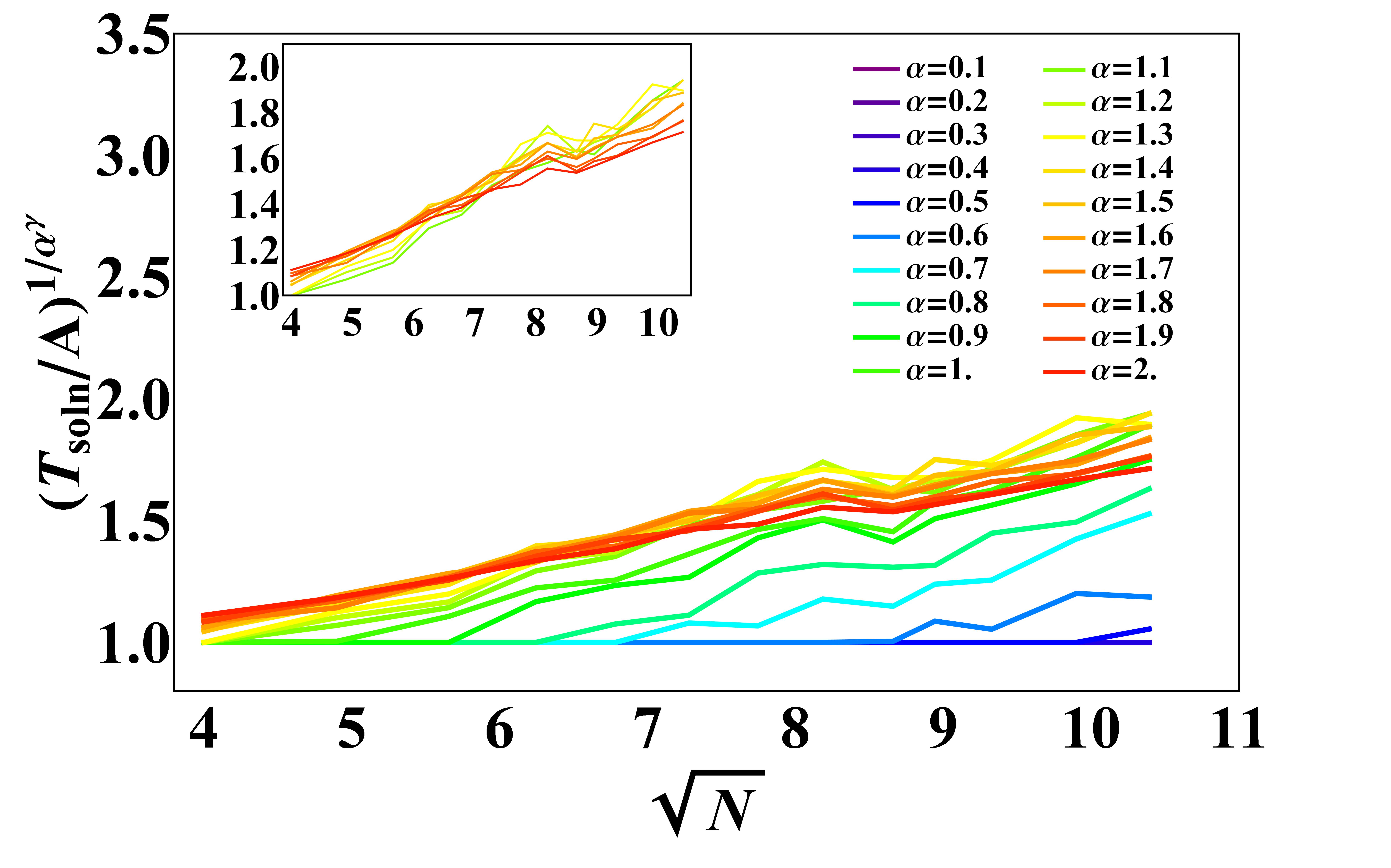}
\caption{Data collapse for solution time vs problem size $N$ for \ak (left) and DW1 (right) for all $\alpha$ using the fitting function given in Eq.~\eqref{eq:exp_fit}. The data collapses well for larger $\a$ values, as can be seen from the inset of each plot where the collapse for $\a\geq1$ is shown. The outlier for \ak is $\a=0.1$. Data collapse for the DW1 data is poor for $\a < \a_c$.}
\label{fig:t-vs-N_collapse}
\end{figure*}

\begin{figure}[tp]
\includegraphics[width=1\linewidth]{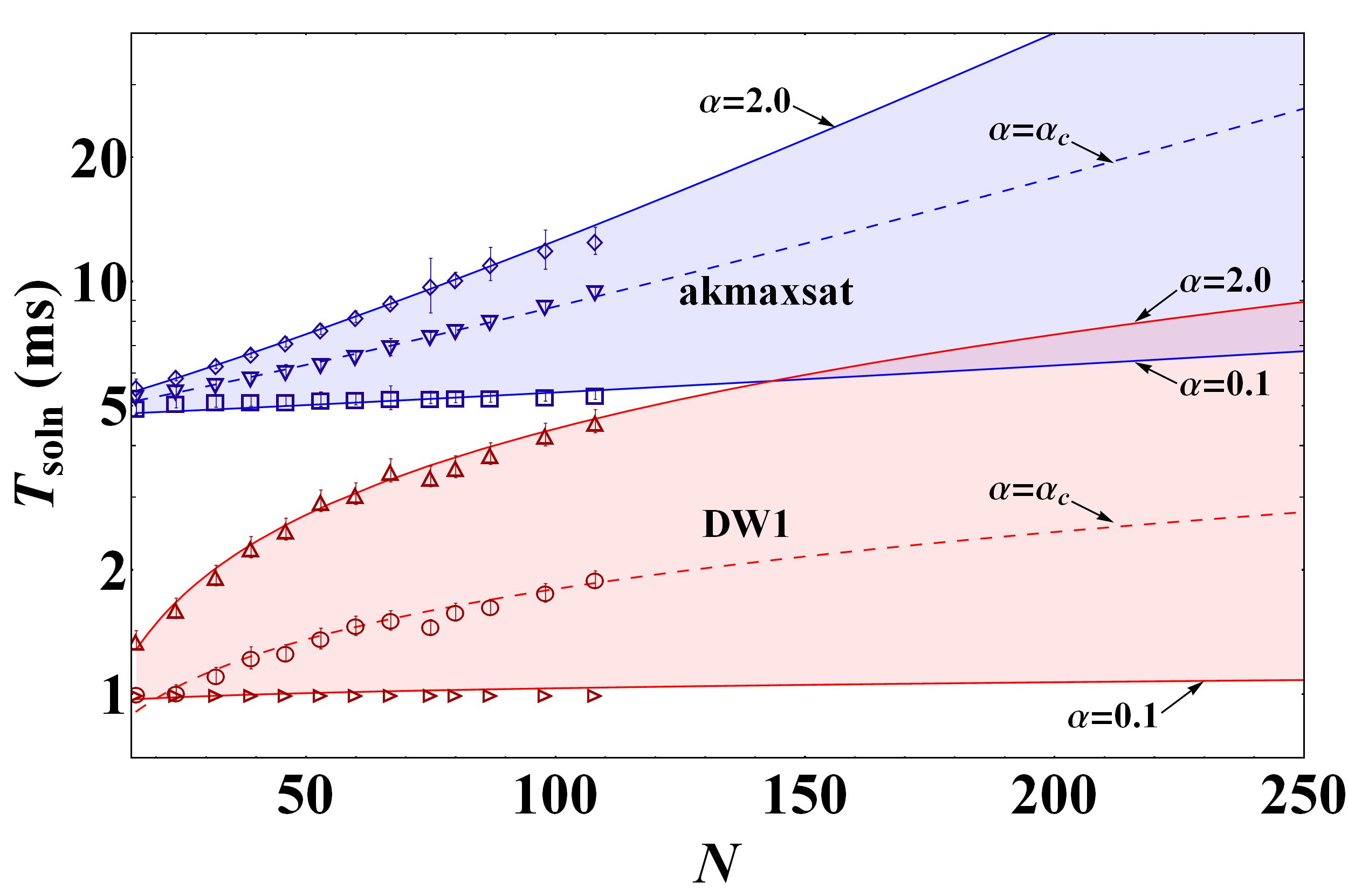}
\caption{(color online) Computational effort comparison between \texttt{akmaxsat} and the DW1 as a function of the problem size $N$ and for different clause densities $\a$. Same as in Fig.~\ref{fig:DW1 processor_vs_ak} except that the fit is given by Eq.~\eqref{eq:exp_fit3} with parameter values from Table~\ref{tbl:fitparams2}. }
\label{fig:fit2}
\end{figure}

\section{Parameter settings for \ak}
\label{app:ak}
The \ak program was run on a Quad-Core AMD $\textrm{Opteron}^{\textrm{TM}}$ Processor 1389 with a clock speed of 2.89 GHz. We chose the version of the algorithm which specifies an initial upper bound of infinity on the solution time as opposed to utilizing the UBCSAT stochastic local search solver to obtain an initial upper bound on the solution. Examining the scaling of the solution time as a function of $N$ for $\alpha=2$, we found that an initial upper bound of infinity results in shorter mean optimal solution times and a more favorable scaling with $N$. Similar results were obtained when varying $\alpha$ for $N=108$. Both comparisons utilized UBCSAT with an Iterated Robust TABU Search (IRoTS) of $10$ iterations.

\section{\texttt{akmaxsat} on random versus DW1-compatible ensembles}
\label{app:t-vs-a}

We compared the performance of \texttt{akmaxsat} on unconstrained versus DW1 processor-compatible ensembles in Section~\ref{subsec:classical} of the main text. Figure~\ref{fig:t-vs-a} presents the same data as in Fig.~\ref{fig:ak_time}, but resolved in more detail for ease of comparison. Solution times are slightly larger for the unconstrained random ensemble $\mathcal{E}(N,\a)$ than for the constrained one $\mathcal{E}_{\textrm{DW}}(N,\a)$, with a noticeable jump at $\alpha=1$ in the former, but not in the latter. This suggests that the constrained ensemble might remove some of the hardest problems, which could be understood as a consequence of the maximum degree $6$ of the Chimera graph.

\section{Rescaling the parameter values}
\label{app:rescale}

The DW1 processor is programmed via a user interface by specifying the qubits and the values of the local fields $h_j$ and couplers $J_{ij}$ as floating point numbers to three bits of precision. The allowed ranges are $h_j \in [-2,2]$ and $J_{ij} \in [-1,1]$. When values outside this range are specified, and the ``autoscale" function is used, as in our case, the DW1 rescales all local fields and couplers by $\max\{|h_j|/2,|J_{ij}|\}$. Control errors result in Gaussian distributed values of the local fields and couplers with respective standard deviations of $\sim 5\%$ and $\sim 10\%$, i.e., $h_j^{\textrm{implemented}} = h_j^{\textrm{specified}} \pm 0.1$ and $J_{ij}^{\textrm{implemented}} = J_{ij}^{\textrm{specified}} \pm 0.1$ \cite{Trevor}. 

Consider a formula $F$ with $M$ clauses, $N$ variables and let us focus on the variable $x_1$ that appears the most, $n_1$ times, in the formula. After we convert the clauses in the formula to local terms in the problem Hamiltonian, suppose that from each of the $n_1$ clauses that $x_1$ participates in, the field contributions are $h_i$, with $i\in\{1,2,\dots,n_1\}$.
The value of the local field for $x_1$, $h^F_1$, corresponding to $F$ is:
\begin{align}
h^F_1=h_1+h_2+\cdots+h_{n_1}
\end{align}
Now we note that $n_1\leq 24$, since the Chimera graph degree of $x_1$ is $\leq 6$ and for each edge $x_1$ can appear in at most $4$ clauses. But if $n_1=24$ then $h^F_1=0$ because it appears negated the same number of times as unnegated. The local field of $x_1$ is maximized if it appears in clauses unnegated for all the couplings that participate in the formula. This means that $n_1=2\times 6$ with $h_1=h_2=\dots =h_{n_1}=+1$, and the maximum possible local field for any of our problems is $h^F_1=12$. Thus rescaling involve division by at most $6$, resulting in fractional coupler and field values, and in such cases the uncertainty of $0.1$ in setting the couplers and local fields could have caused the 2SAT formula to be unfaithfully represented. Such cases probably contributed to lowering the success probability, simply because the ``wrong" problem was solved by the DW1 processor. However, \textit{a priori} for our ensemble of problems the situation is somewhat better, since we imposed a maximum value of $\a=2$ which means that, on an average, each qubit participated in two clauses and the maximum local field strength is $2$, where rescaling is not required. 

To investigate the contribution of such control errors we plot in the left panel of Fig.~\ref{fig:scale-vs-alpha-N} the mean success probability at $N=108$, and for all values of $\alpha$, as a function of the scale factor required to force all local field and coupler values into their allowed ranges. The scale factor is seen to significantly impact the success probability, with an impact that grows with the clause density. At the high end of $\alpha$ values the decrease in success probability is roughly linear with the scale factor. The middle and right panels Figure~\ref{fig:scale-vs-alpha-N} show the impact of the scale factor at $\alpha = 1.0$ and $2.0$ for all values of $N$. From this perspective too it is seen that the larger the scale factor the smaller the success probability, an effect that increases with $N$. Thus rescaling, which results in the $\pm 0.1$ uncertainty in fields and couplings becoming important, explains part of the reduction in the experimental success probability.

\section{Data collapse analysis}
\label{app:collapse}
In Fig.~\ref{fig:DW1 processor_vs_ak} we display best fits for the solution time as a function of problem size for a range of clause densities. Figure~\ref{fig:t-vs-N_dc} shows the complete data set, where the fit parameters are  determined via an initial least squares fit and then a data collapse analysis. 
 As a result, we find that for \texttt{akmaxsat} and DW1 $\g=3/4$ and $\g=3/2$, respectively, correspond to a strong collapse of the data particularly for $\a\geq\a_c$. In Fig.~\ref{fig:t-vs-N_collapse}, the result of the data collapse is shown for \ak (left) and the DW1 (right) with the collapse for $\a\geq\a_c$ shown in the insets. 

To complement the fit given in Eq.~\eqref{eq:exp_fit} and Table~\ref{tbl:fitparams} which use a restricted set of fitting parameters, we present a second, less constrained fit, which includes a purely $\a$-dependent part:
\begin{equation}
T_\text{soln}(\a,N)=A\exp\left(B\a^{\g}N^{\d}\right)+C\exp\left(D\a^{\zeta}\right)+E\a+F
\label{eq:exp_fit3}
\end{equation}

We find the results given in Table~\ref{tbl:fitparams2}, and the fit is shown in Fig.~\ref{fig:fit2}. Note that in this case we did not use a data collapse. The coefficient of determination for the \texttt{akmaxsat} data is identical to that for the first fit (Table~\ref{tbl:fitparams}), while this coefficient improves from $0.991$ for the DW1 data, so the second fit is slightly better. However, this improvement comes at the expense of introducing many more free parameters. Note that again $\d-\g$ is positive for \texttt{akmaxsat} and negative for the DW1 (recall the discussion in Section~\ref{sec:const-M}).  

\begin{table}[th]
\begin{tabular}{cccccc}
 & $\g$ & $\d$  & $\zeta$ & $B$ & $D$ \\ \hline
AK  & $0.673[1]$ & $1.143[3]$  & $0.188[7]$ & $0.003[2]$  & $0.188[7]$ \\
DW1  & $1.589[6]$ & $0.637[8]$  & $0.003[0]$ & $0.003[0]$ & $2.368[7]$
\end{tabular}
\begin{tabular}{cccccc}
 & $A$  & $C$ & $E$ & $F$ & $R^2$\\ \hline
AK & $4.537[4]$ & $0.292[0]$  & $0.097[4]$ & $0.000[6]$ & $0.999[6]$\\
DW1 & $22.943[9]$  &  $-2.079[1]$  &  $-0.270[0]$ & $0.045[5]$ & $0.996[9]$
\end{tabular}
\caption{Numerical values obtained from a least squares fit of the data shown in Fig.~\ref{fig:DW1 processor_vs_ak} for both \texttt{akmaxsat} and the DW1. The parameters are the ones appearing in Eq.~\eqref{eq:exp_fit3}.}
\label{tbl:fitparams2}
\end{table}

\begin{figure}[tp]
\centering
\subfigure{\includegraphics[width=0.5\columnwidth]{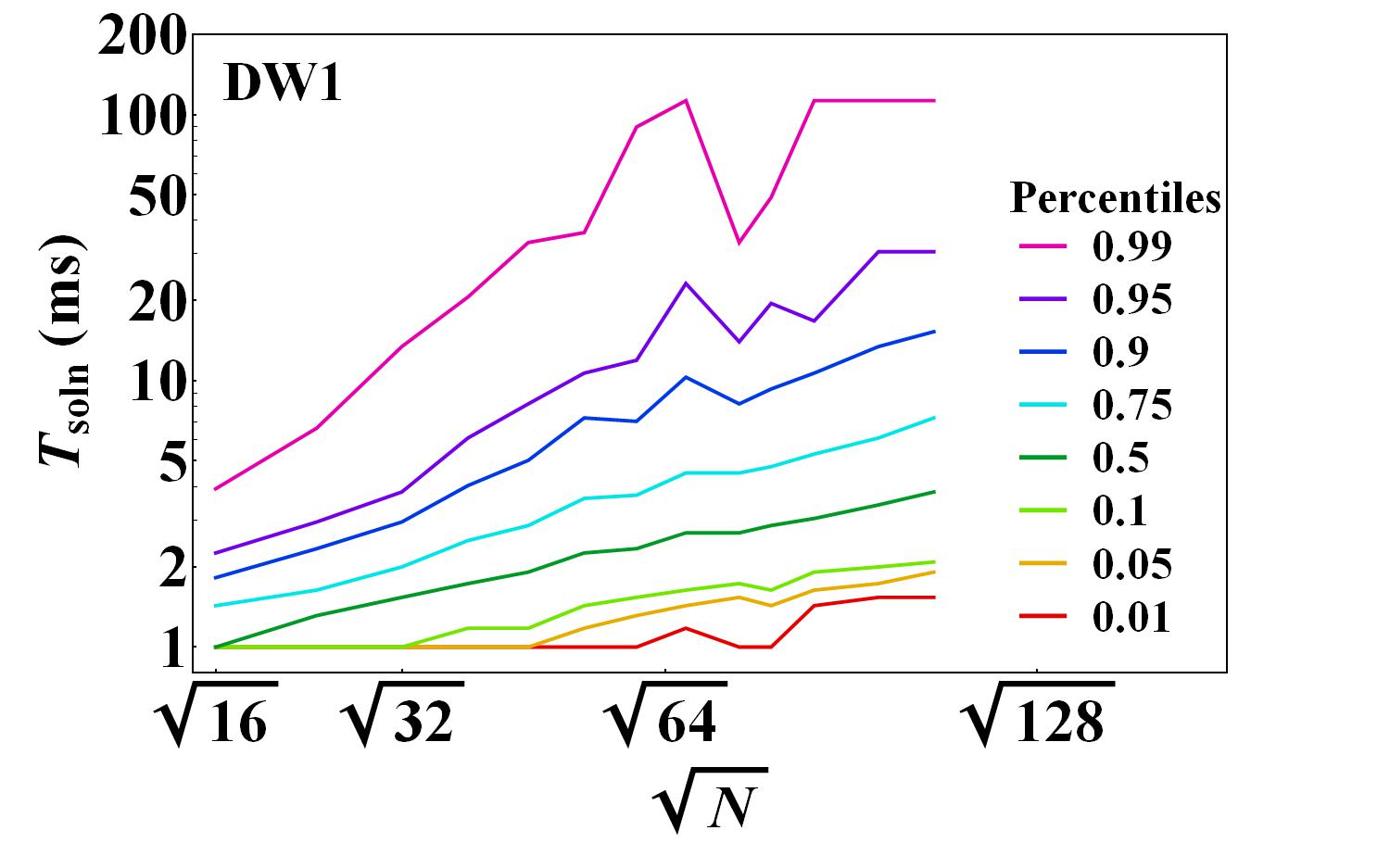}}\hspace*{-0.2cm}
\subfigure{\includegraphics[width=0.5\columnwidth]{QuantilePlot_DW1.jpg}}\hspace*{-0.2cm}
\caption{(color online) Same as Fig.~\ref{fig:quant_plots} but we compare two procedures for extracting the time to solution. Left: Time estimated then percentiles taken. Right: Percentiles taken then time estimated (identical to Fig.~\ref{fig:quant_plots}(b)).}
\label{fig:quant_plots-comp}
\end{figure}

\section{Comparison of different procedures for extracting the time to solution}
\label{app:comp}

In Fig.~\ref{fig:quant_plots-comp} we compare the results of extracting the time to solution in two different orders for $N=108$ and $\a=2.0$. On the left: we first use Eq.~\eqref{eq:time} to compute the time to solution given the success probability $p$ for each instance, create a histogram of the resulting times, and then extract the percentiles. On the right: we first create a histogram of success probabilities for all instances, then calculate the probability $p$ associated with a given percentile, and then use Eq.~\eqref{eq:time} to compute the time to solution for that percentile. It can be seen that the two procedures give essentially identical results.


\begin{thebibliography}{68}%
\makeatletter
\providecommand \@ifxundefined [1]{%
 \@ifx{#1\undefined}
}%
\providecommand \@ifnum [1]{%
 \ifnum #1\expandafter \@firstoftwo
 \else \expandafter \@secondoftwo
 \fi
}%
\providecommand \@ifx [1]{%
 \ifx #1\expandafter \@firstoftwo
 \else \expandafter \@secondoftwo
 \fi
}%
\providecommand \natexlab [1]{#1}%
\providecommand \enquote  [1]{``#1''}%
\providecommand \bibnamefont  [1]{#1}%
\providecommand \bibfnamefont [1]{#1}%
\providecommand \citenamefont [1]{#1}%
\providecommand \href@noop [0]{\@secondoftwo}%
\providecommand \href [0]{\begingroup \@sanitize@url \@href}%
\providecommand \@href[1]{\@@startlink{#1}\@@href}%
\providecommand \@@href[1]{\endgroup#1\@@endlink}%
\providecommand \@sanitize@url [0]{\catcode `\\12\catcode `\$12\catcode
  `\&12\catcode `\#12\catcode `\^12\catcode `\_12\catcode `\%12\relax}%
\providecommand \@@startlink[1]{}%
\providecommand \@@endlink[0]{}%
\providecommand \url  [0]{\begingroup\@sanitize@url \@url }%
\providecommand \@url [1]{\endgroup\@href {#1}{\urlprefix }}%
\providecommand \urlprefix  [0]{URL }%
\providecommand \Eprint [0]{\href }%
\providecommand \doibase [0]{http://dx.doi.org/}%
\providecommand \selectlanguage [0]{\@gobble}%
\providecommand \bibinfo  [0]{\@secondoftwo}%
\providecommand \bibfield  [0]{\@secondoftwo}%
\providecommand \translation [1]{[#1]}%
\providecommand \BibitemOpen [0]{}%
\providecommand \bibitemStop [0]{}%
\providecommand \bibitemNoStop [0]{.\EOS\space}%
\providecommand \EOS [0]{\spacefactor3000\relax}%
\providecommand \BibitemShut  [1]{\csname bibitem#1\endcsname}%
\let\auto@bib@innerbib\@empty
\bibitem [{\citenamefont {Farhi}\ \emph {et~al.}(2000)\citenamefont {Farhi},
  \citenamefont {Goldstone}, \citenamefont {Gutmann},\ and\ \citenamefont
  {Sipser}}]{FarhiAQC:00}%
  \BibitemOpen
  \bibfield  {author} {\bibinfo {author} {\bibfnamefont {E.}~\bibnamefont
  {Farhi}}, \bibinfo {author} {\bibfnamefont {J.}~\bibnamefont {Goldstone}},
  \bibinfo {author} {\bibfnamefont {S.}~\bibnamefont {Gutmann}}, \ and\
  \bibinfo {author} {\bibfnamefont {M.}~\bibnamefont {Sipser}},\ }\href
  {http://arxiv.org/abs/quant-ph/0001106} {\bibfield  {journal} {\bibinfo
  {journal} {ArXiv}\ } (\bibinfo {year} {2000})},\ \Eprint
  {http://arxiv.org/abs/quant-ph/0001106 (2000)} {quant-ph/0001106 (2000)}
  \BibitemShut {NoStop}%
\bibitem [{\citenamefont {Farhi}\ \emph {et~al.}(2001)\citenamefont {Farhi},
  \citenamefont {Goldstone}, \citenamefont {Gutmann}, \citenamefont {Lapan},
  \citenamefont {Lundgren},\ and\ \citenamefont {Preda}}]{Farhi:01}%
  \BibitemOpen
  \bibfield  {author} {\bibinfo {author} {\bibfnamefont {E.}~\bibnamefont
  {Farhi}}, \bibinfo {author} {\bibfnamefont {J.}~\bibnamefont {Goldstone}},
  \bibinfo {author} {\bibfnamefont {S.}~\bibnamefont {Gutmann}}, \bibinfo
  {author} {\bibfnamefont {J.}~\bibnamefont {Lapan}}, \bibinfo {author}
  {\bibfnamefont {A.}~\bibnamefont {Lundgren}}, \ and\ \bibinfo {author}
  {\bibfnamefont {D.}~\bibnamefont {Preda}},\ }\href@noop {} {\bibfield
  {journal} {\bibinfo  {journal} {Science}\ }\textbf {\bibinfo {volume}
  {292}},\ \bibinfo {pages} {472} (\bibinfo {year} {2001})}\BibitemShut
  {NoStop}%
\bibitem [{\citenamefont {Aharonov}\ \emph {et~al.}(2007)\citenamefont
  {Aharonov}, \citenamefont {van Dam}, \citenamefont {Kempe}, \citenamefont
  {Landau}, \citenamefont {Lloyd},\ and\ \citenamefont {Regev}}]{Aharonov:04}%
  \BibitemOpen
  \bibfield  {author} {\bibinfo {author} {\bibfnamefont {D.}~\bibnamefont
  {Aharonov}}, \bibinfo {author} {\bibfnamefont {W.}~\bibnamefont {van Dam}},
  \bibinfo {author} {\bibfnamefont {J.}~\bibnamefont {Kempe}}, \bibinfo
  {author} {\bibfnamefont {Z.}~\bibnamefont {Landau}}, \bibinfo {author}
  {\bibfnamefont {S.}~\bibnamefont {Lloyd}}, \ and\ \bibinfo {author}
  {\bibfnamefont {O.}~\bibnamefont {Regev}},\ }\href@noop {} {\bibfield
  {journal} {\bibinfo  {journal} {SIAM J. Comp.}\ }\textbf {\bibinfo {volume}
  {37}},\ \bibinfo {pages} {166} (\bibinfo {year} {2007})}\BibitemShut
  {NoStop}%
\bibitem [{\citenamefont {Kempe}\ \emph {et~al.}(2006)\citenamefont {Kempe},
  \citenamefont {Kitaev},\ and\ \citenamefont {Regev}}]{kempe:1070}%
  \BibitemOpen
  \bibfield  {author} {\bibinfo {author} {\bibfnamefont {J.}~\bibnamefont
  {Kempe}}, \bibinfo {author} {\bibfnamefont {A.}~\bibnamefont {Kitaev}}, \
  and\ \bibinfo {author} {\bibfnamefont {O.}~\bibnamefont {Regev}},\ }\href
  {\doibase 10.1137/S0097539704445226} {\bibfield  {journal} {\bibinfo
  {journal} {SIAM Journal on Computing}\ }\textbf {\bibinfo {volume} {35}},\
  \bibinfo {pages} {1070} (\bibinfo {year} {2006})}\BibitemShut {NoStop}%
\bibitem [{\citenamefont {Siu}(2005)}]{Siu:2005:062314}%
  \BibitemOpen
  \bibfield  {author} {\bibinfo {author} {\bibfnamefont {M.~S.}\ \bibnamefont
  {Siu}},\ }\href {\doibase 10.1103/PhysRevA.71.062314} {\bibfield  {journal}
  {\bibinfo  {journal} {Phys. Rev. A}\ }\textbf {\bibinfo {volume} {71}},\
  \bibinfo {pages} {062314} (\bibinfo {year} {2005})}\BibitemShut {NoStop}%
\bibitem [{\citenamefont {Oliveira}\ and\ \citenamefont
  {Terhal}(2005)}]{Oliveira:05}%
  \BibitemOpen
  \bibfield  {author} {\bibinfo {author} {\bibfnamefont {R.}~\bibnamefont
  {Oliveira}}\ and\ \bibinfo {author} {\bibfnamefont {B.}~\bibnamefont
  {Terhal}},\ }\href@noop {} {\bibfield  {journal} {\bibinfo  {journal}
  {Quantum Inf. Comput.}\ }\textbf {\bibinfo {volume} {8}},\ \bibinfo {pages}
  {0900} (\bibinfo {year} {2005})}\BibitemShut {NoStop}%
\bibitem [{\citenamefont {Mizel}\ \emph {et~al.}(2007)\citenamefont {Mizel},
  \citenamefont {Lidar},\ and\ \citenamefont
  {Mitchell}}]{PhysRevLett.99.070502}%
  \BibitemOpen
  \bibfield  {author} {\bibinfo {author} {\bibfnamefont {A.}~\bibnamefont
  {Mizel}}, \bibinfo {author} {\bibfnamefont {D.~A.}\ \bibnamefont {Lidar}}, \
  and\ \bibinfo {author} {\bibfnamefont {M.}~\bibnamefont {Mitchell}},\ }\href
  {\doibase 10.1103/PhysRevLett.99.070502} {\bibfield  {journal} {\bibinfo
  {journal} {Phys. Rev. Lett.}\ }\textbf {\bibinfo {volume} {99}},\ \bibinfo
  {pages} {070502} (\bibinfo {year} {2007})}\BibitemShut {NoStop}%
\bibitem [{\citenamefont {Childs}\ \emph {et~al.}(2001)\citenamefont {Childs},
  \citenamefont {Farhi},\ and\ \citenamefont {Preskill}}]{PhysRevA.65.012322}%
  \BibitemOpen
  \bibfield  {author} {\bibinfo {author} {\bibfnamefont {A.~M.}\ \bibnamefont
  {Childs}}, \bibinfo {author} {\bibfnamefont {E.}~\bibnamefont {Farhi}}, \
  and\ \bibinfo {author} {\bibfnamefont {J.}~\bibnamefont {Preskill}},\ }\href
  {\doibase 10.1103/PhysRevA.65.012322} {\bibfield  {journal} {\bibinfo
  {journal} {Phys. Rev. A}\ }\textbf {\bibinfo {volume} {65}},\ \bibinfo
  {pages} {012322} (\bibinfo {year} {2001})}\BibitemShut {NoStop}%
\bibitem [{\citenamefont {Sarandy}\ and\ \citenamefont
  {Lidar}(2005)}]{SarandyLidar:05}%
  \BibitemOpen
  \bibfield  {author} {\bibinfo {author} {\bibfnamefont {M.~S.}\ \bibnamefont
  {Sarandy}}\ and\ \bibinfo {author} {\bibfnamefont {D.~A.}\ \bibnamefont
  {Lidar}},\ }\href@noop {} {\bibfield  {journal} {\bibinfo  {journal} {Phys.
  Rev. Lett.}\ }\textbf {\bibinfo {volume} {95}},\ \bibinfo {pages} {250503}
  (\bibinfo {year} {2005})}\BibitemShut {NoStop}%
\bibitem [{\citenamefont {Roland}\ and\ \citenamefont
  {Cerf}(2005)}]{PhysRevA.71.032330}%
  \BibitemOpen
  \bibfield  {author} {\bibinfo {author} {\bibfnamefont {J.}~\bibnamefont
  {Roland}}\ and\ \bibinfo {author} {\bibfnamefont {N.~J.}\ \bibnamefont
  {Cerf}},\ }\href {\doibase 10.1103/PhysRevA.71.032330} {\bibfield  {journal}
  {\bibinfo  {journal} {Phys. Rev. A}\ }\textbf {\bibinfo {volume} {71}},\
  \bibinfo {pages} {032330} (\bibinfo {year} {2005})}\BibitemShut {NoStop}%
\bibitem [{\citenamefont {Tiersch}\ and\ \citenamefont
  {Sch{\"{u}}tzhold}(2007)}]{PhysRevA.75.062313}%
  \BibitemOpen
  \bibfield  {author} {\bibinfo {author} {\bibfnamefont {M.}~\bibnamefont
  {Tiersch}}\ and\ \bibinfo {author} {\bibfnamefont {R.}~\bibnamefont
  {Sch{\"{u}}tzhold}},\ }\href {\doibase 10.1103/PhysRevA.75.062313} {\bibfield
   {journal} {\bibinfo  {journal} {Phys. Rev. A}\ }\textbf {\bibinfo {volume}
  {75}},\ \bibinfo {pages} {062313} (\bibinfo {year} {2007})}\BibitemShut
  {NoStop}%
\bibitem [{\citenamefont {Amin}\ \emph {et~al.}(2009)\citenamefont {Amin},
  \citenamefont {Averin},\ and\ \citenamefont
  {Nesteroff}}]{PhysRevA.79.022107}%
  \BibitemOpen
  \bibfield  {author} {\bibinfo {author} {\bibfnamefont {M.~H.~S.}\
  \bibnamefont {Amin}}, \bibinfo {author} {\bibfnamefont {D.~V.}\ \bibnamefont
  {Averin}}, \ and\ \bibinfo {author} {\bibfnamefont {J.~A.}\ \bibnamefont
  {Nesteroff}},\ }\href {\doibase 10.1103/PhysRevA.79.022107} {\bibfield
  {journal} {\bibinfo  {journal} {Phys. Rev. A}\ }\textbf {\bibinfo {volume}
  {79}},\ \bibinfo {pages} {022107} (\bibinfo {year} {2009})}\BibitemShut
  {NoStop}%
\bibitem [{\citenamefont {Amin}\ \emph {et~al.}(2008)\citenamefont {Amin},
  \citenamefont {Love},\ and\ \citenamefont {Truncik}}]{TAQC}%
  \BibitemOpen
  \bibfield  {author} {\bibinfo {author} {\bibfnamefont {M.~H.~S.}\
  \bibnamefont {Amin}}, \bibinfo {author} {\bibfnamefont {P.~J.}\ \bibnamefont
  {Love}}, \ and\ \bibinfo {author} {\bibfnamefont {C.~J.~S.}\ \bibnamefont
  {Truncik}},\ }\href {\doibase 10.1103/PhysRevLett.100.060503} {\bibfield
  {journal} {\bibinfo  {journal} {Phys. Rev. Lett.}\ }\textbf {\bibinfo
  {volume} {100}},\ \bibinfo {pages} {060503} (\bibinfo {year}
  {2008})}\BibitemShut {NoStop}%
\bibitem [{\citenamefont {{R. Harris \textit{et al.}}}(2010)}]{Harris2010}%
  \BibitemOpen
  \bibfield  {author} {\bibinfo {author} {\bibnamefont {{R. Harris \textit{et
  al.}}}},\ }\href {\doibase 10.1103/PhysRevB.82.024511} {\bibfield  {journal}
  {\bibinfo  {journal} {Phys. Rev. B}\ }\textbf {\bibinfo {volume} {82}},\
  \bibinfo {pages} {024511} (\bibinfo {year} {2010})}\BibitemShut {NoStop}%
\bibitem [{\citenamefont {{M.W. Johnson \textit{et
  al.}}}(2011)}]{Johnson:2011ys}%
  \BibitemOpen
  \bibfield  {author} {\bibinfo {author} {\bibnamefont {{M.W. Johnson
  \textit{et al.}}}},\ }\href {\doibase DOI 10.1038/nature10012} {\bibfield
  {journal} {\bibinfo  {journal} {Nature}\ }\textbf {\bibinfo {volume} {473}},\
  \bibinfo {pages} {194} (\bibinfo {year} {2011})}\BibitemShut {NoStop}%
\bibitem [{\citenamefont {{Dickson, N. G. \textit{et al.}}}(2013)}]{DWave-16q}%
  \BibitemOpen
  \bibfield  {author} {\bibinfo {author} {\bibnamefont {{Dickson, N. G.
  \textit{et al.}}}},\ }\href {\doibase 10.1038/ncomms2920} {\bibfield
  {journal} {\bibinfo  {journal} {Nat. Commun.}\ }\textbf {\bibinfo {volume}
  {4}},\ \bibinfo {pages} {1903} (\bibinfo {year} {2013})}\BibitemShut
  {NoStop}%
\bibitem [{\citenamefont {Boixo}\ \emph {et~al.}(2013)\citenamefont {Boixo},
  \citenamefont {Albash}, \citenamefont {Spedalieri}, \citenamefont
  {Chancellor},\ and\ \citenamefont {Lidar}}]{q-sig}%
  \BibitemOpen
  \bibfield  {author} {\bibinfo {author} {\bibfnamefont {S.}~\bibnamefont
  {Boixo}}, \bibinfo {author} {\bibfnamefont {T.}~\bibnamefont {Albash}},
  \bibinfo {author} {\bibfnamefont {F.~M.}\ \bibnamefont {Spedalieri}},
  \bibinfo {author} {\bibfnamefont {N.}~\bibnamefont {Chancellor}}, \ and\
  \bibinfo {author} {\bibfnamefont {D.~A.}\ \bibnamefont {Lidar}},\ }\href
  {\doibase 10.1038/ncomms3067} {\bibfield  {journal} {\bibinfo  {journal}
  {Nature Comm.}\ }\textbf {\bibinfo {volume} {4}},\ \bibinfo {pages} {3067}
  (\bibinfo {year} {2013})}\BibitemShut {NoStop}%
\bibitem [{\citenamefont {{S. Boixo \textit{et al.}}}(2013)}]{q100}%
  \BibitemOpen
  \bibfield  {author} {\bibinfo {author} {\bibnamefont {{S. Boixo \textit{et
  al.}}}},\ }\href {http://arxiv.org/abs/1304.4595} {\bibfield  {journal}
  {\bibinfo  {journal} {ArXiv}\ } (\bibinfo {year} {2013})},\ \Eprint
  {http://arxiv.org/abs/arXiv:1304.4595} {arXiv:1304.4595} \BibitemShut
  {NoStop}%
\bibitem [{\citenamefont {{L. Wang \textit{et al.}}}(2013)}]{q100-comment}%
  \BibitemOpen
  \bibfield  {author} {\bibinfo {author} {\bibnamefont {{L. Wang \textit{et
  al.}}}},\ }\href {http://arxiv.org/abs/1305.5837} {\bibfield  {journal}
  {\bibinfo  {journal} {ArXiv}\ } (\bibinfo {year} {2013})},\ \Eprint
  {http://arxiv.org/abs/arXiv:1305.5837} {arXiv:1305.5837} \BibitemShut
  {NoStop}%
\bibitem [{\citenamefont {Brooke}\ \emph {et~al.}(1999)\citenamefont {Brooke},
  \citenamefont {Bitko}, \citenamefont {F.}, \citenamefont {Rosenbaum},\ and\
  \citenamefont {Aeppli}}]{Brooke30041999}%
  \BibitemOpen
  \bibfield  {author} {\bibinfo {author} {\bibfnamefont {J.}~\bibnamefont
  {Brooke}}, \bibinfo {author} {\bibfnamefont {D.}~\bibnamefont {Bitko}},
  \bibinfo {author} {\bibfnamefont {T.}~\bibnamefont {F.}}, \bibinfo {author}
  {\bibnamefont {Rosenbaum}}, \ and\ \bibinfo {author} {\bibfnamefont
  {G.}~\bibnamefont {Aeppli}},\ }\href {\doibase 10.1126/science.284.5415.779}
  {\bibfield  {journal} {\bibinfo  {journal} {Science}\ }\textbf {\bibinfo
  {volume} {284}},\ \bibinfo {pages} {779} (\bibinfo {year}
  {1999})}\BibitemShut {NoStop}%
\bibitem [{\citenamefont {Finnila}\ \emph {et~al.}(1994)\citenamefont
  {Finnila}, \citenamefont {Gomez}, \citenamefont {Sebenik}, \citenamefont
  {Stenson},\ and\ \citenamefont {Doll}}]{finnila_quantum_1994}%
  \BibitemOpen
  \bibfield  {author} {\bibinfo {author} {\bibfnamefont {A.~B.}\ \bibnamefont
  {Finnila}}, \bibinfo {author} {\bibfnamefont {M.~A.}\ \bibnamefont {Gomez}},
  \bibinfo {author} {\bibfnamefont {C.}~\bibnamefont {Sebenik}}, \bibinfo
  {author} {\bibfnamefont {C.}~\bibnamefont {Stenson}}, \ and\ \bibinfo
  {author} {\bibfnamefont {J.~D.}\ \bibnamefont {Doll}},\ }\href {\doibase
  10.1016/0009-2614(94)00117-0} {\bibfield  {journal} {\bibinfo  {journal}
  {Chem. Phys. Lett.}\ }\textbf {\bibinfo {volume} {219}},\ \bibinfo {pages}
  {343} (\bibinfo {year} {1994})}\BibitemShut {NoStop}%
\bibitem [{\citenamefont {Kadowaki}\ and\ \citenamefont
  {Nishimori}(1998)}]{Kadowaki1998}%
  \BibitemOpen
  \bibfield  {author} {\bibinfo {author} {\bibfnamefont {T.}~\bibnamefont
  {Kadowaki}}\ and\ \bibinfo {author} {\bibfnamefont {H.}~\bibnamefont
  {Nishimori}},\ }\href {\doibase 10.1103/PhysRevE.58.5355} {\bibfield
  {journal} {\bibinfo  {journal} {Phys. Rev. E}\ }\textbf {\bibinfo {volume}
  {58}},\ \bibinfo {pages} {5355} (\bibinfo {year} {1998})}\BibitemShut
  {NoStop}%
\bibitem [{\citenamefont {Santoro}\ \emph {et~al.}(2002)\citenamefont
  {Santoro}, \citenamefont {Marto\v{n}\'{a}k}, \citenamefont {Tosatti},\ and\
  \citenamefont {Car}}]{Santoro}%
  \BibitemOpen
  \bibfield  {author} {\bibinfo {author} {\bibfnamefont {G.~E.}\ \bibnamefont
  {Santoro}}, \bibinfo {author} {\bibfnamefont {R.}~\bibnamefont
  {Marto\v{n}\'{a}k}}, \bibinfo {author} {\bibfnamefont {E.}~\bibnamefont
  {Tosatti}}, \ and\ \bibinfo {author} {\bibfnamefont {R.}~\bibnamefont
  {Car}},\ }\href {\doibase 10.1126/science.1068774} {\bibfield  {journal}
  {\bibinfo  {journal} {Science}\ }\textbf {\bibinfo {volume} {295}},\ \bibinfo
  {pages} {2427} (\bibinfo {year} {2002})}\BibitemShut {NoStop}%
\bibitem [{\citenamefont {Das}\ and\ \citenamefont
  {Chakrabarti}(2008)}]{RevModPhys.80.1061}%
  \BibitemOpen
  \bibfield  {author} {\bibinfo {author} {\bibfnamefont {A.}~\bibnamefont
  {Das}}\ and\ \bibinfo {author} {\bibfnamefont {B.~K.}\ \bibnamefont
  {Chakrabarti}},\ }\href {\doibase 10.1103/RevModPhys.80.1061} {\bibfield
  {journal} {\bibinfo  {journal} {Rev. Mod. Phys.}\ }\textbf {\bibinfo {volume}
  {80}},\ \bibinfo {pages} {1061} (\bibinfo {year} {2008})}\BibitemShut
  {NoStop}%
\bibitem [{\citenamefont {Young}\ \emph {et~al.}(2008)\citenamefont {Young},
  \citenamefont {Knysh},\ and\ \citenamefont
  {Smelyanskiy}}]{PhysRevLett.101.170503}%
  \BibitemOpen
  \bibfield  {author} {\bibinfo {author} {\bibfnamefont {A.~P.}\ \bibnamefont
  {Young}}, \bibinfo {author} {\bibfnamefont {S.}~\bibnamefont {Knysh}}, \ and\
  \bibinfo {author} {\bibfnamefont {V.~N.}\ \bibnamefont {Smelyanskiy}},\
  }\href {\doibase 10.1103/PhysRevLett.101.170503} {\bibfield  {journal}
  {\bibinfo  {journal} {Phys. Rev. Lett.}\ }\textbf {\bibinfo {volume} {101}},\
  \bibinfo {pages} {170503} (\bibinfo {year} {2008})}\BibitemShut {NoStop}%
\bibitem [{\citenamefont {Marto\ifmmode~\check{n}\else \v{n}\fi{}\'ak}\ \emph
  {et~al.}(2002)\citenamefont {Marto\ifmmode~\check{n}\else \v{n}\fi{}\'ak},
  \citenamefont {Santoro},\ and\ \citenamefont {Tosatti}}]{sqa1}%
  \BibitemOpen
  \bibfield  {author} {\bibinfo {author} {\bibfnamefont {R.}~\bibnamefont
  {Marto\ifmmode~\check{n}\else \v{n}\fi{}\'ak}}, \bibinfo {author}
  {\bibfnamefont {G.~E.}\ \bibnamefont {Santoro}}, \ and\ \bibinfo {author}
  {\bibfnamefont {E.}~\bibnamefont {Tosatti}},\ }\href {\doibase
  10.1103/PhysRevB.66.094203} {\bibfield  {journal} {\bibinfo  {journal} {Phys.
  Rev. B}\ }\textbf {\bibinfo {volume} {66}},\ \bibinfo {pages} {094203}
  (\bibinfo {year} {2002})}\BibitemShut {NoStop}%
\bibitem [{\citenamefont {Battaglia}\ \emph {et~al.}(2005)\citenamefont
  {Battaglia}, \citenamefont {Santoro},\ and\ \citenamefont
  {Tosatti}}]{Battaglia:2005fk}%
  \BibitemOpen
  \bibfield  {author} {\bibinfo {author} {\bibfnamefont {D.~A.}\ \bibnamefont
  {Battaglia}}, \bibinfo {author} {\bibfnamefont {G.~E.}\ \bibnamefont
  {Santoro}}, \ and\ \bibinfo {author} {\bibfnamefont {E.}~\bibnamefont
  {Tosatti}},\ }\href {http://link.aps.org/doi/10.1103/PhysRevE.71.066707}
  {\bibfield  {journal} {\bibinfo  {journal} {Phys. Rev. E}\ }\textbf {\bibinfo
  {volume} {71}},\ \bibinfo {pages} {066707} (\bibinfo {year}
  {2005})}\BibitemShut {NoStop}%
\bibitem [{\citenamefont {Morita}\ and\ \citenamefont
  {Nishimori}(2008)}]{morita:125210}%
  \BibitemOpen
  \bibfield  {author} {\bibinfo {author} {\bibfnamefont {S.}~\bibnamefont
  {Morita}}\ and\ \bibinfo {author} {\bibfnamefont {H.}~\bibnamefont
  {Nishimori}},\ }\href {\doibase 10.1063/1.2995837} {\bibfield  {journal}
  {\bibinfo  {journal} {J. Math. Phys.}\ }\textbf {\bibinfo {volume} {49}},\
  \bibinfo {pages} {125210} (\bibinfo {year} {2008})}\BibitemShut {NoStop}%
\bibitem [{\citenamefont {Kirkpatrick}\ \emph {et~al.}(1983)\citenamefont
  {Kirkpatrick}, \citenamefont {Gelatt},\ and\ \citenamefont
  {Vecchi}}]{Kirkpatrick1983}%
  \BibitemOpen
  \bibfield  {author} {\bibinfo {author} {\bibfnamefont {S.}~\bibnamefont
  {Kirkpatrick}}, \bibinfo {author} {\bibfnamefont {C.~D.}\ \bibnamefont
  {Gelatt}}, \ and\ \bibinfo {author} {\bibfnamefont {M.~P.}\ \bibnamefont
  {Vecchi}},\ }\href@noop {} {\bibfield  {journal} {\bibinfo  {journal}
  {Science}\ }\textbf {\bibinfo {volume} {220}},\ \bibinfo {pages} {671}
  (\bibinfo {year} {1983})}\BibitemShut {NoStop}%
\bibitem [{\citenamefont {Mezard}\ and\ \citenamefont
  {Montanari}(2009)}]{mezmont}%
  \BibitemOpen
  \bibfield  {author} {\bibinfo {author} {\bibfnamefont {M.}~\bibnamefont
  {Mezard}}\ and\ \bibinfo {author} {\bibfnamefont {A.}~\bibnamefont
  {Montanari}},\ }\href@noop {} {\emph {\bibinfo {title} {Information, Physics,
  and Computation}}}\ (\bibinfo  {publisher} {Oxford University Press},\
  \bibinfo {year} {2009})\BibitemShut {NoStop}%
\bibitem [{\citenamefont {Coppersmith}\ \emph {et~al.}(2004)\citenamefont
  {Coppersmith}, \citenamefont {Gamarnik}, \citenamefont {Hajiaghayi},\ and\
  \citenamefont {Sorkin}}]{Coppersmith}%
  \BibitemOpen
  \bibfield  {author} {\bibinfo {author} {\bibfnamefont {D.}~\bibnamefont
  {Coppersmith}}, \bibinfo {author} {\bibfnamefont {D.}~\bibnamefont
  {Gamarnik}}, \bibinfo {author} {\bibfnamefont {M.}~\bibnamefont
  {Hajiaghayi}}, \ and\ \bibinfo {author} {\bibfnamefont {G.~B.}\ \bibnamefont
  {Sorkin}},\ }\href {\doibase 10.1002/rsa.20015} {\bibfield  {journal}
  {\bibinfo  {journal} {Random Structures and Algorithms}\ }\textbf {\bibinfo
  {volume} {24}},\ \bibinfo {pages} {502} (\bibinfo {year} {2004})}\BibitemShut
  {NoStop}%
\bibitem [{\citenamefont {de~la Vega}(2001)}]{FernandezdelaVega2001131}%
  \BibitemOpen
  \bibfield  {author} {\bibinfo {author} {\bibfnamefont {W.~F.}\ \bibnamefont
  {de~la Vega}},\ }\href {\doibase 10.1016/S0304-3975(01)00156-6} {\bibfield
  {journal} {\bibinfo  {journal} {Theoretical Computer Science}\ }\textbf
  {\bibinfo {volume} {265}},\ \bibinfo {pages} {131 } (\bibinfo {year}
  {2001})}\BibitemShut {NoStop}%
\bibitem [{\citenamefont {McGeoch}\ and\ \citenamefont {Wang}(2013)}]{McGeoch}%
  \BibitemOpen
  \bibfield  {author} {\bibinfo {author} {\bibfnamefont {C.~C.}\ \bibnamefont
  {McGeoch}}\ and\ \bibinfo {author} {\bibfnamefont {C.}~\bibnamefont {Wang}},\
  }in\ \href@noop {} {\emph {\bibinfo {booktitle} {Proceedings of the 2013 ACM
  Conference on Computing Frontiers}}}\ (\bibinfo {year} {2013})\BibitemShut
  {NoStop}%
\bibitem [{Note1()}]{Note1}%
  \BibitemOpen
  \bibinfo {note} {In more detail, the McGeoch and Wang (MW) study \cite
  {McGeoch}, working with the DW2, used a $439$ qubit subgraph of Chimera and
  considered three problems: (1) Chimera-structured QUBO instances (this is
  actually an ensemble of uniform samples from the Ising model on Chimera with
  $J_{ij},h_j \in \protect \{-1,1\protect \}$ \cite {Selby}), (2) Weighted Max
  2-SAT, (3) the Quadratic Assignment Problem. Their main conclusion is that in
  their experiments the DW2 (together with a software layer called Blackbox)
  outperformed the software against which it was tested. In the case of problem
  (1), the DW2 is reported as outperforming its nearest rival (CPLEX), amongst
  those tried, by a factor of $3600$. The times recorded by MW were for the
  first point that CPLEX found the optimal solution, and not the time at which
  it proved it optimality. However, several researchers have reported classical
  implementations for all three problems which outperform the DW2 and in
  particular the MW benchmarks \cite {Selby,Puget}. Our ensemble of MAX 2-SAT
  problems differs from the weighted MAX 2-SAT problems considered by MW, since
  we used uniform weights with the additional constraint of a fixed clause
  density. In addition, unlike MW's case (2), our MAX 2-SAT problem ensemble
  inherits the native Chimera graph structure along with the connectivity
  contraints of the processor by design, which eliminates the need for using
  the Blackbox layer (that uses conventional heuristics along with hardware
  queries), resulting in a more transparent comparison. While we do not tune
  the implementation of \protect \texttt {akmaxsat} to the structure of
  processor-compatible problems, this lets us compare its performance on truly
  random and processor-compatible problems without any bias. Comparison of our
  results to the native (Chimera-structured) case (1) studied by MW is not
  straightforward because the resulting Ising ensemble does not correspond to
  MAX 2-SAT with a fixed clause density. Moreover, our focus is not on the
  absolute time to solution (which is somewhat arbitrary anyway since it
  depends on variables such as the type of processor, its clock speed, and the
  number of cores), but rather on the scaling with problem size.}\BibitemShut
  {Stop}%
\bibitem [{\citenamefont {Jordan}\ \emph {et~al.}(2006)\citenamefont {Jordan},
  \citenamefont {Farhi},\ and\ \citenamefont {Shor}}]{jordan2006error}%
  \BibitemOpen
  \bibfield  {author} {\bibinfo {author} {\bibfnamefont {S.~P.}\ \bibnamefont
  {Jordan}}, \bibinfo {author} {\bibfnamefont {E.}~\bibnamefont {Farhi}}, \
  and\ \bibinfo {author} {\bibfnamefont {P.~W.}\ \bibnamefont {Shor}},\ }\href
  {\doibase 10.1103/PhysRevA.74.052322} {\bibfield  {journal} {\bibinfo
  {journal} {Phys. Rev. A}\ }\textbf {\bibinfo {volume} {74}},\ \bibinfo
  {pages} {052322} (\bibinfo {year} {2006})}\BibitemShut {NoStop}%
\bibitem [{\citenamefont {Lidar}(2008)}]{PhysRevLett.100.160506}%
  \BibitemOpen
  \bibfield  {author} {\bibinfo {author} {\bibfnamefont {D.~A.}\ \bibnamefont
  {Lidar}},\ }\href {\doibase 10.1103/PhysRevLett.100.160506} {\bibfield
  {journal} {\bibinfo  {journal} {Phys. Rev. Lett.}\ }\textbf {\bibinfo
  {volume} {100}},\ \bibinfo {pages} {160506} (\bibinfo {year}
  {2008})}\BibitemShut {NoStop}%
\bibitem [{\citenamefont {Quiroz}\ and\ \citenamefont
  {Lidar}(2012)}]{PhysRevA.86.042333}%
  \BibitemOpen
  \bibfield  {author} {\bibinfo {author} {\bibfnamefont {G.}~\bibnamefont
  {Quiroz}}\ and\ \bibinfo {author} {\bibfnamefont {D.~A.}\ \bibnamefont
  {Lidar}},\ }\href {\doibase 10.1103/PhysRevA.86.042333} {\bibfield  {journal}
  {\bibinfo  {journal} {Phys. Rev. A}\ }\textbf {\bibinfo {volume} {86}},\
  \bibinfo {pages} {042333} (\bibinfo {year} {2012})}\BibitemShut {NoStop}%
\bibitem [{\citenamefont {Paz-Silva}\ \emph {et~al.}(2012)\citenamefont
  {Paz-Silva}, \citenamefont {Rezakhani}, \citenamefont {Dominy},\ and\
  \citenamefont {Lidar}}]{PhysRevLett.108.080501}%
  \BibitemOpen
  \bibfield  {author} {\bibinfo {author} {\bibfnamefont {G.~A.}\ \bibnamefont
  {Paz-Silva}}, \bibinfo {author} {\bibfnamefont {A.~T.}\ \bibnamefont
  {Rezakhani}}, \bibinfo {author} {\bibfnamefont {J.~M.}\ \bibnamefont
  {Dominy}}, \ and\ \bibinfo {author} {\bibfnamefont {D.~A.}\ \bibnamefont
  {Lidar}},\ }\href {\doibase 10.1103/PhysRevLett.108.080501} {\bibfield
  {journal} {\bibinfo  {journal} {Phys. Rev. Lett.}\ }\textbf {\bibinfo
  {volume} {108}},\ \bibinfo {pages} {080501} (\bibinfo {year}
  {2012})}\BibitemShut {NoStop}%
\bibitem [{\citenamefont {{Young}}\ and\ \citenamefont
  {{Sarovar}}(2012)}]{2012arXiv1208.6371Y}%
  \BibitemOpen
  \bibfield  {author} {\bibinfo {author} {\bibfnamefont {K.~C.}\ \bibnamefont
  {{Young}}}\ and\ \bibinfo {author} {\bibfnamefont {M.}~\bibnamefont
  {{Sarovar}}},\ }\href@noop {} {\bibfield  {journal} {\bibinfo  {journal}
  {ArXiv e-prints}\ } (\bibinfo {year} {2012})},\ \Eprint
  {http://arxiv.org/abs/1208.6371} {arXiv:1208.6371 [quant-ph]} \BibitemShut
  {NoStop}%
\bibitem [{\citenamefont {Kuegel}(2012)}]{akmaxsat}%
  \BibitemOpen
  \bibfield  {author} {\bibinfo {author} {\bibfnamefont {A.}~\bibnamefont
  {Kuegel}},\ }in\ \href@noop {} {\emph {\bibinfo {booktitle} {POS-10}}},\
  \bibinfo {series} {EPiC Series}, Vol.~\bibinfo {volume} {8},\ \bibinfo
  {editor} {edited by\ \bibinfo {editor} {\bibfnamefont {D.~L.}\ \bibnamefont
  {Berre}}}\ (\bibinfo  {publisher} {EasyChair},\ \bibinfo {year} {2012})\ pp.\
  \bibinfo {pages} {15--27}\BibitemShut {NoStop}%
\bibitem [{\citenamefont {Choi}(2008)}]{Choi1}%
  \BibitemOpen
  \bibfield  {author} {\bibinfo {author} {\bibfnamefont {V.}~\bibnamefont
  {Choi}},\ }\href {\doibase 10.1007/s11128-008-0082-9} {\bibfield  {journal}
  {\bibinfo  {journal} {Quant. Inf. Proc.}\ }\textbf {\bibinfo {volume} {7}},\
  \bibinfo {pages} {193} (\bibinfo {year} {2008})}\BibitemShut {NoStop}%
\bibitem [{\citenamefont {Choi}(2011)}]{Choi2}%
  \BibitemOpen
  \bibfield  {author} {\bibinfo {author} {\bibfnamefont {V.}~\bibnamefont
  {Choi}},\ }\href {\doibase 10.1007/s11128-010-0200-3} {\bibfield  {journal}
  {\bibinfo  {journal} {Quant. Inf. Proc.}\ }\textbf {\bibinfo {volume} {10}},\
  \bibinfo {pages} {343} (\bibinfo {year} {2011})}\BibitemShut {NoStop}%
\bibitem [{Note2()}]{Note2}%
  \BibitemOpen
  \bibinfo {note} {Specifically, the biqmac algorithm \cite {biqmac} used in
  the spin glass server \cite {sgserver}, exact belief propagation using bucket
  sort \cite {dechter1999bucket} and a related divide-and-conquer
  algorithm.}\BibitemShut {Stop}%
\bibitem [{\citenamefont {Aspvall}\ \emph {et~al.}(1979)\citenamefont
  {Aspvall}, \citenamefont {Plass},\ and\ \citenamefont {Tarjan}}]{lin-2SAT}%
  \BibitemOpen
  \bibfield  {author} {\bibinfo {author} {\bibfnamefont {B.}~\bibnamefont
  {Aspvall}}, \bibinfo {author} {\bibfnamefont {M.~F.}\ \bibnamefont {Plass}},
  \ and\ \bibinfo {author} {\bibfnamefont {R.~E.}\ \bibnamefont {Tarjan}},\
  }\href {\doibase 10.1016/0020-0190(79)90002-4} {\bibfield  {journal}
  {\bibinfo  {journal} {Inf. Proc. Lett.}\ }\textbf {\bibinfo {volume} {8}},\
  \bibinfo {pages} {121 } (\bibinfo {year} {1979})}\BibitemShut {NoStop}%
\bibitem [{\citenamefont {Papadimitriou}(1995)}]{Papadimitriou:book}%
  \BibitemOpen
  \bibfield  {author} {\bibinfo {author} {\bibfnamefont {C.}~\bibnamefont
  {Papadimitriou}},\ }\href@noop {} {\emph {\bibinfo {title} {Computational
  Complexity}}}\ (\bibinfo  {publisher} {Addison Wesley Longman},\ \bibinfo
  {address} {Reading, Massachusetts},\ \bibinfo {year} {1995})\BibitemShut
  {NoStop}%
\bibitem [{\citenamefont {Bollob\'{a}s}\ \emph {et~al.}(2001)\citenamefont
  {Bollob\'{a}s}, \citenamefont {Borgs}, \citenamefont {Chayes}, \citenamefont
  {Kim},\ and\ \citenamefont {Wilson}}]{Bollobas:2001}%
  \BibitemOpen
  \bibfield  {author} {\bibinfo {author} {\bibfnamefont {B.}~\bibnamefont
  {Bollob\'{a}s}}, \bibinfo {author} {\bibfnamefont {C.}~\bibnamefont {Borgs}},
  \bibinfo {author} {\bibfnamefont {J.~T.}\ \bibnamefont {Chayes}}, \bibinfo
  {author} {\bibfnamefont {J.~H.}\ \bibnamefont {Kim}}, \ and\ \bibinfo
  {author} {\bibfnamefont {D.~B.}\ \bibnamefont {Wilson}},\ }\href {\doibase
  10.1002/rsa.1006} {\bibfield  {journal} {\bibinfo  {journal} {Random Struct.
  Algorithms}\ }\textbf {\bibinfo {volume} {18}},\ \bibinfo {pages} {201}
  (\bibinfo {year} {2001})}\BibitemShut {NoStop}%
\bibitem [{\citenamefont {Goerdt}(1996)}]{Goerdt1996469}%
  \BibitemOpen
  \bibfield  {author} {\bibinfo {author} {\bibfnamefont {A.}~\bibnamefont
  {Goerdt}},\ }\href {\doibase 10.1006/jcss.1996.0081} {\bibfield  {journal}
  {\bibinfo  {journal} {J. of Computer and System Sciences}\ }\textbf {\bibinfo
  {volume} {53}},\ \bibinfo {pages} {469} (\bibinfo {year} {1996})}\BibitemShut
  {NoStop}%
\bibitem [{\citenamefont {Monasson}\ \emph {et~al.}(1999)\citenamefont
  {Monasson}, \citenamefont {Zecchina}, \citenamefont {Kirkpatrick},
  \citenamefont {Selman},\ and\ \citenamefont {Troyansky}}]{monasson}%
  \BibitemOpen
  \bibfield  {author} {\bibinfo {author} {\bibfnamefont {R.}~\bibnamefont
  {Monasson}}, \bibinfo {author} {\bibfnamefont {R.}~\bibnamefont {Zecchina}},
  \bibinfo {author} {\bibfnamefont {S.}~\bibnamefont {Kirkpatrick}}, \bibinfo
  {author} {\bibfnamefont {B.}~\bibnamefont {Selman}}, \ and\ \bibinfo {author}
  {\bibfnamefont {L.}~\bibnamefont {Troyansky}},\ }\href
  {http://www.nature.com/nature/journal/v400/n6740/full/400133a0.html}
  {\bibfield  {journal} {\bibinfo  {journal} {Nature}\ }\textbf {\bibinfo
  {volume} {400}},\ \bibinfo {pages} {133} (\bibinfo {year}
  {1999})}\BibitemShut {NoStop}%
\bibitem [{\citenamefont {Shen}\ and\ \citenamefont
  {Zhag}(2003)}]{shen2003empirical}%
  \BibitemOpen
  \bibfield  {author} {\bibinfo {author} {\bibfnamefont {H.}~\bibnamefont
  {Shen}}\ and\ \bibinfo {author} {\bibfnamefont {H.}~\bibnamefont {Zhag}},\
  }\href {http://www.sciencedirect.com/science/article/pii/S1571065304004640}
  {\bibfield  {journal} {\bibinfo  {journal} {Electronic Notes in Discrete
  Mathematics}\ }\textbf {\bibinfo {volume} {16}},\ \bibinfo {pages} {80}
  (\bibinfo {year} {2003})}\BibitemShut {NoStop}%
\bibitem [{\citenamefont {Goemans}\ and\ \citenamefont
  {Williamson}(1995)}]{maxcut}%
  \BibitemOpen
  \bibfield  {author} {\bibinfo {author} {\bibfnamefont {M.}~\bibnamefont
  {Goemans}}\ and\ \bibinfo {author} {\bibfnamefont {D.}~\bibnamefont
  {Williamson}},\ }\href {http://dl.acm.org/citation.cfm?doid=227683.227684}
  {\bibfield  {journal} {\bibinfo  {journal} {J. Assoc. Comput. Mach.}\
  }\textbf {\bibinfo {volume} {42}},\ \bibinfo {pages} {1115} (\bibinfo {year}
  {1995})}\BibitemShut {NoStop}%
\bibitem [{\citenamefont {Lewin}\ \emph {et~al.}(2006)\citenamefont {Lewin},
  \citenamefont {Livnat},\ and\ \citenamefont {Zwick}}]{lewin}%
  \BibitemOpen
  \bibfield  {author} {\bibinfo {author} {\bibfnamefont {M.}~\bibnamefont
  {Lewin}}, \bibinfo {author} {\bibfnamefont {D.}~\bibnamefont {Livnat}}, \
  and\ \bibinfo {author} {\bibfnamefont {U.}~\bibnamefont {Zwick}},\ }\href
  {http://dl.acm.org/citation.cfm?id=660076&CFID=204379201&CFTOKEN=89090407}
  {\bibfield  {journal} {\bibinfo  {journal} {Integer Programming and
  Combinatorial Optimization}\ ,\ \bibinfo {pages} {67}} (\bibinfo {year}
  {2006})}\BibitemShut {NoStop}%
\bibitem [{\citenamefont {H\`{a}stad}(2001)}]{inapprox}%
  \BibitemOpen
  \bibfield  {author} {\bibinfo {author} {\bibfnamefont {J.}~\bibnamefont
  {H\`{a}stad}},\ }\href {http://dl.acm.org/citation.cfm?id=502098} {\bibfield
  {journal} {\bibinfo  {journal} {Journal of the ACM (JACM)}\ }\textbf
  {\bibinfo {volume} {48}},\ \bibinfo {pages} {798} (\bibinfo {year}
  {2001})}\BibitemShut {NoStop}%
\bibitem{maxsatcomp}%
  \BibitemOpen
  \href {http://maxsat.ia.udl.cat/introduction/} {{\bibinfo {title}
  {{Annual Max-SAT Evaluations}}}} (\bibinfo {year} {2013})\BibitemShut
  {NoStop}%
\bibitem [{\citenamefont {Davis}\ and\ \citenamefont {Putnam}(1960)}]{dpl}%
  \BibitemOpen
  \bibfield  {author} {\bibinfo {author} {\bibfnamefont {M.}~\bibnamefont
  {Davis}}\ and\ \bibinfo {author} {\bibfnamefont {H.}~\bibnamefont {Putnam}},\
  }\href {\doibase 10.1145/321033.321034} {\bibfield  {journal} {\bibinfo
  {journal} {J. ACM}\ }\textbf {\bibinfo {volume} {7}},\ \bibinfo {pages} {201}
  (\bibinfo {year} {1960})}\BibitemShut {NoStop}%
\bibitem [{\citenamefont {Feigenbaum}\ and\ \citenamefont
  {Fortnow}(1993)}]{Feigenbaum:1993}%
  \BibitemOpen
  \bibfield  {author} {\bibinfo {author} {\bibfnamefont {J.}~\bibnamefont
  {Feigenbaum}}\ and\ \bibinfo {author} {\bibfnamefont {L.}~\bibnamefont
  {Fortnow}},\ }\href {\doibase 10.1137/0222061} {\bibfield  {journal}
  {\bibinfo  {journal} {SIAM Journal on Computing}\ }\textbf {\bibinfo {volume}
  {22}},\ \bibinfo {pages} {994} (\bibinfo {year} {1993})}\BibitemShut
  {NoStop}%
\bibitem [{\citenamefont {Ajtai}(1999)}]{Ajtai:1999}%
  \BibitemOpen
  \bibfield  {author} {\bibinfo {author} {\bibfnamefont {M.}~\bibnamefont
  {Ajtai}},\ }in\ \href {http://dl.acm.org/citation.cfm?id=646229.681554}
  {\emph {\bibinfo {booktitle} {Proceedings of the 26th International
  Colloquium on Automata, Languages and Programming}}},\ \bibinfo {series and
  number} {ICAL '99}\ (\bibinfo  {publisher} {Springer-Verlag},\ \bibinfo
  {address} {London, UK, UK},\ \bibinfo {year} {1999})\ pp.\ \bibinfo {pages}
  {1--9}\BibitemShut {NoStop}%
\bibitem [{\citenamefont {Austrin}(2007)}]{Austrin:2007:BMM:1250790.1250818}%
  \BibitemOpen
  \bibfield  {author} {\bibinfo {author} {\bibfnamefont {P.}~\bibnamefont
  {Austrin}},\ }in\ \href {\doibase 10.1145/1250790.1250818} {\emph {\bibinfo
  {booktitle} {Proceedings of the thirty-ninth annual ACM symposium on Theory
  of computing}}},\ \bibinfo {series and number} {STOC '07}\ (\bibinfo
  {publisher} {ACM},\ \bibinfo {address} {New York, NY, USA},\ \bibinfo {year}
  {2007})\ pp.\ \bibinfo {pages} {189--197}\BibitemShut {NoStop}%
\bibitem [{\citenamefont {Albash}\ \emph {et~al.}(2012)\citenamefont {Albash},
  \citenamefont {Boixo}, \citenamefont {Lidar},\ and\ \citenamefont
  {Zanardi}}]{Albash:12}%
  \BibitemOpen
  \bibfield  {author} {\bibinfo {author} {\bibfnamefont {T.}~\bibnamefont
  {Albash}}, \bibinfo {author} {\bibfnamefont {S.}~\bibnamefont {Boixo}},
  \bibinfo {author} {\bibfnamefont {D.~A.}\ \bibnamefont {Lidar}}, \ and\
  \bibinfo {author} {\bibfnamefont {P.}~\bibnamefont {Zanardi}},\ }\href
  {http://stacks.iop.org/1367-2630/14/i=12/a=123016} {\bibfield  {journal}
  {\bibinfo  {journal} {New J. of Phys.}\ }\textbf {\bibinfo {volume} {14}},\
  \bibinfo {pages} {123016} (\bibinfo {year} {2012})}\BibitemShut {NoStop}%
\bibitem [{\citenamefont {Amin}\ \emph {et~al.}(2013)\citenamefont {Amin},
  \citenamefont {Dickson},\ and\ \citenamefont {Smith}}]{Amin:13}%
  \BibitemOpen
  \bibfield  {author} {\bibinfo {author} {\bibfnamefont {M.~H.}\ \bibnamefont
  {Amin}}, \bibinfo {author} {\bibfnamefont {N.~G.}\ \bibnamefont {Dickson}}, \
  and\ \bibinfo {author} {\bibfnamefont {P.}~\bibnamefont {Smith}},\ }\href
  {\doibase 10.1007/s11128-012-0480-x} {\bibfield  {journal} {\bibinfo
  {journal} {Quantum Inf. Proc.}\ }\textbf {\bibinfo {volume} {12}},\ \bibinfo
  {pages} {1819} (\bibinfo {year} {2013})}\BibitemShut {NoStop}%
\bibitem [{Note3()}]{Note3}%
  \BibitemOpen
  \bibinfo {note} {We note that Ref.~\cite {q100} found that for an ensemble of
  Ising spin-glass problems the histogram of success probabilities solved using
  the DW1 was bimodal. This result agreed with predictions of a simulated
  quantum annealer, but differed from that of classical simulated annealing.
  The difference is most likely primarily due to the different annealing time,
  $5\protect \tmspace +\thinmuskip {.1667em} \mu $s in Ref.~\cite {q100}
  compared to our $1$ms. While we did not vary the annealing time in the
  present study, Ref.~\cite {q100} observed that a longer annealing time
  resulted in increasingly large fraction of problems being solved with high
  probability, thus shifting some of the weight of the distribution from the
  `hard' (low success probability) problems to the `easy' ones. Moreover, the
  ensemble of Ising spin glass problems considered in Ref.~\cite {q100}
  differed from ours in several important ways. Their coupling strengths were
  randomly allowed to be only $\pm 1$ with no fractional couplings. Another
  difference was that the couplings between two neighboring spins were
  completely independent of the local fields, unlike our Eq.~\protect \textup
  {\hbox {\mathsurround \z@ \protect \normalfont (\ignorespaces \ref
  {eq:handJ}\unskip \@@italiccorr )}}.}\BibitemShut {Stop}%
\bibitem [{\citenamefont {Berkley}\ \emph {et~al.}(2010)\citenamefont
  {Berkley}, \citenamefont {Johnson}, \citenamefont {Bunyk}, \citenamefont
  {Harris}, \citenamefont {Johansson}, \citenamefont {Lanting}, \citenamefont
  {Ladizinsky}, \citenamefont {Tolkacheva}, \citenamefont {Amin},\ and\
  \citenamefont {Rose}}]{Berkley:2010zr}%
  \BibitemOpen
  \bibfield  {author} {\bibinfo {author} {\bibfnamefont {A.~J.}\ \bibnamefont
  {Berkley}}, \bibinfo {author} {\bibfnamefont {M.~W.}\ \bibnamefont
  {Johnson}}, \bibinfo {author} {\bibfnamefont {P.}~\bibnamefont {Bunyk}},
  \bibinfo {author} {\bibfnamefont {R.}~\bibnamefont {Harris}}, \bibinfo
  {author} {\bibfnamefont {J.}~\bibnamefont {Johansson}}, \bibinfo {author}
  {\bibfnamefont {T.}~\bibnamefont {Lanting}}, \bibinfo {author} {\bibfnamefont
  {E.}~\bibnamefont {Ladizinsky}}, \bibinfo {author} {\bibfnamefont
  {E.}~\bibnamefont {Tolkacheva}}, \bibinfo {author} {\bibfnamefont {M.~H.~S.}\
  \bibnamefont {Amin}}, \ and\ \bibinfo {author} {\bibfnamefont
  {G.}~\bibnamefont {Rose}},\ }\href {\doibase 10.1088/0953-2048/23/10/105014}
  {\bibfield  {journal} {\bibinfo  {journal} {Superconductor Science and
  Technology}\ }\textbf {\bibinfo {volume} {23}},\ \bibinfo {pages} {105014}
  (\bibinfo {year} {2010})}\BibitemShut {NoStop}%
\bibitem [{\citenamefont {Johnson}\ \emph {et~al.}(2010)\citenamefont
  {Johnson}, \citenamefont {Bunyk}, \citenamefont {Maibaum}, \citenamefont
  {Tolkacheva}, \citenamefont {Berkley}, \citenamefont {Chapple}, \citenamefont
  {Harris}, \citenamefont {Johansson}, \citenamefont {Lanting}, \citenamefont
  {Perminov}, \citenamefont {Ladizinsky}, \citenamefont {Oh},\ and\
  \citenamefont {Rose}}]{Johnson:2010ys}%
  \BibitemOpen
  \bibfield  {author} {\bibinfo {author} {\bibfnamefont {M.~W.}\ \bibnamefont
  {Johnson}}, \bibinfo {author} {\bibfnamefont {P.}~\bibnamefont {Bunyk}},
  \bibinfo {author} {\bibfnamefont {F.}~\bibnamefont {Maibaum}}, \bibinfo
  {author} {\bibfnamefont {E.}~\bibnamefont {Tolkacheva}}, \bibinfo {author}
  {\bibfnamefont {A.~J.}\ \bibnamefont {Berkley}}, \bibinfo {author}
  {\bibfnamefont {E.~M.}\ \bibnamefont {Chapple}}, \bibinfo {author}
  {\bibfnamefont {R.}~\bibnamefont {Harris}}, \bibinfo {author} {\bibfnamefont
  {J.}~\bibnamefont {Johansson}}, \bibinfo {author} {\bibfnamefont
  {T.}~\bibnamefont {Lanting}}, \bibinfo {author} {\bibfnamefont
  {I.}~\bibnamefont {Perminov}}, \bibinfo {author} {\bibfnamefont
  {E.}~\bibnamefont {Ladizinsky}}, \bibinfo {author} {\bibfnamefont
  {T.}~\bibnamefont {Oh}}, \ and\ \bibinfo {author} {\bibfnamefont
  {G.}~\bibnamefont {Rose}},\ }\href {\doibase 10.1088/0953-2048/23/6/065004}
  {\bibfield  {journal} {\bibinfo  {journal} {Superconductor Science and
  Technology}\ }\textbf {\bibinfo {volume} {23}},\ \bibinfo {pages} {065004}
  (\bibinfo {year} {2010})}\BibitemShut {NoStop}%
\bibitem [{\citenamefont {Lanting}(2013)}]{Trevor}%
  \BibitemOpen
  \bibfield  {author} {\bibinfo {author} {\bibfnamefont {T.}~\bibnamefont
  {Lanting}},\ }\href@noop {} {}\bibinfo {howpublished} {D-Wave Inc., private
  communications} (\bibinfo {year} {2013})\BibitemShut {NoStop}%
\bibitem [{\citenamefont {Selby}(2013)}]{Selby}%
  \BibitemOpen
  \bibfield  {author} {\bibinfo {author} {\bibfnamefont {A.}~\bibnamefont
  {Selby}},\ }\href
  {www.archduke.org/stuff/d-wave-comment-on-comparison-with-classical-computers/}
  {\enquote {\bibinfo {title} {{D-Wave: comment on comparison with classical
  computers}}}} (\bibinfo {year} {2013})\BibitemShut {NoStop}%
\bibitem [{\citenamefont {Puget}(2013)}]{Puget}%
  \BibitemOpen
  \bibfield  {author} {\bibinfo {author} {\bibfnamefont {J.-F.}\ \bibnamefont
  {Puget}},\ }\href
  {https://www.ibm.com/developerworks/community/blogs/jfp/entry/d_wave_vs_cplex_comparison_part_2_qubo}
  {\enquote {\bibinfo {title} {{{D-Wave} vs {CPLEX} Comparison. Part 2:
  QUBO}}}\ } (\bibinfo {year} {2013})\BibitemShut {NoStop}%
\bibitem [{\citenamefont {Rendl}\ \emph {et~al.}(2010)\citenamefont {Rendl},
  \citenamefont {Rinaldi},\ and\ \citenamefont {Wiegele}}]{biqmac}%
  \BibitemOpen
  \bibfield  {author} {\bibinfo {author} {\bibfnamefont {F.}~\bibnamefont
  {Rendl}}, \bibinfo {author} {\bibfnamefont {G.}~\bibnamefont {Rinaldi}}, \
  and\ \bibinfo {author} {\bibfnamefont {A.}~\bibnamefont {Wiegele}},\ }\href
  {\doibase 10.1007/s10107-008-0235-8} {\bibfield  {journal} {\bibinfo
  {journal} {Mathematical Programming}\ }\textbf {\bibinfo {volume} {121}},\
  \bibinfo {pages} {307} (\bibinfo {year} {2010})}\BibitemShut {NoStop}%
\bibitem{sgserver}%
  \BibitemOpen
  \href {http://www.informatik.uni-koeln.de/spinglass/} { {\bibinfo
  {title} {Spin glass server}}}\BibitemShut {NoStop}%
\bibitem [{\citenamefont {Dechter}(1999)}]{dechter1999bucket}%
  \BibitemOpen
  \bibfield  {author} {\bibinfo {author} {\bibfnamefont {R.}~\bibnamefont
  {Dechter}},\ }\href {\doibase
  http://dx.doi.org/10.1016/S0004-3702(99)00059-4} {\bibfield  {journal}
  {\bibinfo  {journal} {Artificial Intelligence}\ }\textbf {\bibinfo {volume}
  {113}},\ \bibinfo {pages} {41} (\bibinfo {year} {1999})}\BibitemShut
  {NoStop}%
\end{thebibliography}

%

\end{document}